\documentclass[11pt]{article} 
\usepackage{amsmath}
\usepackage{amssymb,amsfonts,euscript}
\usepackage{hyperref}
\renewcommand{\baselinestretch}{1.2}
\newcommand\nn{\nonumber}

\newcommand{\Ord}{{\cal{O}}}

\newcommand{\D}{{\cal D}}
\newcommand{\G}{{\cal G}}

\newcommand{\T}{{\cal T}}

\newcommand{\F}{{\cal F}}
\newcommand{\cL}{\cal{L}}
\newcommand{\lr}{{\cal{L}}_{\sgb(\s)} }

\newcommand{\bz}{\bar{z}}

\newcommand{\bM}{\bar{M}}

\def\a{\alpha}
\def\be{\beta}
\def\hb{\hat{\be}}
\def\l{\lambda}
\def\m{\mu}
\def\n{\nu}
\def\r{\rho}
\def\de{\delta}
\def\k{\kappa}

\def\nl{\newline}

\def\hg{\hat{g}}
\def\hgp{\hat{g}_+}
\def\hgm{\hat{g}_-}

\def\part{\partial}
\renewcommand{\sp}{,\hspace{.3in}}

\newcommand{\timestamp}{{\small\vbox{\hbox{\tt\jobname.tex} \today}}}

\newcommand{\sa}{\mathop{\vtop{\ialign{##\crcr
  $\hfil\displaystyle{\longrightarrow}\hfil$\crcr\noalign{\kern-1pt\nointerlineskip}
  \hspace{.12in}$^\sigma$\hskip6pt\crcr\noalign{\kern3pt}}}}}
\newcommand{\slra}{\mathop{\vtop{\ialign{##\crcr
  $\hfil\displaystyle{\longleftrightarrow}\hfil$\crcr\noalign{\kern-1pt\nointerlineskip}
  \hspace{.12in}$^\sigma$\hskip6pt\crcr\noalign{\kern3pt}}}}}
\newcommand{\sat}{\mathop{\vtop{\ialign{##\crcr
  $\hfil\displaystyle{\longrightarrow}\hfil$\crcr\noalign{\kern-1pt\nointerlineskip}
  \hspace{.12in}$^\sigma$\hskip6pt\crcr\noalign{\kern3pt}}}}}
\newcommand{\pa}{\mathop{\vtop{\ialign{##\crcr
  $\hfil\displaystyle{\oplus}\hfil$\crcr\noalign{\kern+1pt\nointerlineskip}
  \hspace{.08in}$^{\alpha=0}$\hskip6pt\crcr\noalign{\kern3pt}}}}}
\newcommand{\pan}{\mathop{\vtop{ialgin{##\crcr
  $\hfil\displaystyle{\oplus}\hfil$\crcr\noaligan{\kern+2pt\nointerlinkeskip}
  \hspace{.03in} $^{\alpha}$\hskip6pt\crcr\noalign{\kern3pit}}}}}
\newcommand{\ka}{\mathop{\vtop{\ialign{##\crcr
  $\hfil\displaystyle{\longleftrightarrow}\hfil$\crcr\noalign{\kern-1pt\nointerlineskip}
  \hspace{.12in}$^K$\hskip6pt\crcr\noalign{\kern3pt}}}}}
\newcommand{\bp}{\mathop{\vtop{ialign{##\crcr
  $\hfil\displaystyle{}\hfil$\crcr\noalign{\kern-13pt\nointerlineskip}
  \big{(}\hskip0pt\crcr\noalign{\kern3pt}}}}}
\newcommand{\cbp}{\mathop{\vtop{ialign{##\crcr
  $\hfil\displaystyle{}\hfil$\crcr\noalign{\kern-13pt\nointerlineskip}
  \big{)}\hskip0pt\crcr\noalign{\kern3pt}}}}}

\newcommand{\+}{\hspace{-.03in}+\hspace{-.02in}}
\newcommand{\s}{\sigma}
\newcommand{\srange}{\sigma=0,...,N_c-1}

\renewcommand{\sp}{,\hspace{.3in}}

\newcommand{\w}{\omega}

\newcommand{\rmod}{\,{\textup{mod}}}

\newcommand{\hc}{$\hat{J}_{\gst}$}

\newcommand{\sgb}{{\mbox{\scriptsize{\gb}}}}
\newcommand{\sgbn}{{\mbox{\scriptsize{\gbn}}}}

\def\gb            {\mbox{$\hat{\mathfrak g}$}}
\def\hb            {\mbox{$\hat{\mathfrak h}$}}
\def\gbn           {\mbox{$\mathfrak g$}}
\def\sm#1      {\mbox{\scriptsize $#1$}}

\def\sz        {\mbox{\scriptsize $\mathbb  Z$}}
\def\z         {\mbox{$\mathbb  Z$}}

\def\d             {\mbox{$\mathbb D$}}

\def\srac#1#2{\smal{\frac{#1}{#2}}}
\def\foot#1{\mbox{\footnotesize $#1$}}

\def\smal#1{\mbox{\small $#1$}}
\def\big#1{\mbox{\large $#1$}}
\def\Big#1{\mbox{\Large $#1$}}
\def\BIG#1{\mbox{\Huge $#1$}}

\def\hsp#1{\hspace{#1in}}

\def\hjb{\hat{\bar{J}}}
\def\comment#1{\hsp{.3}\textup{#1}}
\newcommand{\mnb}{\overline{-n(r)}}
\newcommand{\nb}{\bar n(r)}

\def\hjs{{\hat{j\hspace{.03in}}\hspace{-.03in}}}
\def\hls{\hat{l\hspace{.03in}}}

\def\hms{\hat{m\hspace{.03in}}}

\def\hjbb{ \hat{\bar{J}}^\sharp }

\def\gfrakh{\hat{\mathfrak g}}
\def\su{\mathfrak{su}}
\def\so{\mathfrak{so}}

\newcommand\un{\noindent \underline}
\def\dual{\underset{\s}{\longrightarrow}}

\def\ginv{g^{-1}}
\def\sg{\smal{\EuScript{G}}}
\def\sj{{\cal J}}

\def\hc{^\dagger}
\def\hcj{\dagger}
\def\one{{\mathchoice {\rm 1\mskip-4mu l} {\rm 1\mskip-4mu} {\rm 1\mskip-4.5mu l} {\rm 1\mskip-5mu l}}}
\def\d{\delta}

\def\mmrrs{m+\srac{n(r)}{\r(\s)}}
\def\nnsrs{n+\srac{n(s)}{\r(\s)}}

\def\mmnrrs{{m + \srac{\overline{-n(r)}}{\r (\s)}}}
\def\nrm{{n(r)\m}}
\def\bnrm{{\bar{n}(r)\m}}
\def\mnrm{{-n(r),\m}}

\def\nrn{{n(r)\n}}
\def\mnrn{{-n(r),\n}}
\def\nsn{{n(s)\n}}
\def\bnsn{{\bar{n}(s)\n}}

\def\mnnrnsrsf{{m+n+\frac{n(r)+n(s)}{\r(\s)}}}
\def\mnrrs{{m+\srac{\bar n(r)}{\r(\s)}}}

\def\sG{{\cal G}}
\def\gfrak{\mbox{$\mathfrak g$}}

\def\hj{\hat{J}}
\def\hjh{\hat{J}_{1}}
\def\nrn{{n(r)\n}}
\def\nsn{{n(s)\n}}

\def\schisig{{\foot{\chi(\s)}}}

\def\hc{^\dagger}
\def\st{{\cal T}}
\def\0b{\ }
\def\pl{\partial}

\def\Nrm{{N(r)\m}}

\def\Nsn{{N(s)\n}}

\def\srange{\s=0,\ldots,N_c-1}
\def\sm{{\cal M}}

\def\hG{{\hat G}}
\def\hV{{\hat V}}
\def\hq{{\hat q}}
\def\bhq{\hat{\bar{q}}}
\def\hb{{\hat \beta}}
\def\twcw{\G^{\nrm ; \mnrn} (\s) \srac{\bar{n}(r)}{\r (\s)} \T_\nrm \T_{\mnrn} }
\makeatletter
\renewcommand{\@makefnmark}{\mbox{$^{\ddagger\@thefnmark}$}}
\renewcommand{\subsection}{\@startsection
  {subsection}{2}{0pt
}{-\baselineskip}{0.5\baselineskip}
   {\normalfont\normalsize\bf}}
\renewcommand{\section}{\@startsection
  {section}{2}{0pt
}{-\baselineskip}{0.5\baselineskip}
  {\bf\large}}

\makeatother
\numberwithin{equation}{section}
\numberwithin{table}{section}

\newcommand{\publititle}[8]
{ 
  \vspace*{-3cm}
  \begin{flushright} #1 \\ {\tt #2} \end{flushright}
  \vfill
  \begin{center}{\Large
    \bfseries #3}\end{center}
  \vskip 8mm
  \begin{center}{\large #4}\end{center}
  \begin{center}{\normalsize #5}\end{center}
  \vskip 8mm
  \nopagebreak
  \noindent #6
  \vfill
  \begin{flushleft} #7
  \end{flushleft}
  \hrule width 6.7cm \vskip.1mm
  {\small #8}
  \thispagestyle{empty}
  \clearpage
}

\begin{document}

\publititle{ UCB-PTH-02/08  \\ LBNL-49643 \\ SPIN-2002/07 \\
ITP-UU-02/08 \\ ITFA-2002-06} {hep-th/0203056} {Two Large Examples in Orbifold
Theory: \\ Abelian Orbifolds and the Charge Conjugation Orbifold on $\su(n)$} {
 M.B.Halpern$^{\,a\dagger}$ and N.A.Obers$^{\,b,\,c ,\,d\ddagger{}}$}
 {
$^a$Department of Physics,
     University of California,
     Berkeley, California 94720, USA  \\ {\it and}
Theoretical Physics Group,  Lawrence Berkeley National
Laboratory \\
     University of California,
     Berkeley, California 94720, USA
\\[2mm]
$^b$Spinoza Institute {\it and} Institute for Theoretical Physics
\\ Utrecht University, Leuvenlaan 4, 3584 CE Utrecht, The
Netherlands \\ [2mm]
$^c$Institute for Theoretical Physics, University of Amsterdam \\
 Valckenierstraat 65, 1018 XE Amsterdam, The Netherlands \\ [2mm]
$^d$The Niels Bohr Institute, Blegdamsvej 17, DK-2100 Copenhagen \O,
Denmark
} {Recently the operator algebra and twisted vertex operator equations were
given for each sector of all WZW orbifolds, and a set of twisted KZ
equations for the WZW permutation
orbifolds were worked out as a large example. In this companion paper we
report two further large examples of this development. In the first example we
solve the twisted vertex
operator equations in an abelian limit to obtain the twisted vertex operators
and correlators
of a large class of abelian orbifolds. In the second example, the twisted
vertex operator equations are applied to obtain a set of
twisted KZ equations for the (outer-automorphic) charge conjugation
orbifold on $\su(n \geq 3)$.}
{ $^{\dagger}${\tt halpern@physics.berkeley.edu} \\
$^{\ddagger}${\tt  obers@phys.uu.nl} }

 \clearpage

\renewcommand{\baselinestretch}{.4}\rm
{\footnotesize
\tableofcontents
}

\renewcommand{\baselinestretch}{1.0}\rm

\section{Introduction}

In the last few years there has been a quiet revolution in the
local theory of {\it current-algebraic orbifolds}. Building on the
discovery of orbifold affine algebra
\cite{Borisov:1997nc,Evslin:1999qb} in the cyclic permutation
orbifolds, Refs.~\cite{deBoer:1999na,Halpern:2000vj} gave the
twisted currents and stress tensor in each sector
of any current-algebraic orbifold $A(H)/H$ -- where $A(H)$ is any
current-algebraic conformal field theory [5--9]
with a finite symmetry group $H$. The construction treats all
current-algebraic orbifolds at the same time, using the method
of {\it eigenfields} and the {\it principle of local isomorphisms} to
map OPEs in the symmetric theory to OPEs in the orbifold.
The orbifold results are expressed in terms of a set of
{\it duality transformations}, which are discrete Fourier transforms
constructed from the eigendata of the {\it $H$-eigenvalue problem}.

More recently, the special case of the WZW orbifolds
\begin{equation}
\frac{A_g(H)}{H} \sp H \subset {\rm Aut}(g)
\end{equation}
was worked out in further detail \cite{deBoer:2001nw}, introducing the
{\it linkage relation} and the {\it extended $H$-eigenvalue problem}
in order to include the operator algebra and the {\it twisted
vertex operator equations} of the twisted affine primary fields of the
WZW orbifolds. The twisted vertex operator equations set the stage for
finding the {\it twisted KZ equations} of the WZW orbifolds, and
twisted KZ equations for the WZW permutation orbifolds and the
inner-automorphic WZW orbifolds were worked out as large examples in
Ref.~\cite{deBoer:2001nw}.

In this companion paper, we apply the twisted vertex operator equations
of Ref.~\cite{deBoer:2001nw} to work out the details of two other large
examples. In the first example (see Sec.~\ref{sec2}) we solve the
twisted vertex operator equations in an abelian  limit to obtain the
{\it twisted vertex operators} and correlators of a large class of abelian
orbifolds
\begin{equation}
\frac{A_{{\rm Cartan}\,g}(H)}{H} \sp H \subset {\rm Aut}({\rm Cartan}\,g)
\sp {\rm Cartan}\, g \subset g
\end{equation}
where the ambient algebra $g$ supplies the representation space for the
twisted sectors of each orbifold. In the second example (see Sec.~\ref{sec3}),
we apply the twisted vertex operator equations to obtain a set of
{\it twisted KZ equations} for the (outer-automorphic) charge
conjugation orbifold on $\su (n)$
\begin{equation}
\frac{A_{\su(n)}(\z_2)}{\z_2} \sp n \geq 3
\end{equation}
and some simple solutions of these equations are also discussed.

Subsec.~\ref{kzsec} also notes a {\it more general twisted KZ system}
for the correlators in the ``scalar'' twist-field states of any WZW orbifold.
This result includes as special cases the known twisted KZ system
 for the WZW permutation orbifolds,  our twisted KZ system for the
 charge conjugation orbifold
on $\su (n)$, and a generalization to every outer-automorphic WZW orbifold.

\section{The Abelian Orbifolds  $A_{{\rm Cartan}\,g}(H)/H$ \label{sec2}}

\subsection{An Abelian limit of the WZW orbifolds}

As our first large example, we consider the class of abelian orbifolds
\begin{equation}
\frac{A_{{\rm Cartan}\,g}(H)}{H} \sp H \subset {\rm Aut}({\rm Cartan}\,g)
\end{equation}
where $H$ is any finite group and
$A_{{\rm Cartan}\,g}(H)$ is any $H$-symmetric conformal field theory
constructed from the Cartan subalgebra of a compact semisimple Lie algebra $g$
\begin{equation}
{\rm Cartan}\, g \subset g \sp  g = \oplus_I \gfrak^I \sp
{\rm Cartan}\,g = \oplus_I {\rm Cartan}\, \gfrak^I \  .
\end{equation}
In the composite notation of
Refs.~\cite{Evslin:1999qb,deBoer:1999na,Halpern:2000vj,deBoer:2001nw},
the left-mover sector of the $H$-symmetric CFT is described by the stress
tensor and abelian current algebra
\begin{subequations}
\begin{equation}
T(z) = \frac{1}{2} G^{ab} :
J_a(z) J_b (z): \sp
c = {\rm dim}({\rm Cartan}\,g)
\end{equation}
\begin{equation}
\label{jjalg}
J_a (z) = \sum_{m \in \sz} J_a (m) z^{-m-1} \sp
 [ J_a(m),J_b (n)] =  m G_{ab} \de_{m+n,0}
 \sp a,b = 1 \ldots {\rm dim}({\rm Cartan}\,g)
\end{equation}
\end{subequations}
where $:\cdot :$ is operator product normal ordering and $G_{ab}$ is the
induced metric on ${\rm Cartan}\, g$. The ambient algebra $g$ provides the
induced metric and the representation theory of the symmetric CFT, but is
otherwise inactive. The $H$-symmetry of the system is specified as
\begin{subequations}
\label{HsymCFT}
\begin{equation}
J_a (z)' = \w (h_\s)_a{}^b J_b (z) \sp \w(h_\s)_a{}^c\w(h_\s)_b{}^d
G_{cd} = G_{ab} \sp \w (h_\s) \in H \sp \srange
\end{equation}
\begin{equation}
T(z)'=\frac{1}{2} G^{ab} : J_a (z)' J_b(z)' : \ = T(z)
\end{equation}
\end{subequations}
where $\w (h_\s)$ is the action  of $h_\s \in H$ and $N_c$ is the number of
conjugacy classes of $H$.

When needed, the composite notation can be replaced by the explicit notation
\begin{subequations}
\begin{equation}
a \rightarrow a (I)\sp J_a \rightarrow J_{a(I)} \sp G_{ab} \rightarrow
G_{a(I),b(J)} =\oplus_I k_I \eta^I_{a(I)b(I)} \sp
\end{equation}
\begin{equation}
a(I), b(I) = 1 \ldots
{\rm dim}({\rm Cartan}\,\gfrak^I)
\end{equation}
\end{subequations}
where $\eta^I_{a(I)b(I)}$ is the induced Killing metric of
${\rm Cartan}\,\gfrak^I$ and $x_I = 2k_I/\psi_I^2$
is the invariant level of affine $\gfrak^I$. As an example (see also
Subsec.~\ref{permsec}) permutation-invariant systems satisfy
\begin{subequations}
\begin{equation}
\gfrak^I \simeq \gfrak \sp {\rm Cartan}\,\gfrak^I \simeq {\rm Cartan}\,\gfrak
\end{equation}
\begin{equation}
J_{a(I)} = J_{aI} \sp
k_I =k \sp x_I = x \sp \eta^I_{a(I)b(I)} = \eta_{ab} , \;\;\;
a = 1 \ldots {\rm dim}({\rm Cartan}\,\gfrak)
\end{equation}
\end{subequations}
where $\eta_{ab}$ is the induced Killing metric of ${\rm Cartan}\,\gfrak$
and $H$(permutation) acts among the copies ${\rm Cartan}\,\gfrak^I$
of ${\rm Cartan}\,\gfrak$.

The twisted current-algebraic description of $A_{{\rm Cartan}\,g}(H)/H$
can be easily read as the abelian limit $g \rightarrow {\rm Cartan}\,g$
of the general WZW orbifold in
Refs.~\cite{Halpern:2000vj}  or \cite{deBoer:2001nw}. One finds the stress tensor and twisted
current algebra of sector $\s$
\begin{subequations}
\label{jper0}
\begin{equation}
\hat T_\s (z)  =   \lr^{\nrm ; \mnrn} (\s)
: \hj_{\nrm} (z,\s) \hj_{\mnrn} (z,\s) : \sp \hat c(\s) = c \sp \srange
\end{equation}
\begin{equation}
\label{jper}
\hj_\nrm (z e^{2 \pi i} ,\s) = e^{ -2 \pi i \srac{n(r)}{\r (\s)} }\hj_\nrm (z,\s)
\sp
\hj_{n(r)\pm \r(\s),\m} (z,\s) = \hj_{n(r)\m} (z,\s)
\end{equation}
\begin{equation}
\label{jmodl} \hj_{n(r) \m}(z,\s) = \sum_{m  \in \sz} \hj_{n(r)
\m} ( m + \srac{n(r)}{\rho (\s)} ) z^{-(m+\srac{n(r)}{\rho (\s)})
-1}
\end{equation}
\begin{equation}
\label{malg}
 [\hj_\nrm(\mmrrs),\hj_\nsn(\nnsrs)]=
  (\mmrrs)\de_{\mnnrnsrsf,0}\sG_{\nrm;\mnrn}(\s)
\end{equation}
\begin{equation}
\label{Jper}
\hj_{n(r)\pm \r(\s),\m} (m + \srac{n(r)\pm \r(\s)}{\r(\s)}) =
 \hj_{n(r)\m} (m \pm 1 + \srac{n(r)}{\r (\s)})
\end{equation}
\end{subequations}
where $: \cdot :$ is operator product normal ordering
\cite{Evslin:1999qb,deBoer:1999na,Evslin:1999ve,Halpern:2000vj,deBoer:2001nw} and
the {\it duality transformations}
\cite{deBoer:1999na,Halpern:2000vj,deBoer:2001nw}
\begin{subequations}
\label{Gdual}
\begin{equation}
\label{lab}
\F_{\nrm ; \nsn}{}^{n(t) \de} (\s) =0 \sp
\lr^{\nrm ;\nsn} (\s) = \frac{1}{2} \G^{\nrm ;\nsn} (\s)
\end{equation}
\begin{eqnarray}
\label{sG0}
  \G_{\nrm;\nsn}(\s) & =  & \schisig_\nrm\schisig_\nsn
  U(\s)_\nrm{}^a U(\s)_\nsn{}^b G_{ab} \\
& =  & \d_{n(r) + n(s), 0 \rmod \r(\s)} \G_{\nrm;-n(r),\n} (\s)
 =  \G_{n(r)\pm \r(\s),\m; \nsn}(\s) \label{sG01}
\end{eqnarray}
\begin{eqnarray}
\label{sG1}
 \sG^{\nrm;\nsn}(\s)& = &\schisig^{-1}_\nrm\schisig^{-1}_\nsn  G^{ab}
 U\hc(\s)_a{}^\nrm U\hc(\s)_b{}^\nsn \\
& =  & \d_{n(r) + n(s), 0 \rmod \r(\s)} \G^{\nrm;-n(r),\n} (\s)
 =  \G^{n(r)\pm \r(\s),\m; \nsn} (\s) \label{sG11}
\end{eqnarray}
\end{subequations}
are called respectively the twisted structure constants, the twisted
inverse inertia tensor, the
twisted metric and the twisted inverse metric of sector $\s$. The
Kronecker forms in
\eqref{sG01} and \eqref{sG11} are the solutions of the selection rules
\cite{deBoer:1999na,Halpern:2000vj,deBoer:2001nw} of the twisted tensors.
In further detail, the integer $\r(\s)$ is the order of $h_\s \in H$ while the unitary matrices
$U(\s)$, the spectral integers $n(r) \equiv n(r,\s)$ and the degeneracy
indices $\m \equiv \m (r,\s)$ are determined by solving the
{\it $H$-eigenvalue problem}
\cite{deBoer:1999na,Halpern:2000vj,deBoer:2001nw}
\begin{equation}
\label{Heigvpr}
\w (h_\s) U\hc (\s) = U\hc (\s) E(\s) \sp E(\s) =
e^{-2\pi i \srac{n(r)}{\r(\s)}} \sp \srange \ .
\end{equation}
The quantities $\chi (\s)$ with $\chi (0)=1$ are normalization constants.
As seen in Eqs.~(2.7-9),
all quantities are periodic
$n(r) \rightarrow n(r) \pm \rho (\s)$ in the spectral integers.

We note in particular that the zero modes
$\{ \hj_{0 \m}(0) \}$ of the twisted current algebra \eqref{malg} commute
with themselves and indeed with all the twisted modes
\begin{equation}
[\hj_{0\m}(0),\hj_\nsn(\nnsrs)]= 0 \sp \forall \; \m,\n, n(s) \ .
\end{equation}
As in the case of the WZW orbifolds \cite{deBoer:2001nw} the zero modes of
the twisted current algebra will contribute to  the residual symmetry
 (global Ward identities) of each orbifold sector $\s$
(see Subsecs.~\ref{corsec} and \ref{permsec}).

The ground state (twist-field state) of sector $\s$ satisfies
\begin{equation}
\label{gsc}
\hj_\nrm ( m + \srac{n(r)}{\r (\s)} \geq 0 ) | 0 \rangle_\s
= {}_\s \langle 0 | \hj_\nrm ( m + \srac{n(r)}{\r (\s)} \leq 0 ) = 0 \sp \srange
\end{equation}
and so the $M$ or mode-ordered forms of Refs.~\cite{Halpern:2000vj} or
\cite{deBoer:2001nw} are immediately useful. For example the Virasoro
generators in sector $\s$ of $A_{{\rm Cartan}\,g}(H)/H$
\begin{subequations}
\label{lnll0}
\begin{eqnarray}
L_\s(m) =   \frac{1}{2}\sum_{r,\m,\n} \G^{\nrm; \mnrn }(\s) \!\!\!\!\!\!\!\!\!  & &
\Big\{ \sum_{p \in \sz} :\hat{J}_{\nrm}(p+\srac{n(r)}{\r(\s)})
\hat{J}_{\mnrn} (m-p-\srac{n(r)}{\r(\s)}):_M \nn \\
 & & + \delta_{m,0} \frac{\bar{n}(r)}{2 \r(\s)}\left(1-\frac{\bar{n}(r)}{\r(\s)}
\right)
\G_{\nrm;\mnrn}(\s) \Big\} \label{lnll}
\end{eqnarray}
\begin{equation}
\label{nbdef}
\overline{n(r)} \equiv \bar{n}(r) \equiv n(r) - \r(\s)\lfloor\srac{n(r)}{\r(\s)}\rfloor,\quad \bar{n}(r)
 \in \{0,...,\r(\s)-1\}
 \end{equation}
\end{subequations}
are easily obtained as the abelian limit of the form given for all WZW
orbifolds in these references. Here $\bar{n}(r)$ is the pullback of the
spectral integer $n(r)$ into the fundamental range
$ 0 \leq \bar{n} \leq \r (\s) -1$. Then  the ground state conformal weight
of sector $\s$
\begin{subequations}
\begin{equation}
\Big(L_\s (m \geq 0) -\delta_{m,0} \hat{\Delta}_0 (\s) \Big)| 0 \rangle_\s =
 0
\end{equation}
\begin{eqnarray}
\hat{\Delta}_0 (\s)  & = &
\frac{1}{2} \sum_{r,s,\m,\n} \G^{\nrm ; \nsn} (\s) \G_{\nrm ; \nsn} (\s)
\frac{\nb}{2 \rho (\s)} \left( 1 - \frac{\nb}{\rho (\s)} \right)
 \\
& = & \frac{1}{2} \sum_{r,\m (r)} \frac{\nb}{2 \rho (\s)}
\left( 1 - \frac{\nb}{\rho (\s)} \right) \sp \srange
\label{cw}
\end{eqnarray}
\end{subequations}
is easily computed from \eqref{lnll0} and the duality transformations in \eqref{Gdual}.

Going beyond the general stress tensors of Ref.~\cite{Halpern:2000vj},
Ref.~\cite{deBoer:2001nw} also
discusses the full chiral algebra of all WZW orbifolds, including the twisted
left-mover affine primary fields $\hgp (\T,z,\s)$ of sector $\s$. For our
class of abelian orbifolds the abelian limit $\hgp (\T,z,\s)$ is more
properly called a {\it twisted left-mover vertex operator}. In particular,
we read from Ref.~\cite{deBoer:2001nw} that the twisted left-mover vertex
operators satisfy
\begin{subequations}
\label{sumT}
\begin{equation}
\hj_\nrm  (z,\s) \hgp (\T,w,\s) = \frac{\hgp (\T,w,\s)}{z-w} \T_\nrm (T,\s)
+ \Ord (z-w)^0
\end{equation}
\begin{equation}
\label{hjgpco}
[ \hj_\nrm (m + \srac{n(r)}{\rho(\s)}), \hgp (\T,z,\s)]
= \hgp (\T,z,\s) \T_\nrm (T,\s) z^{m + \srac{n(r)}{\rho(\s)}}
\end{equation}
\begin{equation}
[ L_\s (m), \hgp (\T,z,\s) ] = z^m \Big( z \partial_z + \D (\T,\s)
 (m+1) \Big) \hgp (\T,z,\s)
 \end{equation}
 \begin{equation}
 \D (\T,\s) = \frac{1}{2} \G^{\nrm ; -n (r),\n} (\s)
\T_\nrm (T,\s) \T_{-n(r),\n} (T,\s)
\end{equation}
 \begin{equation}
 \label{Ttw}
 \st_\nrm(T,\s)_\Nrm{}^\Nsn = \schisig_\nrm U(\s)_\nrm{}^a
 \Big{(}\,U(T,\s) T_a U\hc(T,\s)\,\Big{)}{}_\Nrm{}^\Nsn
\end{equation}
\begin{equation}
\label{Tsel}
  = \d_{\frac{n(r)}{\r(\s)}+\srac{N(r)-N(s)}{R (\s)},\,0 \rmod 1 }
  \st_\nrm(T,\s)_\Nrm{}^{N(r)+\frac{R(\s)}{\r(\s)}n(r),\n}
\end{equation}
\begin{equation}
\label{TTcom}
[T_a , T_b]= 0 \sp a ,b = 1 \ldots {\rm dim}({\rm Cartan}\,g)
\end{equation}
\begin{equation}
\label{Tper}
[\T_\nrm (T,\s), \T_\nsn (T,\s)] = 0 \sp \T_{n(r) \pm \r(\s),\m} (T,\s)
= \T_\nrm (T,\s)
\end{equation}
\end{subequations}
where $\{ T_a \}$ are the commuting Cartan matrices of the irreps%
\footnote{Although it is not mentioned explicitly below, one should
choose only those irreps of $g$ consistent with  the affine cutoff
$ T \leq T_{x}$ at invariant level $x = \{ x_I \}$ of affine
$g$.}
of the ambient Lie algebra $g \supset {\rm Cartan}\,g$ and
$\{\T_\nrm (T,\s) \}$ is another set of duality transformations called
the {\it twisted representation matrices}
of sector $\s$. Here $R(\s) \equiv R(T,\s)$ is the order of the action
$W(h_\s;T)$ of $h_\s \in H$ in representation $T$ of $g$, while the unitary
matrices $U(T,\s)$, the spectral integers $N(r)\equiv N(r,T,\s)$
and the degeneracy indices $\m \equiv \m(r,T,\s)$ are obtained as the eigendata
of the {\it extended $H$-eigenvalue problem} \cite{deBoer:2001nw}
\begin{equation}
\label{extHab}
W(h_\s;T) U\hc(T,\s) = U\hc(T,\s) E(T,\s) \sp E(T,\s) = e^{- 2 \pi i \srac{N(r)}{R(\s)}} \ .
\end{equation}
Again all quantities are periodic $N(r) \rightarrow N(r) \pm R(\s)$ in the
spectral integers. These results are essentially unchanged from
Ref.~\cite{deBoer:2001nw}, except that
in our case the twisted representation matrices commute
because the $ T_a$'s commute.
In what follows we generally
abbreviate $\hj (z) \equiv \hj (z,\s)$  and $ \T \equiv \T (T,\s)$.

\subsection{Solution of the twisted left-mover vertex operator equation}

Ref.~\cite{deBoer:2001nw} also gives the operator equations of motion, called
the {\it twisted vertex operator equations}, for the twisted affine-primary
fields in each sector of all WZW orbifolds.
In particular, the abelian limit of the twisted left-mover vertex operator
equation
\begin{subequations}
\label{vopt}
\begin{equation}
\label{vop}
\partial \hgp(\T,z,\s) =  \G^{n(r) \mu; -n(r), \nu}(\s)
\left( : \hat{J}_{n(r) \mu}(z) \hgp (\T,z,\s) :_M -
\srac{\bar n(r)}{\r(\s)} \frac{1}{z} \hgp (\T,z,\s)\T_{n(r) \mu} \right) \T_{-n(r), \nu}
\end{equation}
\begin{equation}
\label{jgMno}
 : \hat{J}_{n(r) \mu}(z) \hgp (\T,z,\s) :_M \  =
\hat J_{n(r) \mu}^-(z) \hgp (\T,z,\s) +  \hgp (\T,z,\s)  \hat J_{n(r) \mu}^+(z)
\end{equation}
\begin{equation}
\label{jm0} \hat J_{n(r) \mu}^-(z)  \equiv  \sum_{m \leq -1}
\hat J_{\bnrm} (m+\srac{\bar n(r)}{\r(\s)})
z^{-(m+\srac{\bar n(r)}{\r(\s)})-1} , \;\;
\hat J_{\nrm}^+(z)  \equiv \sum_{m \geq 0} \hat J_{\bnrm}
(m+\srac{\bar n(r)}{\r(\s)}) z^{-(m+\srac{\bar n(r)}{\r(\s)})-1}
\end{equation}
\begin{equation}
\label{pcurcom}
[ \hj_\nrm^+ (z), \hj_\nsn^+ (w) ]= [ \hj_\nrm^- (z) ,\hj_\nsn^- (w)] = 0
\end{equation}
\end{subequations}
is also easily read from Ref.~\cite{deBoer:2001nw}. Here the form is unchanged
from Ref.~\cite{deBoer:2001nw} except for the vanishing commutators \eqref{pcurcom}
of the partial currents $\hj^{\pm}$, which reflect the abelian character of the
present case.

The central task of this section is to solve the twisted vertex operator
equation \eqref{vop}, following the general strategy given for the
untwisted WZW vertex operator equation and an abelian limit of this
equation in Ref.~\cite{Halpern:1996et}. In what
follows, we will assume that the twisted vertex operator  $\hgp$ is a square
matrix and that the twisted representation matrices
$\T$ commute with everything, including the solution:
\begin{equation}
\label{as1}
\hgp (\T,z,\s)_{N(r)\m}{}^{N(s) \n}
\sp [ \T_\nrm , \hgp (\T,z,\s) ] = 0 \ .
\end{equation}
This assumption  will be justified a posteriori by our solution below.

The first step in the solution is the direct integration
\begin{eqnarray}
\label{dirint}
\hgp (\T,z,\s) & = & \left( \frac{z}{z_0} \right)^{-\twcw}  \nn \\
 & & \times \exp \left( \int_{z_0}^z d z' \G^{\nrm ; \mnrn} (\s) \hj_\nrm^- (z')
\T_\mnrn \right)\nn \\
& & \times \hgp (\T,z_0,\s) \exp\left(\int_{z_0}^z d z' \G^{\nrm ; \mnrn} (\s)
\hj_\nrm^+ (z')
\T_\mnrn \right)
\end{eqnarray}
which follows using \eqref{pcurcom} and \eqref{as1}. The reference point
$z_0$ in \eqref{dirint} is arbitrary.

The integrals in \eqref{dirint} are easily performed and the results
rearranged in the form
\begin{subequations}
\label{fint0}
\begin{eqnarray}
\hgp (\T,z,\s) &= &   z^{-\twcw} \hV_- (\T,z,\s) \hG_+ (\T,z_0,\s) \nn
\\
 & & \times z^{\hj_{0\m} (0) \G^{0 \m ; 0 \n} (\s) \T_{0 \n}}
\hV_+^{(0)}(\T,z,\s) \hV_+ (\T,z,\s) \label{fint}
\end{eqnarray}
\begin{equation}
\hV_- (\T,z,\s) \equiv \exp \left\{ - \G^{\nrm; \mnrn} (\s)
\sum_{m \leq -1} \hj_\bnrm (\mnrrs) \frac{z^{-(\mnrrs)}}{\mnrrs} \T_\mnrn
\right\}
\end{equation}
\begin{equation}
\hV_+ (\T,z,\s) \equiv \exp \left\{ - \G^{\nrm; \mnrn} (\s) ( 1- \d_{\nb,0})
\sum_{m \geq 0 } \hj_\bnrm (\mnrrs) \frac{z^{-(\mnrrs)}}{\mnrrs} \T_\mnrn
\right\}
\end{equation}
\begin{equation}
\hV_+^{(0)} (\T,z,\s) \equiv \exp \left\{ - \G^{0 \m; 0 \n} (\s)
\sum_{m \geq 1} \hj_{0 \m} (m) \frac{z^{-m}}{m } \T_{ 0 \n}
\right\}
\sp
[ \hV_+^{(0)} (\T,z,\s) , \hV_+ (\T,z,\s) ]
= 0
\end{equation}
\end{subequations}
where we have collected all $z_0$ dependence in the quantity
$\hG_+(\T,z_0,\s)$
\begin{subequations}
\label{intres}
\begin{equation}
\hG_+ (\T,z_0,\s) \equiv \hV_-(\T,z_0,\s)^{-1} \hgp (\T,z_0,\s)
z_0^{-\hat F(\T,\s)} \hV_+^{(0)} (\T,z_0,\s)^{-1}  \hV_+ (\T,z_0,\s)^{-1}
\end{equation}
\begin{equation}
\label{Fdef}
\hat F(\T,\s) \equiv - \twcw + \G^{0 \m ; 0 \n} (\s) \hj_{0\m} (0)  \T_{0 \n}
\end{equation}
\begin{equation}
[ \hat F(\T,\s), \hV_- (\T,z,\s) ]
=[ \hat F(\T,\s), \hV_+^{(0)} (\T,z,\s) ]
=[ \hat F(\T,\s), \hV_+ (\T,z,\s) ] = 0 \ .
\end{equation}
\end{subequations}
Following Ref.~\cite{Halpern:1996et}, we observe that $\hG_+ (\T,z_0,\s)$
is in fact independent of the reference point
\begin{equation}
\partial_{z_0} \hG_+ (\T,z_0,\s) = 0 \quad \rightarrow \quad
 \hG_+ (\T,z_0,\s) = \hG_+ (\T,\s) \ .
\end{equation}
To verify this one uses \eqref{intres}, the explicit forms of
$\hat V_\pm, \hat V_+^{(0)}$  and the equation
\begin{eqnarray}
\partial_{z_0} \hgp (\T,z_0,\s) & = &
 \G^{n(r) \mu; -n(r), \nu}(\s)
 \left( : \hat{J}_{n(r) \mu}(z_0) \hgp (\T,z_0,\s) :_M \right. \hskip 3cm\nn \\
& & \hskip 3cm \left. -
\srac{\bar n(r)}{\r(\s)} \frac{1}{z_0} \hgp (\T,z_0,\s)\T_{n(r) \mu} \right) \T_{-n(r), \nu}
\end{eqnarray}
which is the twisted vertex operator equation \eqref{vop} evaluated at the
reference point.

The next step is to invert Eq.~\eqref{fint}, which allows us to express the
quantity $\hG_+ (\T,\s)$ in terms of the twisted vertex operator
$\hgp (\T,z,\s)$
\begin{equation}
\label{zmvop}
\hG_+ (\T,\s) \equiv \hV_-(\T,z,\s)^{-1} \hgp (\T,z,\s)
\hV_+^{(0)} (\T,z,\s)^{-1}  \hV_+ (\T,z,\s)^{-1}z^{-\hat F(\T,\s)} \ .
\end{equation}
Using the twisted vertex operator equation \eqref{vop}, it can be checked that
the right side of \eqref{zmvop} is independent of $z$ as it should be.

Continuing to follow the strategy of Ref.~\cite{Halpern:1996et}, we note that the relation
\eqref{zmvop} allows us to compute the commutator of the twisted current
modes with the quantity $\hG_+ (\T,\s)$. In particular, the twisted
current algebra \eqref{malg} as well as the relations
\begin{subequations}
\begin{equation}
\label{jhgp}
[ \hj_\bnrm (m +\srac{\nb}{\r(\s)} ), \hgp (\T,z,\s) ] =
 \hgp (\T,z,\s) \T_\bnrm z^{m + \srac{\nb}{\r(\s)}}
\end{equation}
\begin{equation}
\sum_\d \G_{\nrm ; -n(r),\d} \G^{-n(r),\d ; n(r) \n}
= \sum_{s ,\d} \G_{\nrm ; n(s) \d } \G^{n(s) \d ; n(r) \n} = \d_\m{}^\n
\sp \forall \; n(r),\m
\end{equation}
\end{subequations}
are needed for this computation, and the result
\begin{equation}
\label{jGalg}
[ \hj_\bnrm (m + \srac{\nb}{\r(\s)} ), \hG_+ (\T,\s) ] =
\hG_+(\T,\s) \d_{m,0} \d_{\nb, 0} \T_{ 0 \m}
\end{equation}
is obtained after some algebra.

The algebra \eqref{jGalg} can in fact be solved to obtain the explicit form
of the quantity $\hG_+ (\T,\s)$
\begin{subequations}
\begin{equation}
\label{hGsol}
\hG_+ (\T,\s) = \hat \Gamma (\T,\hj_0 (0),\s) e^{i \hq^\m (\s) \T_{0\m} }
\end{equation}
\begin{equation}
\label{hqcom}
[ \hq^\m (\s), \hj_{\bnsn} (n + \srac{\bar{n}(s)}{\r (\s)} ) ] =
i \d_\n^\m \d_{n,0} \d_{\bar{n}(s) , 0} \sp
[\hq^\m (\s),\hq^\n (\s)] =0
\end{equation}
\begin{equation}
[\hat \Gamma (\T,\hj_0(0),\s),\T_\nrm] =
 [ \hat \Gamma (\T,\hj_0(0),\s),\hj_{\nsn} ( n + \srac{n(s)}{\r(\s)}) ] =0
\end{equation}
\end{subequations}
where $\hq^\m (\s)$ is a ``coordinate'' conjugate to the
``momentum'' $\hj_{0\m} (0)$, and $\hat \Gamma$ is a sector-dependent
Klein transformation (cocycle)
which is a so-far undetermined function of $\T_{\nrm}$ and $\hj_{0 \m}(0)$.

The periodic forms of Eqs.~\eqref{hqcom} and \eqref{jGalg} are
\begin{subequations}
\begin{equation}
\label{covhqcom}
[ \hq^\m (\s), \hj_{\nsn} (n + \srac{n(s)}{\r (\s)} ) ] =
i \d_\n^\m \d_{n + \srac{n(s)}{\r(\s)},0}
\end{equation}
\begin{equation}
\label{covjGalg}
[ \hj_\nrm (m + \srac{n(r)}{\r(\s)} ), \hG_+ (\T,\s) ] =
\hG_+(\T,\s) \d_{m + \srac{n(r)}{\r(\s)} ,0}  \T_{ \nrm}
\end{equation}
\end{subequations}
where \eqref{covjGalg} follows from \eqref{covhqcom}, \eqref{hGsol} and
the periodicity \eqref{Tper} of the  twisted representation matrices.
The relations \eqref{covhqcom} and \eqref{covjGalg} are  consistent
with the periodicity \eqref{Jper} of the twisted current modes, and
reduce to \eqref{hqcom} and \eqref{jGalg} in the fundamental
range $n \rightarrow \bar n$.

\subsection{Summary of the twisted left-mover vertex operators \label{sumsec}}

Assembling our results above, we summarize our solution for the twisted
vertex operators \linebreak
$\{ \hgp (\T(T,\s),z,\s),\forall \; T)\}$
in sector $\s$ of $A_{{\rm Cartan}\,g}(H)/H$
\begin{subequations}
\label{sum}
\begin{eqnarray}
\hgp (\T,z,\s) & = &  z^{-\twcw} \hat \Gamma (\T,\hj_0 (0),\s) e^{i \hq^\m (\s) \T_{0 \m} }
\nn \\ & & \times z^{\hj_{ 0 \m} (0) \G^{0\m; 0 \n}(\s) \T_{0 \n} }
 \hV_- (\T,z,\s) \hV_+^{(0)}(\T,z,\s)  \hV_+(\T,z,\s) \label{vop2}
\end{eqnarray}
\begin{equation}
\hV_- (\T,z,\s) \equiv \exp \left\{ - \G^{\nrm; \mnrn} (\s)
\sum_{m \leq -1} \hj_\bnrm (\mnrrs) \frac{z^{-(\mnrrs)}}{\mnrrs} \T_\mnrn
\right\}
\end{equation}
\begin{equation}
\hV_+^{(0)} (\T,z,\s) \equiv \exp \left\{ - \G^{0 \m; 0 \n} (\s)
\sum_{m \geq 1} \hj_{0 \m} (m) \frac{z^{-m}}{m } \T_{ 0 \n}
\right\}
\end{equation}
\begin{equation}
\hV_+ (\T,z,\s) \equiv \exp \left\{ - \G^{\nrm; \mnrn} (\s) ( 1- \d_{\nb,0})
\sum_{m \geq 0 } \hj_\bnrm (\mnrrs) \frac{z^{-(\mnrrs)}}{\mnrrs} \T_\mnrn
\right\}
\end{equation}
\begin{equation}
\label{malgr2}
 [\hj_\nrm(\mmrrs),\hj_\nsn(\nnsrs)]=
  (\mmrrs)\de_{\mnnrnsrsf,0}\sG_{\nrm;\mnrn}(\s)
\end{equation}
\begin{equation}
[ \hq^\m (\s), \hj_{\nsn} (n + \srac{n(s)}{\r (\s)} ) ] =
i \d_\n^\m \d_{n + \srac{n(s)}{\r(\s)},0} \sp
[\hq^\m (\s),\hq^\n (\s)] =0
\end{equation}
\begin{equation}
\s = 0 , \ldots  , N_c-1
\end{equation}
\end{subequations}
where $\G_\cdot(\s)$, $\G^\cdot (\s)$ and the twisted representation matrices
$\T = \T (T,\s)$ are
given respectively in \eqref{sG0}, \eqref{sG1} and \eqref{sumT}. The
commutators of the twisted current modes and the Virasoro operators
with the twisted vertex operators
are found in Eq.~\eqref{sumT}. Using this solution, our assumption
\eqref{as1} is easily verified.

Another form of the prefactor in \eqref{vop2} is obtained with the relation
\begin{equation}
\label{id1}
 \G^{n(r) \mu; -n(r), \nu}(\s)
\srac{\bar n(r)}{\r(\s)} \T_{n(r) \mu}  \T_{-n(r), \nu}
=\frac{1}{2} \G ^{n(r) \mu; -n(r), \nu}(\s)  \T_{n(r) \mu}
\T_{-n(r), \nu} (1 - \delta_{\bar n(r),0} )
\end{equation}
which is the abelian limit of the consistency relation (6.20a) in
Ref.~\cite{deBoer:2001nw}.

Twisted vertex operators are very old \cite{Corrigan:1975sn,Lepowsky:1978jk}
and twisted vertex operators related to ours have been studied in mathematics
(see  e.g. Ref.~\cite{Lepowsky:1985,Lepowsky:1996}) and physics
(see e.g. Ref.~\cite{Borisov:1997nc}). We emphasize however that our
solution \eqref{sum} tells us exactly which twisted vertex operators
 $\{\hgp (\T(T,\s),z,\s), \forall\; T \}$ appear
in each sector $\s$ of all the abelian orbifolds in our class.

\subsection{Properties of the twisted vertex operators \label{propsec}}

The twisted vertex operators \eqref{sum} of $A_{{\rm Cartan}\,g}(H)/H$
satisfy the following properties.

\noindent $\bullet$ {\bf Untwisted vertex operators} \nl
In the untwisted sector $\s=0$ of each orbifold we know that
\begin{subequations}
\begin{equation}
 U(0) = U(T,0) = 1 \sp \bar{n} =0 \sp  \chi (0) = 1
\end{equation}
\begin{equation}
\hj = J \sp \G = G \sp \T = T \sp \hat \Gamma = \Gamma \sp \hq = q
\sp \hgp = g_+
\end{equation}
\end{subequations}
and so the twisted result \eqref{sum} reduces to the Fubini-Veneziano
vertex operator \cite{Fubini:1970}
\begin{subequations}
\begin{eqnarray}
g_+ (T,z,0) & =& \Gamma (T,J(0)) e^{iq^a(0) T_a} z^{J_a (0) G^{ab} J_b (0)} \nn\\
& & \times \exp \left( G^{ab}
\sum_{m \geq 1} J_a (-m)  \frac{z^{m}}{m} T_b \right)
 \exp \left( - G^{ab}
\sum_{m \geq 1} J_a (m)  \frac{z^{-m}}{m} T_b \right) \hskip 1cm
\end{eqnarray}
\begin{equation}
[J_a (m) ,J_b (n)] = m G_{ab} \de_{m+n,0} \sp
[q^a, J_b (n)] = i \de_a^b \de_{n,0} \sp [q^a , q^b] =0
\end{equation}
\begin{equation}
[T_a,T_b] =0
\sp a ,b= 1  \ldots {\rm dim}({\rm Cartan}\,g)
\end{equation}
\end{subequations}
where the representations $\{T_a\}$ are the Cartan matrices of the ambient
Lie algebra $g$.

\noindent $\bullet$ {\bf Alternate form} \nl
Returning to \eqref{sum}, the alternate form  of the twisted vertex operators
\begin{subequations}
\label{FV}
\begin{equation}
\label{FV0}
\hgp (\T,z,\s) = z^{-\twcw} \hat \Gamma (\T,\hj_0 (0),\s)
 : e^{i \hb^\nrm (z,\s)  \T_{\nrm} } :_M
\end{equation}
\begin{eqnarray}
\hb^\nrm (z,\s) & = & \left\{ \hq^\m (\s) + \G^{0 \m; 0\n}(\s)
\left( \hj_{0 \n} (0) \ln z + i \sum_{m \neq 0} \hj_{0 \n} (m) \frac{z^{-m}}{m} \right)
\right\} \d_{\nb , 0} \nn \\
& +& \!\!\!\! \left\{ i \G^{\nrm ; \mnrn} (\s)
\sum_{m \in \sz} \hj_{\mnrn} ( m + \srac{\overline{-n(r)}}{\r (\s)} )
\frac{z^{-(\mmnrrs)}}{\mmnrrs}
\right\} (1 - \d_{\nb, 0} ) \hskip 1.3cm
\end{eqnarray}
\begin{equation}
\label{mnbdef}
\overline{-n(r)}=\overline{-\bar{n}(r)}=\left\{
\begin{array}{cc}
\r(\s)-\bar{n}(r) \textup{ when } \bar{n}(r)\neq 0 \\
0 \comment{\hsp{.4} when $\bar{n}(r)= 0$}
\end{array} \right.
\end{equation}
\end{subequations}
defines a twisted Fubini-Veneziano field $\hb$ for each sector $\s$ of
$A_{{\rm Cartan}\,g}(H)/H$. The mode ordering $M$ defined here in \eqref{FV0}
is the left to right ordering $\{ \hq ,\hj (0), - , +\}$ given
explicitly in \eqref{vop2}.

\noindent $\bullet$ {\bf Intrinsic monodromy} \nl
For general WZW models \cite{Halpern:1996et} and WZW orbifolds \cite{deBoer:2001nw}
the {\it intrinsic monodromy} of the twisted affine primary fields is quite
complicated, but this property is easily computable in our abelian limit.
{} From Eqs.~\eqref{sum} or \eqref{FV} we find the intrinsic monodromy of
the twisted vertex operators
\begin{subequations}
\label{intmon}
\begin{eqnarray}
\hgp (\T,z e^{2 \pi i},\s) & = &  E(T,\s) \hgp (\T,z,\s) E (T,\s)^\ast
\nn \\
& & \times
e^{2 \pi i ( - \twcw + \hj_{0\m }(0) \G^{0\m; 0 \n} (\s) \T_{0 \n} ) }
\label{intmon0}
\end{eqnarray}
\begin{equation}
\label{Eeig0}
E(T,\s)_{N(r) \m}{}^{N (s) \n} = \d_{N(r) \m}{}^{N(s) \n} E_{N(r)}(T,\s)
 \sp E_{N(r)}(T,\s) = e^{- 2 \pi i \srac{N(r)}{R (\s)} } \ .
\end{equation}
\end{subequations}
To obtain this result, we used the solution \eqref{Tsel} of the $\T$-selection
rule in the form
\begin{subequations}
\begin{equation}
\label{Tselal}
e^{- 2 \pi i \srac{n(r)}{\r (\s)} } \T_\mnrn (T,\s)=
E(T,\s) \T_\mnrn  (T,\s) E (T,\s)^\ast
\end{equation}
\begin{equation}
\label{sr2}
\T_{0\m} (T,\s) = E(T,\s) \T_{0\m} (T,\s) E(T,\s)^\ast
\end{equation}
\begin{equation}
\hV_\pm(\T,z e^{2\pi i},\s)  = E(T,\s) \hV_\pm(\T,z,\s) E(T,\s)^\ast
\end{equation}
\begin{equation}
\label{sr3}
\hV_+^{(0)}(\T,z e^{2\pi i},\s)  = E(T,\s) \hV_+^{(0)}(\T,z,\s) E(T,\s)^\ast
\end{equation}
\begin{equation}
\label{sr4}
[E(T,\s), \hat \Gamma (\T,\hj_0(0),\s)] =[ E(T,\s),\hG_+ (\T,\s)]= 0
\end{equation}
\end{subequations}
where the relations \eqref{sr3},\eqref{sr4} follow from the special
case of the $\T$-selection rule in \eqref{sr2}.
Similarly we find the intrinsic monodromy
\begin{equation}
\hb^\nrm (z e^{2 \pi i} ,\s) = \hb^\nrm (z,\s) e^{2 \pi i \srac{n(r)}{\r (\s)}}
+ \d_{\nb ,0} 2 \pi i \G^{0\m;0\n} (\s) \hj_{0\n}(0)
\end{equation}
of the twisted Fubini-Veneziano field $\hb $ in \eqref{FV}, and the
$\T$-selection rule \eqref{Tselal}
guarantees the consistency of this monodromy with that given in \eqref{intmon}
for the twisted vertex operators themselves.

\noindent $\bullet$ {\bf Braid relations} \nl
The following exchange operations
\begin{subequations}
\label{braid}
\begin{equation}
z^{\hj_{0 \m} (0) \G^{0\m; 0\n} (\s) \T_{0\n}^{(1)}} e^{i \hq^\m (\s) \T_{0\m}^{(2)}}
=e^{i \hq^\m (\s) \T_{0\m}^{(2)}} z^{\hj_{0 \m} (0) \G^{0\m; 0\n} (\s) \T_{0\n}^{(1)}}
z^{\T_{0\m}^{(2)} \G^{0\m; 0\n} (\s) \T_{0\n}^{(1)} }
\end{equation}
\begin{equation}
\hV_+^{(0)} (\T^{(1)},z_1,\s) \hV_-(\T^{(2)},z_2,\s) =
 \hV_-(\T^{(2)},z_2,\s) \hV_+^{(0)} (\T^{(1)},z_1,\s)
 \left( 1 -\frac{z_2}{z_1} \right)^{\T_{0\m}^{(2)} \G^{0\m; 0\n}(\s)  \T_{0\n}^{(1)}}
\end{equation}
\begin{eqnarray}
\hV_+ (\T^{(1)},z_1,\s) \hV_-(\T^{(2)},z_2,\s)\!\!\!\!\! \!\!\!\! & & =
 \hV_-(\T^{(2)},z_2,\s) \hV_+ (\T^{(1)},z_1,\s)  \\
 \times \!\!\!\!\! \!\!\!\!\! & &
  \exp \left\{ \T_{\nrm}^{(2)} \G^{\nrm; \mnrn} (\s) \T_{\mnrn}^{(1)}
( 1- \d_{\nb,0}) I_{\srac{\nb}{\r(\s)}} ( \srac{z_1}{z_2},\infty) \right\} \nn
\end{eqnarray}
\begin{equation}
\label{Iint}
I_{\srac{\nb}{\r(\s)}} (y,\infty) \equiv \int_{\infty}^y
\frac{dx}{x-1} x^{-\srac{\nb}{\r (\s)}} = - \sum_{n=0}^\infty
\frac{y^{-( n + \srac{\nb}{\r(\s)})}}{ n + \srac{\nb}{\r(\s)} } \sp |y| > 1
\end{equation}
\end{subequations}
hold for $|z_1| > |z_2|$ and $ \srange$.
 The integrals $I_{\nb/\rho(\s)}$ were
encountered in Ref.~\cite{deBoer:2001nw} and
evaluated as indefinite integrals in the last appendix of that reference.

\noindent $\bullet$ {\bf Operator products} \nl
Using \eqref{braid} we find that the exact operator products of the twisted
vertex operators for $\hat \Gamma =1$
\begin{subequations}
\label{opprod}
\begin{equation}
\hgp (\T^{(1)},z_1,\s) \hgp (\T^{(2)},z_2,\s) = :\hgp (\T^{(1)},z_1,\s)
\hgp (\T^{(2)},z_2,\s):_M
B (\T^{(1)},\T^{(2)};z_1,z_2)
\end{equation}
\begin{eqnarray}
& & B (\T^{(1)},\T^{(2)}; z_1,z_2)  \nn \\
 && \equiv
z_{12}^{ \T^{(2)}_{0\m} \G^{0 \m; 0\n} (\s) \T^{(1)}_{0 \n} }
 \exp \left\{ \T^{(2)}_\nrm \G^{\nrm; \mnrn} (\s) \T^{(1)}_\mnrn
( 1 - \d_{\nb ,0} ) I_{\srac{\nb}{\r (\s)}} \left( \srac{z_1}{z_2},\infty \right)
\right\} \hskip 1.5cm \\
 & &  :\hgp (\T^{(1)},z_1,\s)  \hgp (\T^{(2)},z_2,\s):_M
 \qquad  \nn \\
    & & \equiv  e^{i \hq^\m (\s) (\T^{(1)}_{0 \m} + \T^{(2)}_{0 \m})}
z_1^{\hat F (\T^{(1)},\s)} z_2^{\hat F (\T^{(2)},\s)}
 \hV_- (\T^{(1)},z_1,\s) \hV_- (\T^{(2)},z_2,\s)
 \nn \\
    & &    \hskip 1cm \times 
  \hV_+^{(0)} (\T^{(1)},z_1,\s) \hV_+^{(0)} (\T^{(2)},z_2,\s)
   \hV_+ (\T^{(1)},z_1,\s) \hV_+ (\T^{(2)},z_2,\s)
\end{eqnarray}
\end{subequations}
also hold for $|z_1| > |z_2|$,  $\s = 0 , \ldots N_c -1$.
The quantity $\hat F(\T,\s)$ is defined in \eqref{Fdef}.

\subsection{Correlators of the twisted vertex operators \label{corsec}}

The ground state conditions in Eq.~\eqref{gsc} imply the further conditions
on the partial currents and the components of the twisted vertex operators
\begin{subequations}
\begin{equation}
\hj_\nrm^+ (z) | 0 \rangle_\s = {}_\s \langle 0 | \hj_\nrm^- (z) =0
\end{equation}
\begin{equation}
\hV_+^{(0)} (\T,z,\s) | 0 \rangle_\s =
\hV_+ (\T,z,\s) | 0 \rangle_\s =| 0 \rangle_\s \sp
{}_\s \langle 0 | \hV_- (\T,z,\s) ={}_\s \langle 0 |
\end{equation}
\end{subequations}
and these conditions allow us to evaluate the correlators in sector
$\s$ of $A_{{\rm Cartan}\,g} (H)/H$:
\begin{subequations}
\label{abcor0}
\begin{eqnarray}
\hat A_+ (\T,z,\s) & \equiv  &
\langle \hgp (\T^{(1)},z_1,\s)  \cdots \hgp (\T^{(N)},z_N,\s) \rangle_\s
\sp \srange
\\
 & = &  C_+ (\T,\s) \left( \prod_{\r} z_\r^{- \G^{\nrm; \mnrn} (\s)
\srac{\nb }{\r (\s)} \T_\nrm^{(\r)} \T_\mnrn^{(\r)} } \right)
 \left( \prod_{\r < \k} z_{\r \k}^{ \T_{0 \m}^{(\k)} \G^{0 \m; 0 \n } (\s)
 \T_{0 \n}^{(\r)} } \right) \nn \\
 & & \times  \prod_{\r < \k} \exp \left\{  \T_\nrm^{(\k)} \G^{\nrm; \mnrn} (\s)
 \T_\mnrn^{(\r)} ( 1 - \d_{\nb ,0} )
I_{\srac{\nb}{\r (\s)}} \left( \srac{z_\r}{z_\k},\infty \right) \right\}
\label{abcor}
\hskip 1cm
\end{eqnarray}
\begin{equation}
C_+ (\T,\s) \equiv \langle \hG_+ (\T^{(1)},\s) \cdots \hG_+ (\T^{(N)},\s)
\rangle_\s
\end{equation}
\begin{equation}
\label{gwcor}
\langle [ \hj_{0\m} (0),\hG_+ (\T^{(1)},\s) \cdots \hG_+ (\T^{(N)},\s)]
\rangle_\s =0 \quad \Rightarrow \quad
 C_+ (\T,\s) \left( \sum_{\r =1}^N \T_{0 \m}^{(\r)} \right) =0 \sp
 \forall \; \m  \ .
\end{equation}
\end{subequations}
Here we have used the residual
symmetry generators $\hj_{0\m} (0)$ to include the global Ward identities
in \eqref{gwcor}. Note that the global Ward identities can also be written as
\begin{equation}
\left( \sum_{\r=1}^N \T_{0\m}^{(\r)}\right) C_+ (\T,\s) =  0 \sp \forall \;\m
\end{equation}
because the twisted representation matrices  commute with $\hgp$. Similarly,
the constant factor $C_+(\T,\s)$ can be moved to the right in \eqref{abcor}.

Another relation on the constant factor $C_+ (\T,\s)$ follows from \eqref{sr4}
\begin{subequations}
\begin{equation}
[ E(\{ T\},\s), \hG_+ (\T^{(1)},\s) \cdots \hG_+ (\T^{(N)},\s)]=0
\quad \rightarrow [E(\{ T \},\s), C_+ (\T,\s)]=0
 \end{equation}
\begin{equation}
E(\{T\},\s) \equiv \otimes_{\r =1}^N E(T^{(\r)},\s)
\end{equation}
\end{subequations}
where the eigenvalue matrix $E(T,\s)$ is given in \eqref{Eeig0}.

We finally note that the orbifold correlators \eqref{abcor0} satisfy the twisted KZ -like equations
\begin{subequations}
\label{twkz0}
\begin{equation}
\part_\k \hat A_+(\T,z,\s) = \hat A_+ (\T,z,\s) \hat W_\k (\T,z,\s) \sp
\s = 0, \ldots ,N_c -1
\end{equation}
\begin{equation}
\label{kzct}
\hat W_{\k}(\T,z,\s) = \G^{\nrm ; \mnrn } (\s)
 \left[ \sum_{\r\neq \k}\left( \frac{z_{\r}}{z_{\k}}
\right)^{ \srac{\bar n(r)}{\r(\s)}} \frac{1}{z_{\k \r}}
\T_{\nrm }^{(\r)} \T_{\mnrn }^{(\k)}
- \srac{\bar n(r)}{\r(\s)} \frac{1}{z_{\k}}
\T_{\nrm }^{(\k)} \T_{\mnrn }^{(\k)} \right]
\end{equation}
\begin{equation}
\T^{(\r)} \T^{(\k)} \equiv
\T^{(\r)} \otimes \T^{(\k)} \sp z_{\k\r } \equiv z_\k - z_\r
\sp \sum_{\r \neq \k} \equiv \sum_{ {\r =1 \atop \r \neq \k} }^N
\end{equation}
\end{subequations}
as they must because these equations follow directly from the twisted vertex
operator equation \eqref{vop} and the ground state condition \eqref{gsc}.

This completes our general discussion of the left-mover sectors of the abelian
orbifolds $A_{{\rm Cartan}\,g}(H)/H$, and we turn in the following sections
to specific examples in $A_{{\rm Cartan}\,g}(H)/H$. The right-mover sectors
of $A_{{\rm Cartan}\,g}(H)/H$ are considered in Subsecs.~\ref{rmsec} and
\ref{ncsec}.

\subsection{Example: Abelian permutation orbifolds \label{permsec}}

In this subsection we study the abelian permutation orbifolds
\begin{equation}
\label{permorb}
\frac{A_{{\rm Cartan}\,g}(H)}{H} \sp H(\mbox{permutation}) \subset
{\rm Aut}({\rm Cartan}\,g)
\end{equation}
as a special case of our  large example $A_{{\rm Cartan}\,g}(H)/H$.
In the permutation-symmetric theory $A_{{\rm Cartan}\,g}(H)$, the permutations act among
the copies ${\rm Cartan}\,\gfrak^I$ of ${\rm Cartan}\,\gfrak$
\begin{subequations}
\begin{equation}
{\rm Cartan}\,g = \oplus_{I=0}^{K-1} {\rm Cartan}\, \gfrak^I \sp
{\rm Cartan}\,\gfrak^I \simeq {\rm Cartan}\,\gfrak\sp K \leq N
\end{equation}
\begin{equation}
H(\mbox{permutation}) \subset S_N  \sp k_I =k \sp x_I = x
\sp \eta^I_{a(I)b(I)} = \eta_{ab} \sp
T^I \simeq T
\end{equation}
\begin{equation}
[T_a, T_b] = 0 \sp a, b = 1\ldots {\rm dim}({\rm Cartan}\, \gfrak)
\end{equation}
\end{subequations}
where $T_a$ are the Cartan matrices of ${\rm Cartan}\, \gfrak \subset \gfrak$.

For the WZW permutation orbifolds it is conventional \cite{Halpern:2000vj,deBoer:2001nw}
to use the notation $\m = aj$, $a=1 \ldots {\rm dim}\,\gfrak$ for the
degeneracy indices of the $H$-eigenvalue problem,
and it is known \cite{deBoer:2001nw} for the extended $H$-eigenvalue problem
\eqref{extHab} that $R(\s)= \r(\s)$ and $N(r)=n(r)$.
For the abelian permutation orbifolds we know that
$\gfrak \rightarrow {\rm Cartan}\,\gfrak$ so the corresponding
relabelling for the abelian permutation orbifolds is
\begin{equation}
n(r) \m \rightarrow n(r) aj \sp N(r) \m \rightarrow n(r) \a j
\sp a =1 \ldots {\rm dim}({\rm Cartan}\, \gfrak)
\sp \a = 1 \ldots {\rm dim}\,T
\end{equation}
and then the explicit form of all the required duality transformations can be read off
as the abelian limit of those given for the WZW permutation orbifolds in
Ref.~\cite{deBoer:2001nw}.
It will be helpful here to introduce the further spectral index relabelling
$ n(r) \rightarrow \hjs$
\begin{subequations}
\begin{equation}
n(r) aj  \quad \rightarrow \quad \hjs a j \sp N(r) aj  \quad\rightarrow \quad
\hjs \a  j
\end{equation}
\begin{equation}
\label{Nin}
\frac{N(r)}{R(\s)} = \frac{n(r)}{\r(\s)} =
\frac{\hjs}{f_j(\s)} \sp
  \bar{\hjs} = 0,1 , \ldots , f_j (\s)-1\sp \sum_j f_j (\s) = K \leq N
\end{equation}
\begin{equation}
\label{sran}
\srange
\end{equation}
\end{subequations}
where the correspondence \eqref{Nin}, \eqref{sran} was
also given in Ref.~\cite{deBoer:2001nw}. In this notation, each
element $h_\s \in H (\mbox{permutation})$ is expressed in terms of disjoint cycles
$j$ of size and order $f_j (\s)$, and the hatted index $\hjs$ runs inside
the disjoint cycle $j$. The disjoint cycles are labelled periodically
$\hjs \rightarrow \hjs \pm f_j (\s)$ so that $\bar{\hjs}$
in \eqref{Nin} is the pullback of $\hjs$ to  the fundamental range.
As an example we recall \cite{Halpern:2000vj} the data for
$H(\mbox{permutation})=S_N$
\begin{equation}
K = N \sp f_j(\s) = \s_j \sp  \s_{j+1} \leq \s_j \sp j = 0,\ldots, n(\vec{\s})-1
\sp \sum_{j=0}^{n(\vec{\s})-1} \s_j = N
\end{equation}
and the data for $H(\mbox{permutation})= \z_\l$ is included in
Subsec.~\ref{cyclsec}.

Then we may read the following results directly from Ref.~\cite{deBoer:2001nw}:
\begin{subequations}
\begin{equation}
\hj_{n(r)a j} (z)\rightarrow \hj_{\hjs a j}(z) \sp
\hj_{\hjs a j} (ze^{2\pi i}) = e^{-2\pi i \srac{\hjs}{f_j(\s)}} \hj_{\hjs aj}(z)
\sp \hj_{\hjs \pm f_j(\s), aj}(z) = \hj_{\hjs aj}(z)
\end{equation}
\begin{equation}
\hj_{\hjs aj}(z) = \sum_{m \in \sz} \hj_{\hjs a j} (m + \srac{\hjs}{f_j(\s)})
z^{-(m + \srac{\hjs}{f_j(\s)}) -1} \sp
\label{jperio}
\hj_{\hjs \pm f_j(\s), aj}(m + \srac{\hjs\pm f_j(\s)}{f_j(\s)})
=\hj_{\hjs a j}(m \pm 1 + \srac{\hjs}{f_j(\s)})
\end{equation}
\begin{equation}
\G_{\nrm ; \nsn} (\s) \quad \rightarrow \quad \G_{\hjs aj; \hls bl}
= \hat{k}_j (\s) \eta_{ab} \d_{jl}  \d_{\hjs + \hls,\,0 \rmod f_j(\s)} \sp
\hat{k}_j (\s)  = k f_j (\s)
\end{equation}
\begin{equation}
\G^{\nrm ; \nsn} (\s) \quad \rightarrow \quad \G^{\hjs a j;\hls b l}(\s)
= \frac{\eta^{ab}}{\hat k_j(\s)} \d^{jl} \d_{\hjs + \hls, 0 \,\rmod f_j(\s)}
\end{equation}
\begin{equation}
\label{twaper}
[ \hj_{\hjs a j}(m + \srac{\hjs}{f_j(\s)}),
\hj_{\hls b l}(n + \srac{\hls}{f_j(\s)})] =
\d_{jl} \hat{k}_j (\s) \eta_{ab} (m + \srac{\hjs}{f_j(\s)})
\d_{m+n+\srac{\hjs+\hls}{f_j(\s)},0}
\end{equation}
\begin{equation}
\label{trepp}
\T_\nrm =T_a t_\nrm(\s)\quad \rightarrow \quad \T_{\hjs a j} = T_a t_{\hjs j}(\s)
\sp [ \T_{\hjs a j}(T,\s), \T_{\hls b l} (T,\s) ] = 0
\end{equation}
\begin{equation}
\label{ttrep}
[T_a , T_b] = 0 \sp
t_{\hjs j} (\s) t_{\hls l} (\s) = \d_{jl} t_{\hjs + \hls,j}(\s) \sp
t_{\hjs \pm f_j (\s),j} (\s) = t_{\hjs j} (\s)
\end{equation}
\begin{equation}
\label{ttex}
t_{\hjs j}(\s)_{\hls l}{}^{\hms m} = \d_{jl} \d_l^m
\d_{\hjs + \hls - \hms,0\,\rmod f_j (\s)}
\end{equation}
\begin{equation}
[\hj_{\hjs a j} (m +\srac{\hjs}{f_j(\s)}), \hgp (\T,z,\s)]
= \hgp (\T,z,\s)  T_a t_{\hjs j} (\s)z^{m +\srac{\hjs}{f_j(\s)}}
\end{equation}
\begin{equation}
\label{Lhpcom}
[ L_\s (m), \hgp (\T,z,\s) ] = z^m \Big( z \partial_z + \Delta (\T)
 (m+1) \Big) \hgp (\T,z,\s)
 \sp \frac{(\eta^{ab} T_a T_b)_\a{}^\be}{2k} =
\Delta (T) \de_\a^\be
 \end{equation}
\begin{equation}
a=1 \ldots \text{dim }({\rm Cartan}\,\gfrak) \sp
\bar{\hjs} = 0 , \ldots , f_j (\s)-1 \sp
\bar{\hls} = 0 , \ldots , f_l (\s)-1 \ .
\end{equation}
\end{subequations}
Relative to Ref.~\cite{deBoer:2001nw} the
only differences here are the abelian form of the twisted current algebra
\eqref{twaper} and
the fact that the Cartan matrices $T_a$ of $\gfrak$ commute. The
quantity $\Delta (T)$ in \eqref{Lhpcom} is the conformal weight of
untwisted representation $T$ under untwisted ${\rm Cartan}\,\gfrak$.
Note that, as in the nonabelian
case, $j$ and $l$ are semisimplicity indices for the twisted current algebra.
See also App.~\ref{Heigp}, which
solves the $H$-eigenvalue problem for all permutation groups in this notation.

The spectral index relabelling helps to simplify many of the sums
encountered in our general discussion above. For example, we obtain for the Virasoro operator and
ground state conformal weight $\hat{\Delta}_0 (\s)$ of sector $\s$
\begin{subequations}
\begin{equation}
{\cL}^{n(r)  aj;n(s),bl} (\s) \quad \rightarrow \quad
{\cL}^{\hjs a j,\hls b l } (\s) = \frac{\eta^{ab} \de^{jl}}{2 \hat{k}_j(\s)}
\de_{\hjs + \hls, \, 0 \, \rmod f_j(\s)}
\end{equation}
\begin{equation}
L_\s(m) =    \frac{1}{2k} \sum_j \sum_{\hjs =0}^{f_j(\s) -1}
 \frac{1}{ f_j (\s)}
 \sum_{p \in \sz} \eta^{ab} :\hat{J}_{\hjs aj}(p+\srac{\hjs}{f_j(\s)})
\hat{J}_{-\hjs,bj} (m-p-\srac{\hjs}{f_j(\s)}):_M +\delta_{m,0}
\hat{\Delta}_0 (\s)
\end{equation}
\begin{equation}
\label{cwper}
\hat c (\s) = c = K {\rm dim}({\rm Cartan}\,\gfrak) \sp
\hat{\Delta}_0 (\s) =
\frac{{\rm dim}({\rm Cartan}\,\gfrak)}{24} \sum_j
\left(f_j(\s) - \frac{1}{f_j(\s)}\right)
\end{equation}
\end{subequations}
where $: \cdot :_M$ is mode normal ordering
\cite{Evslin:1999qb,deBoer:1999na,Evslin:1999ve,Halpern:2000vj,deBoer:2001nw}.
Here we used $\m (r) = \m = a j$ and the identity
\begin{equation}
\sum_{r,\m(r)} \frac{\nb}{2 \rho (\s)} \left( 1 - \frac{\nb}{\rho (\s)} \right)
 = {\rm dim}({\rm Cartan}\,\gfrak)
 \sum_j \sum_{\hjs=0}^{f_j(\s)-1} \frac{\hjs}{2 f_j(\s)}
\left( 1- \frac{\hjs}{f_j(\s)} \right)
\end{equation}
to obtain \eqref{cwper} from \eqref{cw}. The result \eqref{cwper} is the
abelian limit of the formula given for each sector of all WZW permutation
orbifolds in Refs.~\cite{Halpern:2000vj} and \cite{deBoer:2001nw}.

Similarly, we find that the following identities
\begin{subequations}
\label{usid}
\begin{equation}
\sum_j \sum_{\hjs=0}^{f_j(\s)-1} \frac{1}{f_j (\s)} t_{0j}(\s) =
\sum_j t_{0j} (\s) = \one \sp (\one)_{\hjs j}{}^{\hls l} = \d_j^l
\d_{\hjs - \hls, 0 \, \rmod f_j(\s)}
\end{equation}
\begin{equation}
\label{cw0}
\G^{n(r) aj; - n(r) bl}(\s) \T_{n(r)aj} \T_{-n(r),bl} =
2 \Delta (T) \one \sp
\G^{0 aj; 0 bl}(\s) \T_{0aj} \T_{0bl} =
2 \Delta (T)  \sum_j \frac{1}{f_j (\s)} t_{0j}(\s)
\end{equation}
\begin{equation}
  \sum_j \frac{1}{f_j (\s)} t_{0j}(\s)_{\hls l}{}^{\hms m}
  =\frac{1}{f_l (\s)} (\one)_{\hls l}{}^{\hms m}
  = (\one)_{\hls l}{}^{\hms m} \frac{1}{f_m(\s)}
  \end{equation}
\begin{eqnarray}
\G^{n(r) aj; - n(r) bl}(\s) \srac{\nb}{\r(\s)}\T_{n(r)aj} \T_{-n(r),bl} & = &
\frac{1}{2} \G^{n(r) aj; - n(r) bl}(\s) \T_{n(r)aj} \T_{-n(r),bl} (1- \d_{\nb,0})
\nn \\ && =\Delta (T) \left( \one- \sum_j \frac{1}{f_j (\s)} t_{0j}(\s) \right)
\end{eqnarray}
\end{subequations}
 are useful in simplifying the general twisted vertex operators and
 correlators above.  Closely related identities and simplifications are
given  for the nonabelian case in App.~\ref{nakz}. There we also include the
spectral integer relabelling $n(r) \rightarrow \hjs$ of the twisted KZ equations given in
Ref.~\cite{deBoer:2001nw} for the WZW permutation orbifolds.

Then we obtain from Eq.~\eqref{sum}  the explicit form of the twisted vertex
operators in each sector $\s$ of the abelian permutation orbifolds
\begin{subequations}
\label{tvop2}
\begin{eqnarray}
\hgp (\T,z,\s) &=& z^{-\Delta (T) (\one - \sum_j \srac{t_{0j}(\s)}{f_j(\s) })}
\hat \Gamma (\T,\hj_0 (0),\s) e^{i T_a \sum_j \hq^{aj} (\s) t_{0j} (\s)}
z^{\srac{\eta^{ab}T_b}{k} \sum_j \hj_{0aj}(0) t_{0j}(\s)/f_j(\s)} \nn \\
& & \times \exp \left( -\frac{\eta^{ab}T_b}{k} \sum_j
\sum_{\hjs=0}^{f_j(\s)-1} \frac{1}{f_j(\s)} \sum_{m \leq -1} \hj_{\hjs a j}
( m + \srac{\hjs}{f_j(\s)})
\frac{ z^{-(m+ \srac{\hjs}{f_j(\s)} ) }} {m+ \srac{\hjs}{f_j(\s)} }
t_{-\hjs,j} (\s) \right) \nn \\
& & \times \exp \left( -\frac{\eta^{ab}T_b}{k} \sum_j \frac{1}{f_j(\s)}
\sum_{m \geq 1} \hj_{0 a j} (m) \
\frac{z^{-m}}{m} t_{0j} (\s) \right) \nn \\
& & \times \exp \left( -\frac{\eta^{ab}T_b}{k} \sum_j
\sum_{\hjs =1}^{f_j(\s)-1} \frac{1}{f_j(\s)} \sum_{m \geq 0} \hj_{\hjs a j}
( m + \srac{\hjs}{f_j(\s)})
\frac{z^{-(m+ \srac{\hjs}{f_j(\s)})}}{m+ \srac{\hjs}{f_j(\s)}}
t_{-\hjs,j} (\s) \right) \nn \\
\end{eqnarray}
\begin{equation}
[\hq^{aj} (\s), \hj_{\hls bl} ( m + \srac{\hls}{f_l(\s)})]
= i \d_l^j\d_b^a \d_{m + \srac{\hls}{f_l(\s)},0}  \sp [ \hq^{aj}(\s),\hq^{bl}(\s)] = 0
\end{equation}
\begin{equation}
\label{twaper2}
[ \hj_{\hjs a j}(m + \srac{\hjs}{f_j(\s)}),
\hj_{\hls b l}(n + \srac{\hls}{f_j(\s)})] =
\d_{jl} k f_j (\s) \eta_{ab} (m + \srac{\hjs}{f_j(\s)})
\d_{m+n+\srac{\hjs+\hls}{f_j(\s)},0}
\end{equation}
\begin{equation}
a = 1 \ldots {\rm dim}({\rm Cartan}\, \gfrak) \sp \bar{\hjs} = 0, \ldots f_j(\s)-1
\sp \sum_j f_j(\s) = K \leq N
\end{equation}
\begin{equation}
 \srange \ .
\end{equation}
\end{subequations}
The commutator of the twisted current modes with the twisted vertex operators
is given in Eq.~\eqref{twaper}. The twisted vertex operators \eqref{tvop2}
can be written in a $j$-factorized form
\begin{subequations}
\label{facform0}
\begin{equation}
\hgp (\T,z,\s) =  \hat \Gamma (\T,\hj_0(0),\s)
\prod_j \hat V_j (\T,z,\s)
\end{equation}
\begin{eqnarray}
\hat V_j (\T,z,\s) & \equiv &  z^{-\Delta(T) ( 1- \srac{1}{f_j(\s)})t_{0j}(\s)}
e^{i T_a \hq^{aj} (\s) t_{0j}(\s)} z^{\srac{\eta^{ab}T_b}{kf_j(\s)} \hj_{0aj}(0)
t_{0j}(\s) } \nn \\
& & \times \exp \left( -\frac{\eta^{ab}T_b}{k f_j(\s)}
\sum_{\hjs=0}^{f_j(\s)-1}  \sum_{m \leq -1} \hj_{\hjs a j}
( m + \srac{\hjs}{f_j(\s)})
\frac{ z^{-(m+ \srac{\hjs}{f_j(\s)} ) }} {m+ \srac{\hjs}{f_j(\s)} }
t_{-\hjs,j} (\s) \right) \nn \\
& & \times \exp \left( -\frac{\eta^{ab}T_b}{k f_j(\s)}
\sum_{m \geq 1} \hj_{0 a j} (m) \
\frac{z^{-m}}{m} t_{0j} (\s) \right) \nn \\
& & \times \exp \left( -\frac{\eta^{ab}T_b}{k f_j(\s)}
\sum_{\hjs =1}^{f_j(\s)-1} \sum_{m \geq 0} \hj_{\hjs a j}
( m + \srac{\hjs}{f_j(\s)})
\frac{z^{-(m+ \srac{\hjs}{f_j(\s)})}}{m+ \srac{\hjs}{f_j(\s)}}
t_{-\hjs,j} (\s) \right) \nn \\
\end{eqnarray}
\begin{equation}
[\hat V_j (\T,z,\s), \hat V_l (\T,z,\s) ] =0 \qquad {\rm when} \; j \neq l
\end{equation}
\end{subequations}
which reflects the semisimplicity of the abelian orbifold affine algebra
\eqref{twaper2}.

Using \eqref{opprod} and \eqref{usid}, we also give the explicit form of the
operator product of these twisted vertex operators at $\hat \Gamma =1$
\begin{eqnarray}
\hgp (\T^{(1)},z_1,\s) \hgp (\T^{(2)},z_2,\s) & = & :\hgp (\T^{(1)},z_1,\s)
\hgp (\T^{(2)},z_2,\s):_M \nn \\
& \times & \!\!\! z_{12}^{ \srac{T_a^{(2)} \eta_{ab} T_b^{(1)}}{k} \sum_j \srac{1}{f_j(\s)}
t_{0j}^{(2)} (\s) t_{0j}^{(1)} (\s) } \nn \\
&  \times  & \!\!\! \exp
\left\{\frac{T_a^{(2)} \eta_{ab} T_b^{(1)}}{k} \sum_j
\sum_{\hjs=1}^{f_j(\s)-1} \frac{t_{\hjs j}^{(2)} (\s) t_{-\hjs ,j}^{(1)}(\s)}{f_j(\s)}
 I_{\srac{\hjs}{f_j (\s)}} \left( \srac{z_1}{z_2},\infty \right)
\right\}   \nn \\
\end{eqnarray}
where the integrals $I_{\nb/\r(\s)}$ are defined in Eq.~\eqref{Iint}.
Similarly, we obtain the explicit form of the twisted left-mover
correlators
\begin{subequations}
\label{Nptaper}
\begin{eqnarray}
\hat A_+ (\T,z,\s) \!\!\!\! \!\!\!\!\! & & \equiv
\langle \hgp (\T^{(1)},z_1,\s)  \cdots \hgp (\T^{(N)},z_N,\s) \rangle_\s
\nn \\ &  & \!\!\!\!\!\! = C_+ (\T,\s) \left( \prod_{\r} z_\r^{- \Delta(T^{(\r)})
( \one- \sum_j \srac{1}{f_j (\s)} t_{0j}^{(\r)}(\s) )} \right)
\left(   \prod_{\r < \k} z_{\r \k}^{ \srac{T_a^{(\k)} \eta^{ab} T_b^{(\r)}}{k}
 \sum_j \srac{1}{f_j (\s)}
t_{0j}^{(\k)}(\s)  t_{0j}^{(\r)} (\s) } \right) \nn \\
 & & \times \prod_{\r < \k}
\exp \left\{ \frac{T_a^{(\k)} \eta^{ab}T_b^{(\r)} }{k}
 \sum_j \sum_{\hjs=1}^{f_j(\s)-1}
  \frac{1}{f_j(\s)}t_{\hjs j}^{(\k)} (\s) t_{-\hjs, j}^{(\r)} (\s)
I_{\srac{\hjs}{f_j (\s)}} \left( \srac{z_\r}{z_\k},\infty \right) \right\}
\hskip 1cm  \\
& = & C_+ (\T,\s)  \prod_j \left\{
\left( \prod_\r z_\r^{ -\Delta (T^{(\r)}) (1- \srac{1}{f_j (\s) })t_{0j}^{(\r)}(\s) }
\right) \left(
   \prod_{\r < \k} z_{\r \k}^{ \srac{T_a^{(\k)} \eta^{ab} T_b^{(\r)}}{kf_j(\s)}
t_{0j}^{(\k)}(\s)  t_{0j}^{(\r)} (\s) } \right) \right. \nn \\
 & & \left. \times \prod_{\r < \k}
\exp \left( \frac{T_a^{(\k)} \eta^{ab}T_b^{(\r)} }{k f_j(\s)}
  \sum_{\hjs=1}^{f_j(\s)-1} t_{\hjs j}^{(\k)} (\s) t_{-\hjs, j}^{(\r)} (\s)
I_{\srac{\hjs}{f_j (\s)}} \left( \srac{z_\r}{z_\k},\infty \right) \right)
\right\} \label{facper}
\hskip 1cm
\end{eqnarray}
\begin{equation}
C_+ (\T,\s) \left( \sum_{\r =1}^N T_{a}^{(\r)} t_{0j}^{(\r)}(\s) \right)=0 \sp
a = 1 \ldots {\rm dim}({\rm Cartan}\,\gfrak) \sp \forall \;j \sp \srange
\end{equation}
\end{subequations}
as a special case of \eqref{abcor0}. The $j$-factorized form
\eqref{facper} of the correlators again reflects the semisimplicity of the abelian
orbifold affine algebra \eqref{twaper2}.

\subsection{Subexample: Abelian cyclic permutation orbifolds \label{cyclsec}}

We may be more explicit for the case $H(\mbox{permutation})=\z_\l$
by substitution of the data
\begin{subequations}
\begin{equation}
K = \l \sp f_j (\s) = \r (\s) \sp  \hjs = \bar{r} = 0 , \ldots, \r(\s)  -1
\sp j = 0 , \ldots , \srac{\l}{\r(\s)} - 1
\end{equation}
\begin{equation}a = 1 \ldots {\rm dim}({\rm Cartan}\,
\gfrak)  \sp \s=0,\ldots,\l-1
\end{equation}
\end{subequations}
into the more general results above. Here $\r(\s)$ is the order of $h_\s \in \z_\l$,
and we have also changed $\hjs \rightarrow r$ in accord with the prior convention
\cite{Borisov:1997nc,Evslin:1999qb,deBoer:1999na,Evslin:1999ve,Halpern:2000vj,deBoer:2001nw}
for cyclic orbifolds.
This gives the results for each sector $\s$ of the abelian cyclic permutation
orbifolds $A_{{\rm Cartan}\,g}(\z_\l)/\z_\l$:
\begin{subequations}
\begin{equation}
\hj_{raj}(ze^{2\pi i}) = e^{-2\pi i \srac{r}{\r(\s)}} \hj_{raj}(z) \sp
\hj_{r\pm \r(\s),aj}(z) = \hj_{raj}(z)
\end{equation}
\begin{equation}
L_\s(m) =    \frac{1}{2k \rho(\s)} \sum_{j=0}^{\srac{\l}{\r(\s)}-1}
\sum_{r=0}^{\rho(\s)-1}
 \sum_{p \in \sz} \eta^{ab} :\hat{J}_{raj}(p+\srac{r}{\rho(\s)})
\hat{J}_{-r,bj} (m-p-\srac{r}{\rho(\s)}):_M +\delta_{m,0}
\hat{\Delta}_0 (\s)
\end{equation}
\begin{equation}
\hat c (\s) = c = \l {\rm dim}({\rm Cartan}\,\gfrak) \sp
\hat{\Delta}_0 (\s) =\frac{\l}{24}
{\rm dim}({\rm Cartan}\,\gfrak)
\left(1- \frac{1}{\r^2(\s)}\right)
\end{equation}
\end{subequations}
\begin{subequations}
\begin{equation}
\T_{raj} (T,\s) = T_a t_{rj} (\s) \sp [\T_{raj}(T,\s),\T_{sbl}(T,\s)]=0
\end{equation}
\begin{equation}
[T_a, T_b] =0 \sp t_{r \pm \r(\s),j}(\s) = t_{rj} (\s)\sp t_{rj} (\s)_{sl}{}^{tm} =
\d_{jl} \d_{l}^m \d_{r + s -t , 0 \rmod \r (\s)}
\end{equation}
\begin{eqnarray}
\hgp (\T,z,\s) &=& z^{-\Delta (T) \srac{\r (\s) -1}{\r(\s)}}
\hat \Gamma (\T,\hj_0 (0),\s) e^{i T_a \sum_{j=0}^{\srac{\l}{\r(\s)}-1}
 \hq^{aj} (\s) t_{0j} (\s)}
z^{\srac{\eta^{ab}T_b}{\r  (\s) k} \sum_{j=0}^{\srac{\l}{\r(\s)}-1} \hj_{0aj}(0)
t_{0j}(\s)} \nn \\
& & \times \exp \left( -\frac{\eta^{ab}T_b}{\r (\s) k} \sum_{r=0}^{\r (\s)-1}
\sum_{j=0}^{\srac{\l}{\r(\s)}-1} \sum_{m \leq -1} \hj_{r a j}
( m + \srac{r}{\r(\s)})
\frac{ z^{-(m+ \srac{r}{\r(\s)} ) }} {m+ \srac{r}{\r(\s)} }
t_{-r,j} (\s) \right) \nn \\
& & \times \exp \left( -\frac{\eta^{ab}T_b}{\r (\s) k}
\sum_{j=0}^{\srac{\l}{\r(\s)}-1} \sum_{m \geq 1} \hj_{0 a j} (m)
\frac{z^{-m}}{m} t_{0j} (\s) \right) \nn \\
& & \times \exp \left( -\frac{\eta^{ab}T_b}{\r (\s)k}\sum_{r=1}^{\r (\s)-1}
\sum_{j=0}^{\srac{\l}{\r(\s)}-1} \sum_{m \geq 0} \hj_{r a j}
( m + \srac{r}{\r(\s)})
\frac{ z^{-(m+ \srac{r}{\r(\s)} ) }} {m+ \srac{r}{\r(\s)} }
t_{-r,j} (\s) \right) \nn \\
\end{eqnarray}
\begin{equation}
[\hq^{aj} (\s), \hj_{r bl} ( m + \srac{r}{\r(\s)})]
= i \d_l^j\d_b^a \d_{m + \srac{r}{\r(\s)},0}  \sp [ \hq^{aj} (\s), \hq^{bl}(\s)]=0
\end{equation}
\begin{equation}
[ \hj_{r a j}(m + \srac{r}{\r(\s)}), \hj_{s b l}(n + \srac{s}{\r(\s)})] =
 \d_{jl}  k \r(\s)  \eta_{ab}(m + \srac{r}{\r(\s)})\d_{m+n+\srac{r+s}{\r(\s)},0}
\end{equation}
\begin{equation}
\hj_{r\pm \r(\s),aj}(m + \srac{r\pm \r(\s)}{\r(\s)}) = \hj_{raj}(m \pm 1
+\srac{r}{\r(\s)})
\end{equation}
\end{subequations}
\begin{subequations}
\label{Nptf}
\begin{eqnarray}
\langle \hgp (\T^{(1)},z_1,\s)  &&\!\!\!\!\!\!\! \!\! \cdots
\hgp (\T^{(N)},z_N,\s) \rangle_\s  \nn \\
& & =C_+ (\T,\s) \left( \prod_{\r} z_\r^{- \srac{\r(\s)-1}{\r (\s)}
\Delta (T^{(\r)}) } \right)
\left(  \prod_{\r < \k} z_{\r \k}^{ \srac{T_{a}^{(\k)} \eta^{ab}
 T_{b}^{(\r)}}{ \r (\s) k } \sum_{j=0}^{\srac{\l}{\r (\s)}-1} t_{0j}^{(\k)} (\s)
 t_{0j}^{(\r)} (\s) } \right) \nn \\
 & & \times \prod_{\r < \k} \exp \left\{ \frac{T_a^{(\k)} \eta^{ab}
 T_b^{(\r)}}{\r(\s)k} \sum_{r=1}^{\r (\s)-1}\sum_{j=0}^{\srac{\l}{\r (\s)}-1}
  t_{rj}^{(\k)} (\s) t_{-r,j}^{(\r)} (\s)
I_{\srac{r}{\r (\s)}} \left( \srac{z_\r}{z_\k},\infty \right) \right\}
\hskip 1cm
\end{eqnarray}
\begin{equation}
C_+ (\T,\s) \left( \sum_{\r =1}^N T_{a}^{(\r)} t_{0j}^{(\r)}(\s)\right)=0
\sp a = 1 \ldots {\rm dim}({\rm Cartan}\,\gfrak)
\sp j =0, \ldots \frac{\l}{\r(\s)}-1 \ .
\end{equation}
\end{subequations}
The $j$-factorized forms (see Eqs.~\eqref{facform0} and \eqref{facper})
of these twisted vertex operators and correlators are also easily obtained.

When $\l$=prime, these results simplify in the twisted sectors because
\begin{equation}
\r (\s) = \l = \mbox{prime} \sp \bar{r},\bar{s} = 0, \ldots ,\l-1 \sp j,l= 0 \sp \s = 1 ,
\ldots ,\l-1 \ .
\end{equation}
In this case it is conventional to suppress the degeneracy indices $j,l=0$,
and we obtain:
\begin{subequations}
\begin{equation}
\hj_{ra} \equiv \hj_{ra0} \sp \T_{ra0} \equiv \T_{ra} = T_a t_r \sp
t_r \equiv t_r (\s)
\end{equation}
\begin{equation}
\hj_{ra}(ze^{2\pi i}) = e^{- 2\pi i \srac{r}{\l} }\hj_{ra}(z) , \;\;\;
\hj_{r \pm \l,a} (z) = \hj_{ra}(z)
\end{equation}
\begin{equation}
\label{twcacy}
[ \hj_{r a }(m + \srac{r}{\l}), \hj_{s b }(n + \srac{s}{\l})] = k \l
\eta_{ab}(m + \srac{r}{\l}) \d_{m+n+\srac{r+s}{\l},0} \sp
\hj_{r \pm \l,a}( m +\srac{r\pm \l}{\l})= \hj_{ra}(m \pm 1 + \srac{r}{\l})
\end{equation}
\begin{equation}
L_\s(m) =    \frac{1}{2k \l} \sum_{r=0}^{\l -1}
 \sum_{p \in \sz} \eta^{ab} :\hat{J}_{r a}(p+\srac{r}{\l})
\hat{J}_{-r,b} (m-p-\srac{r}{\l}):_M +\delta_{m,0}
\hat{\Delta}_0 (\s)
\end{equation}
\begin{equation}
\hat{\Delta}_0 (\s)  =
\frac{1}{24} {\rm dim}({\rm Cartan}\,\gfrak) \left( \l -\frac{1}{\l} \right)
\end{equation}
\begin{equation}
[\hj_{r a } (m +\srac{r}{\l}), \hgp (\T,z,\s)]
= \hgp (\T,z,\s)  T_a t_{r}z^{m +\srac{r}{\l}}
\end{equation}
\begin{equation}
\label{ttrel}
 t_r t_s = t_{r+s} \sp
 t_{r \pm \l} = t_r \sp (t_r)_s{}^t = \d_{r+s-t,\,0 \rmod \l} \sp t_0 = \one \ .
\end{equation}
\begin{equation}
 a = 1 \ldots {\rm dim}({\rm Cartan}\,\gfrak) \sp
\bar r , \bar s = 0 , \ldots, \l - 1 \sp
 \s =1 ,\ldots , \l-1 \ .
 \end{equation}
\end{subequations}
Note that these relations are independent of $\s$.
This extends a well-known fact \cite{Borisov:1997nc} about $ \l$=prime
cyclic permutation orbifolds - abelian or nonabelian - that their twisted
current-algebraic formulation is independent of $\s$ in the twisted sectors.

Indeed, defining $\hat q \equiv \hat q(\s)$ and taking a $\s$-independent
Klein transformation $\hat \Gamma$ we find the $\s$-independent twisted
vertex operators and correlators
\begin{subequations}
\begin{eqnarray}
\hgp (\T,z,\s) &=& z^{-\Delta (T) \srac{\l -1}{\l}}
\hat \Gamma (\T,\hj_0 (0)) e^{i  \hq^{a} T_a}
z^{\srac{\eta^{ab}T_b}{\l k}  \hj_{0a}(0)} \nn \\
& & \times \exp \left( -\frac{\eta^{ab}T_b}{\l k} \sum_{r=0}^{\l-1}
 \sum_{m \leq -1} \hj_{r a } ( m + \srac{r}{\l})
\frac{ z^{-(m+ \srac{r}{\l} ) }} {m+ \srac{r}{\l} }
t_{-r} \right) \nn \\
& & \times \exp \left( -\frac{\eta^{ab}T_b}{\l k}
 \sum_{m \geq 1} \hj_{0 a } (m)
\frac{z^{-m}}{m}  \right) \nn \\
& & \times \exp \left( -\frac{\eta^{ab}T_b}{\l k}\sum_{r=1}^{\l-1}
 \sum_{m \geq 0} \hj_{r a}
( m + \srac{r}{\l})
\frac{ z^{-(m+ \srac{r}{\l} ) }} {m+ \srac{r}{\l} }t_{-r}  \right)
\end{eqnarray}
\begin{equation}
[\hq^{a}  , \hj_{r b} ( m + \srac{r}{\l})]
= i \d_b^a \d_{m + \srac{r}{\l},0} \sp [ \hq^a ,\hq^b ] = 0 \sp
\sp \s =1 ,\ldots , \l-1
\end{equation}
\end{subequations}
\begin{subequations}
\begin{eqnarray}
\langle \hgp (\T^{(1)},z_1,\s)  &&\!\!\!\!\!\!\! \!\! \cdots
\hgp (\T^{(N)},z_N,\s) \rangle_\s  \nn \\
& & =C_+ (\T) \left( \prod_{\r} z_\r^{- \srac{\l-1}{\l}
\Delta (T^{(\r)}) } \right)
\left(  \prod_{\r < \k} z_{\r \k}^{ \srac{T_{a}^{(\k)} \eta^{ab}
 T_{b}^{(\r)}}{ \l  k }  } \right) \nn \\
 & & \times \prod_{\r < \k} \exp \left\{ \frac{T_a^{(\k)} \eta^{ab}
 T_b^{(\r)}}{\l k} \sum_{r=1}^{\l-1} t_{r}^{(\k)}  t_{-r}^{(\r)}
I_{\srac{r}{\l}} \left( \srac{z_\r}{z_\k},\infty \right) \right\}
\hskip 1cm
\end{eqnarray}
\begin{equation}
C_+ (\T) \left( \sum_{\r =1}^N T_{a}^{(\r)}\right)=0
\sp a= 1 \ldots {\rm dim}({\rm Cartan}\, \gfrak) \sp \s =1 ,\ldots , \l -1
\end{equation}
\end{subequations}
in each twisted sector of these orbifolds.

Finally for the special case $H(\mbox{permutation}) =\z_2$ we find for the single
twisted sector $\s=1$
\begin{subequations}
\begin{equation}
\hj_{0a} (ze^{2\pi i}) = \hj_{0a}(z) \sp \hj_{1a} (ze^{2\pi i})=-\hj_{1a}(z)
\sp\hat{\Delta}_0  =\frac{1}{16} {\rm dim}({\rm Cartan}\,\gfrak)
\end{equation}
\begin{eqnarray}
\langle \hgp (\T^{(1)},z_1)\cdots \hgp (\T^{(N)},z_N) \rangle
 &=& C_+ (\T) \left( \prod_{\r} z_\r^{-\srac{1}{2} \Delta (T^{(\r)}) }
 \right) \left( \prod_{\r < \k} z_{\r \k}^{\frac{1}{2k}T_{a}^{(\k)} \eta^{ab}
 T_{b}^{(\r)} } \right) \nn \\
 & & \times \prod_{\r < \k}
\left( \frac{ \sqrt{z_\r} - \sqrt{z_\k}}{\sqrt{z_\r} + \sqrt{z_\k}}
\right)^{ \srac{T_a^{(\k)} \eta^{ab}
 T_b^{(\r)}}{2k}t_{1}^{(\k)}  t_{1}^{(\r)} }
\hskip 1cm
\end{eqnarray}
\begin{equation}
\label{z2glob}
C_+ (\T) \left(\sum_{\r =1}^N T_{a}^{(\r)} \right) = 0 \sp
a= 1 \ldots {\rm dim}({\rm Cartan}\, \gfrak)  \sp t_1^{(\k)} = t_{-1}^{(\k)}
 =\tau_1^{(\k)}
\end{equation}
\end{subequations}
where $\tau_1$ is the first Pauli matrix. Here we used the identities
\begin{equation}
\label{Iid0}
I_{\srac{1}{2}} (y,\infty)= \int_{\infty}^y \frac{dx}{x-1} x^{-\srac{1}{2}}
= -\sum_{n=0}^\infty
\frac{1}{n + \srac{1}{2}} y^{-(n+ \srac{1}{2})}
\sp
e^{I_{\srac{1}{2}} (y,\infty)} =
\frac{ \sqrt{y} - 1}{\sqrt{y} + 1}
\end{equation}
given in Ref.~\cite{deBoer:2001nw}.

\subsection{Example: The inversion orbifold
$A_{{\rm Cartan}\,\su (2)}(\z_2)/\z_2$ \label{invsec}}

The orbifold we consider here is a version of the outer-automorphic
inversion orbifold
\begin{equation}
\label{invout}
\frac{\mathfrak{u}(1)}{\z_2} : \qquad   J(z)' = - J(z) \sp \s=1 \ .
\end{equation}
In particular, we study the realization of this orbifold obtained
by embedding the $\mathfrak{u}(1)$ in an ambient $\su(2)$
\begin{equation}
\frac{A_{{\rm Cartan}\, \su(2)} (\z_2)}{\z_2} :
\qquad J_3(z)' = - J_3(z) \sp  \s=1
\end{equation}
which is a simple example in our class of abelian orbifolds. Note in
particular that this embedding promotes the outer automorphism
$J(z)' = -J(z)$ of the abelian current algebra into an inner automorphism of
the ambient $\su(2)$ current algebra.

In this case the $c=1$ stress tensor and untwisted currents of the symmetric CFT are
\begin{equation}
T(z) = \frac{1}{2k} : J_3 (z) J_3 (z) : \sp
J_3(z) J_3 (w) = \frac{k}{(z-w)^2} + \Ord (z-w)^0
\end{equation}
where $k$ is the level of the ambient affine $\su(2)$.
The solution to the {\it $H$-eigenvalue problem} in the twisted sector is very simple
\begin{equation}
\label{Heiginv}
\w U\hc = U\hc E \quad : \quad
 E = \omega  = -1 \sp \r = 2 \sp \nb = \mnb = 1 \sp U\hc =1
\end{equation}
and, choosing the normalization $\chi =1$, we obtain the simple duality
transformations
\begin{equation}
\label{datinv}
\G_{1,-1} = k \sp \G^{1,-1} = \frac{1}{k}
\sp {\cL}_{1,-1} = \frac{1}{2k} \ .
\end{equation}
Then the twisted sector of the orbifold is described by the $\hat c =1$
stress tensor and twisted currents
\begin{subequations}
\begin{equation}
\hat T (z) = \frac{1}{2k} : \hjh (z) \hj_{-1}(z) : \sp
\hjh (z e^{2 \pi i}) = -\hjh(z) \sp
\hj_{1\pm2 } (z) = \hjh (z)
\end{equation}
\begin{equation}
\hjh(z)=\sum_{m\in \sz} \hjh(m+\srac{1}{2})
z^{-(m+\frac{1}{2})-1} \sp
\hj_{1\pm2 } (m + \srac{1 \pm 2}{2}) = \hjh (m \pm 1 + \srac{1}{2})
\end{equation}
\begin{equation}
L (m)=\frac{1}{2k} \sum_{p\in \sz}
:\hjh(p+\srac{1}{2})\hj_{-1} (m-p-\srac{1}{2}):_M
+\delta_{m,0}\frac{1}{16}
\end{equation}
\begin{equation}
 [\hjh(m+\srac{1}{2}),
\hjh(n+\srac{1}{2})]=k (m+\srac{1}{2}) \ \delta_{m+n+1,0} ,\;\;\;
\hj_1 (m + \srac{1}{2})\hc = \hj_{-1}( - m -\srac{1}{2}) =
\hj_1 (-m-1+\srac{1}{2})
\end{equation}
\begin{equation}
\hjh( \foot{(} m+\srac{1}{2} \foot{)} \geq 0)|0\rangle =
\langle 0 | \hjh( \foot{(} m+\srac{1}{2} \foot{)} \leq 0) = 0
\sp (L (m\geq 0)-\delta_{m,0}\frac{1}{16})|0\rangle=0 \ .
\end{equation}
\end{subequations}
as a special case of \eqref{lnll0}.
This much is standard \cite{Halpern:1971qj,Halpern:2000vj} for any version
of the inversion orbifold $\mathfrak{u}(1)/\z_2$.

In current-algebraic orbifold theory, the representation theory of the
orbifold begins by finding the action $W(h_\s;T)$ of $h_\s \in H$ in
representation $T$, which solves the {\it linkage relation} \cite{deBoer:2001nw}
\begin{equation}
 W^\hcj(h_\s;T) T_a W(h_\s;T) = w(h_\s)_a{}^b T_b \sp
 \w (h_\s), W(h_\s;T) \in H
  \end{equation}
given the action $\w (h_\s)$ of $h_\s \in H$ in the adjoint
(see Eq.~\eqref{HsymCFT}). For our  realization
$A_{{\rm Cartan}\,\su(2)} (\z_2)/\z_2$ of the inversion orbifold, we
introduce the weight basis of the ambient $\su (2)$
\begin{subequations}
\begin{equation}
[T_A , T_B] = i \epsilon_{ABC} T_C \sp \sp A= 1,2,3 \sp \epsilon_{123} = 1
\end{equation}
\begin{equation}
(T_A)_{m'}{}^m = \langle j m'| J_A (0) | j m \rangle \sp
(T_3)_{m'}{}^m = m \de_{m'}^m \sp |m'|,|m| \leq j \sp
j = 0 , \frac{1}{2}, 1, \ldots
\end{equation}
\end{subequations}
where we have chosen root length $\alpha^2 = 1$. Then the
linkage relation and its solution for any spin $j$ are
\begin{subequations}
\label{linkinv}
\begin{equation}
W\hc (T) T_3 W(T) = - T_3\sp W\hc(T) W(T) = \one
\end{equation}
\begin{equation}
\label{lri}
W(T) = e^{i \pi T_2}
\sp R(T) = R(j) = \left\{
\begin{array}{ll}
 2 & \mbox{for integer} \; \, j \\
 4 & \mbox{for half  integer}\;\, j
 \end{array} \right.
 \end{equation}
 \end{subequations}
 where $R(T)$ is the order of $W(T)$.
Note that $W(T) \in SU(2)$, which reflects our promotion of the outer
automorphism \eqref{invout} to an inner automorphism of the ambient $\su(2)$.
The affine cutoff is $j \leq x/2$, $x=2k$ in this case.

The  next step is to solve the {\it extended $H$-eigenvalue problem}
\eqref{extHab} for representation $T$. In this case, the extended
$H$-eigenvalue problem and its solution are
\begin{subequations}
\label{exHeig}
\begin{equation}
\sum_{n'=-j}^j W(T)_n{}^{n'} U\hc(T)_{n'}{}^m = U\hc (T)_n{}^m E_m (T)
\sp |n|,|m| \leq j
\end{equation}
\begin{equation}
U\hc(T) = e^{- i \srac{\pi}{2} T_1 } \sp
E_m (T) = e^{i \pi m} = e^{-2 \pi i \srac{N(r)}{R(T)}}
\sp N(r) = - 2 m R(T)
\end{equation}
\end{subequations}
where $U\hc (T)$ is again in $SU (2)$.
Then we may use \eqref{Ttw} to obtain the twisted representation matrices
$\T$ in the twisted sector
\begin{subequations}
\label{exHeig1}
\begin{equation}
\hjh (z) \hgp (\T,w) = \frac{ \hgp (\T,w)}{z-w} \T (T) +\Ord(z-w)^0
\end{equation}
\begin{equation}
\T (T) \equiv \T_{\pm 1 } (T) =  U(T) T_3 U\hc (T) = - T_2 \sp
E(T)_{m'}{}^m= \de_{m'}^m e^{i \pi m}
\end{equation}
\begin{equation}
\label{selspj}
\T (T) = - E (T) \T (T) E^\ast (T)
\quad \leftrightarrow \quad (T_2)_{m'}{}^m ( 1 + e^{i \pi ( m'-m)}) =0
\end{equation}
\end{subequations}
where the $\T$-selection rule in \eqref{selspj} is self-evident because
$(T_2)_{m'}{}^m$ satisfies $|\Delta m| = 1$.

Using this data, the twisted vertex operator equation and its solution,
the twisted vertex operators
\begin{subequations}
\begin{equation}
\partial \hgp(\T,z) = \frac{\T}{k} : \hjh (z) \hgp (\T,z) :_M -
\frac{\T^2}{2kz} \hgp (\T,z) \sp \T \equiv \T(T) = - T_2
\end{equation}
\begin{equation}
\label{exlmvop}
\hgp (\T,z) = z^{-\srac{\T^2}{2k}}
\exp \left( - \frac{\T}{k} \sum_{m \leq -1} \hjh (m + \srac{1}{2})
\frac{z^{-(m+\srac{1}{2})}}{m +\srac{1}{2}} \right)
\exp \left( - \frac{\T}{k} \sum_{m \geq 0} \hjh (m + \srac{1}{2})
\frac{z^{-(m+\srac{1}{2})}}{m +\srac{1}{2}} \right)
\end{equation}
\begin{equation}
[ \hj_1 ( m +\srac{1}{2}),\hgp (\T,z)]= \hgp (\T,z) \T z^{m + \srac{1}{2}}
, \;\;\; \;[ L (m), \hgp (\T,z) ] = z^m \left( z \partial_z +
\frac{\T^2}{2k} (m+1) \right) \hgp (\T,z)
\end{equation}
\begin{equation}
\hgp (\T^{(1)},z_1) \hgp (\T^{(2)},z_2) = :\hgp (\T^{(1)},z_1)
\hgp (\T^{(2)},z_2):_M
\left( \frac{ \sqrt{z_1} - \sqrt{z_2}}{\sqrt{z_1} + \sqrt{z_2}}
\right)^{ \srac{\T^{(2)} \T^{(1)}}{k} }
\end{equation}
\end{subequations}
can be read off (ignore all structure with zero twist class $\bar{n}(r) =0$)
from Eq.~\eqref{vop}, the general results of
Subsecs.~\ref{sumsec}, \ref{propsec} and the identity \eqref{Iid0}.

Similarly, the correlators of the twisted vertex operators
\begin{equation}
\langle \hgp (\T^{(1)},z_1)  \cdots \hgp (\T^{(N)},z_N) \rangle =
 \left( \prod_{\r} z_\r^{- \srac{1}{2k} \T^{(\r)} \T^{(\r)} } \right)
  \prod_{\r < \k}
\left( \frac{ \sqrt{z_\r} - \sqrt{z_\k}}{\sqrt{z_\r} + \sqrt{z_\k}}
\right)^{\srac{\T^{(\k)} \T^{(\r)}}{k}}
\end{equation}
are easily read from the data and Eq.~\eqref{abcor0}. The absence of a global
Ward identity is another consequence of the absence of zero twist class
in the inversion orbifold.

\subsection{The twisted right-mover sectors of $A_{{\rm Cartan}\,g }(H)/H$
\label{rmsec} }

We turn now to the twisted right-mover sectors of our class
$A_{{\rm Cartan}\,g }(H)/H$ of abelian orbifolds, whose description may be read
off as the abelian limit of the right-mover sectors of the general WZW
orbifold in Ref.~\cite{deBoer:2001nw}.

One finds in Ref.~\cite{deBoer:2001nw} that the twisted right-mover currents
\begin{subequations}
\begin{equation}
\label{jmodr}
\hjb_\nrm (\bz e^{-2 \pi i},\s) = e^{-2 \pi i \srac{n(r)}{\r(\s)}}
\hjb_\nrm (\bz,\s) \sp
 \hjb_{n(r) \m}(\bz,\s) = \sum_{m  \in \sz} \hjb_{n(r)
\m} ( m + \srac{n(r)}{\rho (\s)} ) \bz^{(m+\srac{n(r)}{\rho (\s)})-1}
\end{equation}
\begin{equation}
\hjb_{n(r) \pm \r(\s),\m} (\bz,\s) = \hjb_\nrm (\bz ,\s) \sp
\hjb_{n(r) \pm \r (\s),\m} (m + \srac{n(r) \pm \r(\s)}{\r(\s)})
=\hjb_\nrm( m \pm 1 + \srac{n(r)}{\r(\s)})
\end{equation}
\end{subequations}
have the same monodromies as the twisted left-mover currents when the same
path is followed. Moreover, the twisted right-mover Virasoro operators, twisted right-mover
current algebra and ground state conformal weights are easily read from
Ref.~\cite{deBoer:2001nw}
\begin{subequations}
\begin{eqnarray}
\bar L_\s (m)  =   \frac{1}{2} \G^{\nrm;\mnrn}(\s) \!\!\!\!\!\!\!\!\! &
& \Big\{ \sum_{p \in \sz} :\hjb_{\nrm}(p+\srac{n(r)}{\r(\s)})
\hjb_{\mnrn}(-m-p-\srac{n(r)}{\r(\s)}):_{\bM} \nn \\ & & +
\delta_{m,0}\frac{\mnb}{2
\r(\s)}\left(1-\frac{\mnb}{\r(\s)}\right) \G_{\nrm;\mnrn}(\s)
\Big\}  \label{lnlr}
\end{eqnarray}
\begin{equation}
\label{malgr}
 [\hjb_\nrm(\mmrrs),\hjb_\nsn(\nnsrs)]=
  -(\mmrrs)\de_{\mnnrnsrsf,0}\sG_{\nrm;\mnrn}(\s)
\end{equation}
\begin{equation}
\label{gscr}
\hjb_\nrm ( m + \srac{n(r)}{\r (\s)} \leq 0 ) | 0 \rangle_\s
= {}_\s \langle 0 | \hjb_\nrm ( m + \srac{n(r)}{\r (\s)} \geq 0 ) = 0
\end{equation}
\begin{equation}
\Big( \bar{L}_\s (m \geq 0) - \delta_{m,0} \hat{\bar{\Delta}}_0
(\s) | 0 \rangle_\s = 0 \sp
\hat{\bar{\Delta}}_0 (\s) = \hat{\Delta}_0 (\s)
\end{equation}
\end{subequations}
where $\bar M$ normal ordering is defined in that reference and
$\overline{-n(r)}$ and ${\hat{\Delta}}_0(\s)$ are given respectively in
\eqref{mnbdef} and \eqref{cw}. The central charge is $\bar{\hat{c}} (\s) =
\hat c (\s) =c$. Similarly, the setup for the twisted
right-mover vertex operators $\hgm (\T,\bz,\s)$
\begin{subequations}
\begin{equation}
\label{hjbgc}
[ \hjb_\nrm (m + \srac{n(r)}{\r(\s)}),\hgm (\T,\bz,\s)] =
-\bz^{-(m+ \srac{n(r)}{\r(\s)})} \T_\nrm \hgm (\T,\bz,\s)
\end{equation}
\begin{equation}
[ L_\s (m), \hgm (\T,\bz,\s) ] = \bz^m \Big( \bz \partial_{\bz} +
\D (\T,\s) (m+1) \Big) \hgm (\T,\bz,\s)
 \end{equation}
\begin{equation}
\label{vopr}
\bar \partial \hgm(\T,\bz,\s) =  - \G^{n(r) \mu; -n(r), \nu}(\s)\T_{-n(r), \nu}
\left( : \hat{\bar{J}}_{n(r) \mu}(z) \hgm (\T,\bz,\s) :_{\bM} +
\srac{\overline{-n(r)}}{\r(\s)} \frac{1}{\bz} \T_{n(r) \mu} \hgm (\T,\bz,\s)\right)
\end{equation}
\begin{equation}
\label{jgMnor}
 : \hat{\bar{J}}_{n(r) \mu}(\bz) \hgm (\T,\bz,\s) :_{\bM} \  =
\hat{\bar{J}}_{n(r) \mu}^+(\bz) \hgm (\T,\bz,\s) +
\hgm (\T,\bz,\s)  \hat{\bar{J}}_{n(r) \mu}^-(\bz)
\end{equation}
\begin{equation}
\label{jm0r} \hat{\bar{J}}_{n(r) \mu}^+(\bz)  \equiv   \sum_{m > 0}
\hat{\bar{J}}_{n(r) \mu} (m-\srac{\mnb}{\r(\s)})
\bz^{(m-\srac{\mnb}{\r(\s)})-1}
\end{equation}
\begin{equation}
\label{jp0r}
\hat{\bar{J}}_{n(r) \mu}^-(\bz)  \equiv   \sum_{m\leq 0}
\hat{\bar{J}}_{n(r) \mu} (m-\srac{\mnb}{\r(\s)})
\bz^{(m-\srac{\mnb}{\r(\s)})-1}
\end{equation}
\end{subequations}
is easily read from Ref.~\cite{deBoer:2001nw}. The twisted representation
matrices $\T(T,\s)$ and the twisted conformal
weight matrix $\D (\T,\s)$ are given in \eqref{sumT}.
Again, we assume that the
twisted right-mover vertex operators $\hgm (\T,\bz,\s)$ are square matrices
which commute with the twisted representation matrices $\T = \T(T,\s)$.

Note the minus sign on the right side of the twisted right-mover
current algebra \eqref{malgr}. This sign change (relative to the
twisted left-mover current algebra \eqref{malg}) is a universal
phenomenon \cite{deBoer:2001nw} in the  right-mover sectors of
all current-algebraic orbifolds. It is known for permutation
orbifolds (and we will check below for the inversion orbifold)
that the twisted right-mover current algebra is rectifiable
\cite{deBoer:2001nw} to a copy of the twisted left-mover current
algebra (see also  Subsec~\ref{recsec}). More generally, it is known
\cite{deBoer:2001nw} for all current-algebraic orbifolds that the
twisted right-mover current algebra of sector $\s \leftrightarrow
h_\s$ is isomorphic to the twisted left-mover current algebra of
sector $h_\s^{-1}$. Here we exploit this isomorphism to solve the
twisted right-mover vertex operator equation \eqref{vopr}.

Following Ref.~\cite{deBoer:2001nw}, we first define the mode-number reversed
right-mover currents
\begin{subequations}
\label{double-bar-alg}
\begin{equation}
 \hjb^{\; \rm R}_\nrm(m +\srac{n(r)}{\r(\s)}) \equiv \hjb_{-n(r),\m} (-m-\srac{n(r)}{\r(\s)})
\end{equation}
\begin{equation}
[\hjb^{\;{\rm R}}_\nrm(m\+\srac{n(r)}{\r(\s)}), \hjb^{\; \rm R}_\nsn(n\+\srac{n(s)}{\r(\s)}) ] =
 (m+\srac{n(r)}{\r(\s)})\,\d_{m+n+\frac{n(r)+n(s)}{\r(\s)},\,0}\;
\tilde \sG_{n(r),\m;-n(r),\n}(\s) \ .
\end{equation}
\end{subequations}
Here $\tilde \sG (\s)$ is the twisted metric of sector $h_\s^{-1}$, which satisfies
\begin{equation}
\label{quar}
\tilde \sG_{\nrm; n(s)\n}(\s) \equiv
 \sG_{\mnrm; -n(s),\n}(\s) \propto \de_{n(r)+n(s), 0 \rmod \r(\s)} ,
\;\;\;
\tilde \sG_{\nrm; -n(r),\n}(\s) =  \sG_{ n(r)\n; -n(r),\m}(\s) \ .
\end{equation}
We will also need the inverse of $\tilde \sG (\s)$ which satisfies
\begin{subequations}
\begin{equation}
\tilde\sG_{\nrm; n(t)\de} (\s) \tilde \sG^{n(t)\de;\nsn}(\s)=\de_\nrm{}^\nsn
\sp \tilde \sG^{\nrm; -n(r),\n}(\s) =  \sG^{ n(r)\n; -n(r),\m}(\s)
\end{equation}
\begin{equation}
\sum_\de \tilde \sG_{\nrm;-n(r)\de} (\s) \tilde \sG^{-n(r)\de;n(r)\n}(\s)
= \de_\m{}^\n \ .
\end{equation}
\end{subequations}
When the twisted right-mover current algebra is rectifiable one finds that
$\tilde \sG (\s) = \sG (\s)$, but we will not need this fact here

We can also extend this isomorphism to the twisted representation matrices:
\begin{equation}
\label{twrmr}
[\hjb^{\; \rm R}_\nrm( m +\srac{n(r)}{\r(\s)}),\hgm(\T,\bz,\s)]
= \bz^{m+\srac{n(r)}{\r(\s)}} \tilde \T_\nrm \hgm(\T,\bz,\s)
\sp \tilde \T_\nrm (\s) \equiv - \T_\mnrm (\s)
\end{equation}
by rewriting Eq.~\eqref{hjbgc} in terms of $\hjb^{\; \rm R}$ and the
matrices $\tilde \T$.
This relation is isomorphic to the analogous left-mover result \eqref{jhgp} because
$\tilde \T$ commutes with $\hgm$.

Finally, the twisted right-mover vertex operator equation takes the form
\begin{subequations}
\begin{equation}
\label{ivopr}
\bar \partial \hgm(\T,\bz,\s) =  \tilde \G^{n(r) \mu; -n(r), \nu}(\s)
\tilde \T_{-n(r), \nu}
\left( : \hjb^{\; \rm R}_{n(r) \mu}(z) \hgm (\T,\bz,\s) :_{M}
-\srac{\nb}{\r(\s)} \frac{1}{\bz} \tilde \T_{n(r) \mu} \hgm (\T,\bz,\s)\right)
\end{equation}
\begin{equation}
\label{ijgMnor}
 : \hjb^{\; \rm R}_{n(r) \mu}(\bz) \hgm (\T,\bz,\s) :_{M} \  =
\hjb^{\; \rm R\,-}_{n(r) \mu}(\bz) \hgm (\T,\bz,\s) +
\hgm (\T,\bz,\s) \hjb^{\; \rm R \, + }_{n(r) \mu} (\bz)
\end{equation}
\begin{equation}
\label{ijm0r} \hjb^{\; \rm R\, -}_{n(r) \mu}(\bz) \equiv
\hjb_\nrm^+ (\bz)   =   \sum_{m \leq -1}
\hjb^{\; \rm R }_{\bnrm} (m+\srac{\nb}{\r(\s)})
\bz^{-(m+\srac{\nb}{\r(\s)})-1}
\end{equation}
\begin{equation}
\label{ijp0r}
\hjb^{\; \rm R \, +}_{n(r) \mu}(\bz)  \equiv \hjb_\nrm^- (\bz)=   \sum_{m\geq 0}
\hjb^{\; \rm R}_{\bnrm} (m+\srac{\nb}{\r(\s)})
\bz^{-(m+\srac{\nb}{\r(\s)})-1}
\end{equation}
\end{subequations}
where we have changed variable $n(r) \rightarrow -n(r)$ to obtain
\eqref{ivopr} from \eqref{vopr}.

This completes the isomorphism with Eq.~\eqref{vopt}
and allows us to read off the twisted
right-mover vertex operators of $A_{{\rm Cartan}\,g}(H)/H$
on inspection from the left-mover results
\begin{subequations}
\label{rmres}
\begin{equation}
z\rightarrow \bz \sp \T \rightarrow \tilde \T \sp
\hj \rightarrow \hjb^{\; \rm R} \sp \hat \Gamma \rightarrow \hat{\bar{\Gamma}}
\end{equation}
\begin{eqnarray}
\hgm (\T,\bz,\s) & = &  \bz^{-\tilde \G^{\nrm;\mnrn} (\s) \srac{\nb}{\r(\s)}
\tilde \T_\nrm \tilde \T_\mnrn}  \hat{\bar{\Gamma}} (\tilde \T,\hjb_0^{\; \rm R}
(0),\s) e^{i \bhq^\m (\s) \tilde \T_{0 \m} }
\nn \\ & & \times \bz^{\hjb_{ 0 \m}^{\; \rm R} (0) \tilde \G^{0\m; 0 \n}(\s)
\tilde \T_{0 \n} }
\hat{\bar{V}}_- (\tilde \T,\bz,\s)  \hat{\bar{V}}_+^{(0)}(\tilde \T,\bz,\s)
\hat{\bar{V}}_+(\tilde \T,\bz,\s)
  \label{rvop2}
\end{eqnarray}
\begin{equation}
\hat{\bar{V}}_- (\tilde \T,\bz,\s) \equiv \exp \left\{ - \tilde \G^{\nrm; \mnrn} (\s)
\sum_{m \leq -1} \hjb_\bnrm^{\; \rm R} (\mnrrs) \frac{\bz^{-(\mnrrs)}}{\mnrrs}
\tilde \T_\mnrn
\right\}
\end{equation}
\begin{equation}
\hat{\bar{V}}_+^{(0)} (\tilde \T,\bz,\s) \equiv \exp \left\{ - \tilde \G^{0 \m; 0 \n} (\s)
\sum_{m \geq 1} \hjb_{0 \m}^{\; \rm R} (m) \frac{\bz^{-m}}{m }\tilde \T_{ 0 \n}
\right\}
\end{equation}
\begin{equation}
\hat{\bar{V}}_+ (\tilde \T,\bz,\s) \equiv \exp \left\{ -\tilde \G^{\nrm; \mnrn} (\s)
 ( 1- \d_{\nb,0}) \sum_{m \geq 0 } \hjb_\bnrm^{\; \rm R} (\mnrrs)
 \frac{\bz^{-(\mnrrs)}}{\mnrrs} \tilde \T_\mnrn \right\}
\end{equation}
\begin{equation}
[ \bhq^\m (\s), \hjb_{\nsn}^{\; \rm R} (n + \srac{n(s)}{\r (\s)} ) ] =
i \d_\n^\m \d_{n + \srac{n(s)}{\r(\s)} , 0} \sp
[\bhq^\m (\s),\bhq^\n (\s)] =0 \sp \srange
\end{equation}
\end{subequations}
where $\hat{\bar{\Gamma}}$ is an undetermined right-mover Klein transformation.
The twisted mode algebra of $\hjb^{\; \rm R}$ and the quantities
$\tilde \sG$, $\tilde \T$ are given in Eqs.~\eqref{double-bar-alg},
\eqref{quar} and \eqref{twrmr}.
At this point we verify our initial assumption that $\hgm$ is a square matrix
which commutes with the twisted representation matrices $\T$.

Using \eqref{rmres} we may now discuss for the twisted right-mover vertex operators all the properties
given for the twisted left-mover vertex operators in Subsec.~\ref{propsec}.
For brevity we confine ourselves here to the following three topics.

\noindent $\bullet$ {\bf Intrinsic monodromy of the right-mover
vertex operators} \nl
The monodromies of the twisted right-mover vertex operators follow the same
line as given above for the left movers. The result in sector $\s$
of $A_{{\rm Cartan}\,g}(H)/H$ is
\begin{subequations}
\begin{eqnarray}
\hgm (\T,\bz e^{-2 \pi i},\s) & = &  E(T,\s) \hgm (\T,\bz,\s) E (T,\s)^\ast
\nn \\
& & \times
e^{2 \pi i ( \tilde \G^{\nrm ; \mnrn} (\s) \srac{\nb}{\r(\s)}
\tilde \T_\nrm \tilde \T_\mnrn - \hjb_{0\m }^{\; \rm R}(0) \tilde \G^{0\m; 0 \n}
 (\s) \tilde \T_{0 \n} ) }
\label{intmonr}
\end{eqnarray}
\begin{equation}
\label{Tselr}
e^{ 2 \pi i \srac{n(r)}{\r (\s)} } \tilde \T_\mnrn =
E(T,\s) \tilde \T_\mnrn  E (T,\s)^\ast \sp \srange
\end{equation}
\end{subequations}
where the $\T$-selection rule \eqref{Tselal} was used in the form \eqref{Tselr}.

\noindent $\bullet$ {\bf Right-mover orbifold correlators} \nl
The right-mover correlators of sector $\s$ of $A_{{\rm Cartan}\,g}(H)/H$ are:
\begin{subequations}
\begin{eqnarray}
\hat A_- (\T,\bz,\s) & & \equiv
\langle \hgm (\T^{(1)},\bz_1,\s)  \cdots \hgm (\T^{(N)},\bz_N,\s) \rangle_\s
\sp \srange \\
& & = C_- (\tilde \T,\s) \left( \prod_{\r} \bz_\r^{- \tilde \G^{\nrm; \mnrn} (\s)
\srac{\nb }{\r (\s)} \tilde \T_\nrm^{(\r)} \tilde \T_\mnrn^{(\r)} } \right)
\left(  \prod_{\r < \k} \bz_{\r \k}^{\tilde \T_{0 \m}^{(\k)} \tilde \G^{0 \m; 0 \n } (\s)
 \tilde \T_{0 \n}^{(\r)} } \right) \nn \\
 & & \times \prod_{\r < \k} \exp \left\{  \tilde \T_\nrm^{(\k)} \tilde
 \G^{\nrm; \mnrn} (\s) \tilde \T_\mnrn^{(\r)} ( 1 - \d_{\nb ,0} )
I_{\srac{\nb}{\r (\s)}} \left( \srac{\bz_\r}{\bz_\k},\infty \right) \right\}
\hskip 1cm
\end{eqnarray}
\begin{equation}
C_- (\tilde \T,\s) \equiv \langle \hG_-(\tilde \T^{(1)},\s) \cdots
 \hG_-(\tilde\T^{(N)},\s)
\rangle_\s \sp \hG_- (\tilde \T,\s) \equiv
\hat{\bar{\Gamma}}(\tilde \T,\hjb_0^{\; \rm R} (0),\s) e^{i \bhq^\m(\s) \tilde \T_{0 \m}}
\end{equation}
\begin{equation}
\label{rgwi}
\langle [ \hjb_{0\m}^{\; \rm R} (0),\hG_- (\tilde \T^{(1)},\s) \cdots \hG_-
(\tilde \T^{(N)},\s)]
\rangle_\s =0 \quad \Rightarrow \quad
 \left(\sum_{\r =1}^N \tilde \T_{0 \m}^{(\r)}\right) C_- (\tilde \T,\s)
 =0 \sp \forall \; \m \ .
\end{equation}
\end{subequations}
As for the twisted left movers, the sidedness of the right-mover global Ward
identity is irrelevant and we may equivalently write the right-mover
identity as
\begin{equation}
C_- (\tilde \T,\s)\left(\sum_{\r =1}^N \tilde \T_{0 \m}^{(\r)}\right) =0
\sp \forall \; \m
\end{equation}
because $\T $ and $\tilde \T$ commute with the twisted vertex operators.
Similarly, the constant factor $C_- (\tilde \T,\s)$ can be moved to the right of the
$\bz$-dependent factor in the twisted vertex operators.

\noindent $\bullet$ {\bf In terms of untilded quantities} \nl
Using \eqref{double-bar-alg}, \eqref{quar} and \eqref{twrmr}, all the twisted
right-mover results can be expressed in terms of the original untilded
quantities $\hjb$, $\G$ and $\T$. For example, one finds that
\begin{equation}
\label{untid}
\tilde \G^{\nrm ; \mnrn} (\s) \srac{\nb}{\r(\s)}
\tilde \T_\nrm \tilde \T_\mnrn = \G^{\nrm ; \mnrn} (\s) \srac{\nb}{\r(\s)}
 \T_\nrm  \T_\mnrn
\end{equation}
because the $\T$'s commute and $\G$ is symmetric. It follows that, under
$\bz \rightarrow z$, the first factor in the twisted right-mover vertex
operator \eqref{rvop2} is the same factor we found for the twisted
left-mover vertex operators in \eqref{vop2}. As another example, we give
the result for the right-mover orbifold correlators
\begin{subequations}
\begin{eqnarray}
\hat A_- (\T,\bz,\s) & =& C_- ( -\T,\s)
\left( \prod_{\r} \bz_\r^{-  \G^{\nrm; \mnrn} (\s)
\srac{\nb }{\r (\s)} \T_\nrm^{(\r)}  \T_\mnrn^{(\r)} } \right)
\left(  \prod_{\r < \k} \bz_{\r \k}^{ \T_{0 \m}^{(\k)}  \G^{0 \m; 0 \n } (\s)
  \T_{0 \n}^{(\r)} } \right) \nn \\
 & & \times \prod_{\r < \k} \exp \left\{ \T_\nrm^{(\r)}
 \G^{\nrm; \mnrn} (\s)  \T_\mnrn^{(\k)} ( 1 - \d_{\nb ,0} )
I_{\srac{\nb}{\r (\s)}} \left( \srac{\bz_\r}{\bz_\k},\infty \right) \right\}
\hskip 1cm
\end{eqnarray}
\begin{equation}
\left(\sum_{\r =1}^N \T_{0 \m}^{(\r)} \right) C_- (-\T,\s) =
C_- (-\T,\s)\left(\sum_{\r =1}^N \T_{0 \m}^{(\r)} \right) = 0 \sp \forall \; \m
\sp \srange
\end{equation}
\end{subequations}
in each sector $\s$ of all $A_{{\rm Cartan}\,g }(H)/H$.

\subsection{Full non-chiral results \label{ncsec} }

Combining the twisted left- and right-mover results above, we obtain the full
twisted vertex operators $\hg = \hgm \hgp $ and their intrinsic monodromy
in each sector $\s$ of $A_{{\rm Cartan}\,g }(H)/H$:
\begin{subequations}
\begin{equation}
\hg (\T,\bz,z,\s) = \hgm (\T,\bz,\s) \hgp (\T,z,\s) \sp \srange
\end{equation}
\begin{equation}
\label{monnc}
\hg (\T,\bz e^{-2 \pi i} ,z e^{2 \pi i},\s) =
E(T,\s)\hg (\T,\bz ,z ,\s) E(T,\s)^\ast
e^{2 \pi i (\hj_{0\m}(0) + \hjb_{0\m} (0)) \G^{0\m;0\n}(\s)\T_{0\n}} \ .
\end{equation}
\end{subequations}
Here we used Eqs.~\eqref{intmon0}, \eqref{intmonr} and \eqref{untid}
 to obtain the result \eqref{monnc}. Except for
the last factor, which is a quantum correction, the monodromy \eqref{monnc}
is consistent with the {\it classical} monodromy
\begin{equation}
\label{clasmon}
\hg (\T,\bz e^{-2 \pi i} ,z e^{2 \pi i},\s) =
E(T,\s)\hg (\T,\bz ,z ,\s) E(T,\s)^\ast \sp \srange
\end{equation}
obtained for classical group orbifold elements in Ref.~\cite{deBoer:2001nw}.

For the full orbifold correlators $\hat A = \hat A_- \hat A_+$ we obtain
\begin{subequations}
\label{orbcor}
\begin{eqnarray}
 & &  \hat A  (\T,\bz,z,\s) =   \langle \hg(\T^{(1)},\bz_1,z_1,\s) \cdots
\hg(\T^{(N)},\bz_N,z_N,\s) \rangle_\s \sp \srange \;\;\;\;\; \;\;\;\;\;\;\\
&  & =  C(\T,\s)\prod_{\r} |z_\r|^{- 2 \G^{\nrm; \mnrn} (\s)
\srac{\nb }{\r (\s)} \T_\nrm^{(\r)}  \T_\mnrn^{(\r)} }
   \prod_{\r < \k} |z_{\r \k}|^{ 2 \T_{0 \m}^{(\k)}  \G^{0 \m; 0 \n } (\s)
  \T_{0 \n}^{(\r)} } \prod_{\r < \k} e^{F(\r,\k) } \nn \\
\end{eqnarray}
\begin{eqnarray}
F(\r,\k) & \equiv & \G^{\nrm; \mnrn} (\s)( 1 - \d_{\nb ,0} ) \nn \\
& & \times \left( \T_\nrm^{(\r)} \T_\mnrn^{(\k)}
I_{\srac{\nb}{\r (\s)}} \left( \srac{z_\r}{z_\k},\infty \right)
+\T_\nrm^{(\k)} \T_\mnrn^{(\r)}
I_{\srac{\nb}{\r (\s)}} \left( \srac{\bz_\r}{\bz_\k},\infty \right) \right)
\end{eqnarray}
\begin{equation}
\label{glwifc}
C(\T,\s) \equiv C_- (-\T,\s) C_+ (\T,\s)
\sp\left(\sum_{\r =1}^N \T_{0 \m}^{(\r)} \right) C(\T,\s) =
C (\T,\s)\left(\sum_{\r =1}^N \T_{0 \m}^{(\r)} \right) = 0
\end{equation}
\end{subequations}
where the function $F(\r,\k)$ is defined only for $\r \neq \k$ and the
global Ward identities \eqref{glwifc} hold for all degeneracy indices $\m$.
As expected, the orbifold correlators \eqref{orbcor} are symmetric under the
exchange of any two full twisted vertex operators
\begin{equation}
\hg (\T^{(\r)},\bz_\r,z_\r,\s) \leftrightarrow \hg (\T^{(\k)},\bz_\k,z_\k,\s) \ .
\end{equation}
This is checked explicitly in App.~\ref{idnc}, where the identities
\begin{subequations}
\label{symid0}
\begin{equation}
\label{Iid}
I_{\srac{\nb}{\r (\s)}} (y,y_0) = I_{\srac{\r(\s)-\nb}{\r (\s)}} (y^{-1},y_0^{-1})
\sp I_{\srac{\nb}{\r (\s)}} (y,y_0) \equiv \int_{y_0}^y \frac{dx}{x-1}
 x^{-\srac{\nb}{\r (\s)}}
 \end{equation}
\begin{equation}
\label{symid}
F(\k,\r) = F(\r,\k) + \Delta
\end{equation}
\begin{equation}
\Delta =\G^{\nrm; \mnrn} (\s)( 1 - \d_{\nb ,0} )
\left( \T_\nrm^{(\r)} \T_\mnrn^{(\k)} +\T_\nrm^{(\k)} \T_\mnrn^{(\r)} \right)
I_{\srac{\nb}{\r (\s)}} (\infty,0) = 0 \ .
\end{equation}
\end{subequations}
are used to establish the required symmetry of $F$. The identity in
\eqref{Iid} was given in Ref.~\cite{deBoer:2001nw}.

\subsection{Example: More about the inversion orbifold}

The twisted left-mover sector of the inversion orbifold
$A_{{\rm Cartan}\,\su (2)}(\z_2)/\z_2$ was studied in Subsec.~\ref{invsec}.
Here we illustrate the discussion above by providing the corresponding
twisted right-mover and non-chiral results for this simple case.

Using \eqref{vopr}, \eqref{rvop2} and the data \eqref{Heiginv}, \eqref{datinv}
we find the twisted
right-mover vertex operator equation and the twisted right-mover vertex
operators
\begin{subequations}
\begin{equation}
\bar \partial \hgm(\T,\bz) = -\frac{\T}{k} : \hjb_1 (\bz) \hgm (\T,\bz) :_{\bM}
- \frac{\T^2}{2k\bz} \hgm (\T,\bz) \sp \T = \T(T) = - T_2
\end{equation}
\begin{equation}
\label{exrmvop}
\hgm (\T,\bz) = \bz^{-\srac{\T^2}{2k}}
\exp \left(  \frac{\T}{k} \sum_{m \leq -1} \hjb_1^{\; \rm R} (m + \srac{1}{2})
\frac{\bz^{-(m+\srac{1}{2})}}{m +\srac{1}{2}} \right)
\exp \left( \frac{\T}{k} \sum_{m \geq 0} \hjb_1^{\; \rm R} (m + \srac{1}{2})
\frac{\bz^{-(m+\srac{1}{2})}}{m +\srac{1}{2}} \right) \ .
\end{equation}
\end{subequations}
In this case, it is easy to check from Eq.~\eqref{malgr} that the
mode-reversed currents satisfy a copy of the twisted left-mover current
algebra
\begin{subequations}
\begin{equation}
[\hjb_1 (m+\srac{1}{2}),\hjb_1 (n+\srac{1}{2})]
= -k (m+\srac{1}{2}) \de_{m+n+1,0}
\end{equation}
\begin{equation}
\hjb_1^{\; \rm R} (m+\srac{1}{2}) = \hjb_{-1}(-m -\srac{1}{2})
= \hjb_1(-m-1 + \srac{1}{2})
\end{equation}
\begin{equation}
[\hjb_1^{\; \rm R} (m+\srac{1}{2}),\hjb_1^{\; \rm R} (n+\srac{1}{2})]
= k (m+\srac{1}{2}) \de_{m+n+1,0}
\end{equation}
\end{subequations}
and hence that the mode-reversed currents provide a realization of the
rectified right-mover currents $\hjbb$ of Ref.~\cite{deBoer:2001nw}.

Combining the result \eqref{exrmvop} with the twisted left-mover vertex
operators in \eqref{exlmvop}, we find the full twisted vertex operators
$\hg  = \hgm  \hgp $:
\begin{subequations}
\begin{eqnarray}
\hg (\T,\bz,z)
& = & |z|^{-\srac{\T^2}{k}}
\exp \left(  \frac{\T}{k} \sum_{m \leq -1} \frac{1}{m +\srac{1}{2}}
\Big( \hjb_1^{\; \rm R} (m + \srac{1}{2})\bz^{-(m+\srac{1}{2})}
-\hj_1 (m + \srac{1}{2})z^{-(m+\srac{1}{2})} \Big) \right) \nn \\
& & \times
\exp \left(  \frac{\T}{k} \sum_{m \geq 0} \frac{1}{m +\srac{1}{2}}
\Big( \hjb_1^{\; \rm R} (m + \srac{1}{2})\bz^{-(m+\srac{1}{2})}
-\hj_1 (m + \srac{1}{2})z^{-(m+\srac{1}{2})} \Big) \right) \\
 & = &|z|^{-\srac{\T^2}{k}} : e^{i \hb^1 (\bz,z) \T} :_M
\end{eqnarray}
\begin{equation}
i \hb^1 (\bz,z) = \frac{1}{k}\sum_{m \in \sz} \frac{1}{m +\srac{1}{2}}
\Big( \hjb_1^{\; \rm R} (m + \srac{1}{2})\bz^{-(m+\srac{1}{2})}
-\hj_1 (m + \srac{1}{2})z^{-(m+\srac{1}{2})} \Big) \ .
\end{equation}
\end{subequations}
Because the inversion orbifold has no zero twist class $\nb = 0$, we find
that the intrinsic monodromy of the full twisted vertex operators
\begin{subequations}
\label{intfvop}
\begin{equation}
\label{Tselg}
\hb^1(\bz e^{-2 \pi i},z e^{2\pi i}) = -\hb^1 (\bz,z) \sp
\T(T) = - E(T) \T(T) E(T)^\ast
\end{equation}
\begin{eqnarray}
\hg (\T,\bz e^{-2 \pi i},z e^{2\pi i}) & = &  E(T) \hg (\T,\bz,z) E(T)^\ast
\nn \\
  &  \leftrightarrow  & \quad
\hg (\T,\bz e^{- 2 \pi i}, z e^{2 \pi i})_{m'}{}^m =
e^{ i \pi (m' -m)} \hg (\T,\bz,z)_{m'}{}^m ,\;\; |m'| , |m| \leq j \nn \\
\label{Tselg2}
\end{eqnarray}
\end{subequations}
is in agreement with the classical monodromy \eqref{clasmon} of the corresponding
group orbifold elements.

In this connection we also use the results of Ref.~\cite{deBoer:2001nw} to
give the explicit forms of the group elements $g(T)$, the eigengroup
elements $\sg (\T)$ and the group orbifold elements $\hg (\T)$
\begin{subequations}
\begin{equation}
g (T,\bz,z) = e^{i \be (\bz,z) T_3} \sp \be  (\bz,z)' = - \be (\bz,z)
\end{equation}
\begin{equation}
g (T,\bz,z)' = W(T) g (T,\bz,z) W\hc (T) = e^{i \be (\bz,z)' T_3}
= e^{-i \be (\bz,z) T_3}
\end{equation}
\begin{equation}
\sg (\T,\bz,z) = U(T) g (T,\bz,z) U\hc(T) = e^{i \be (\bz,z) \T} =
e^{-i \be (\bz,z) T_2}
\end{equation}
\begin{equation}
 \sg (T,\bz,z)'= E(T) \sg (T,\bz,z) E^\ast (T)
\end{equation}
\begin{equation}
\hg (\T,\bz,z) =e^{i \hb (\bz,z) \T} = e^{-i \hb (\bz,z) T_2}
\end{equation}
\begin{equation}
\hb (\bz e^{- 2\pi i  },z e^{2 \pi i}) = - \hb (\bz,z) \sp
\hg (\T,\bz e^{- 2\pi i },z e^{2 \pi i}) = E (T) \hg (\T,\bz,z) E^\ast(T)
\end{equation}
\end{subequations}
in the classical theory of the inversion orbifold. Here the primed quantities
are the responses to the inversion in the symmetric theory, and
$W(T)$, $U\hc(T)$ and $E(T)$ are given in Eqs.~\eqref{linkinv} and
\eqref{exHeig1}.

Continuing with the quantum theory, we give the right-mover and the full
non-chiral correlators of the inversion orbifold:
\begin{subequations}
\begin{equation}
\langle \hgm (\T^{(1)},\bz_1)  \cdots \hgm (\T^{(N)},\bz_N) \rangle=
 \left( \prod_{\r} \bz_\r^{- \srac{1}{2k} \T^{(\r)} \T^{(\r)} } \right)
  \prod_{\r < \k}
\left( \frac{ \sqrt{\bz_\r} - \sqrt{\bz_\k}}{\sqrt{\bz_\r} + \sqrt{\bz_\k}}
\right)^{\srac{\T^{(\k)} \T^{(\r)}}{k}}
\end{equation}
\begin{equation}
\label{npcornc}
\langle \hg (\T^{(1)},\bz_1,z_1)  \cdots \hg (\T^{(N)},\bz_N,z_N) \rangle =
\left( \prod_{\r} |z_\r|^{- \srac{1}{k} \T^{(\r)} \T^{(\r)} } \right)
  \prod_{\r < \k}
\left| \frac{ \sqrt{z_\r} - \sqrt{z_\k}}{\sqrt{z_\r} + \sqrt{z_\k}}
\right|^{\srac{2}{k} \T^{(\k)} \T^{(\r)} } \ .
\end{equation}
\end{subequations}
Note that the intrinsic monodromy \eqref{Tselg2} of the full vertex operators is
consistent with the form of the full correlators, e.g.
\begin{subequations}
\begin{eqnarray}
 & & \langle \hg (\T^{(1)},\bz_1e^{-2 \pi i},z_1e^{2 \pi i})  \cdots
\hg (\T^{(N)},\bz_Ne^{-2 \pi i},z_N e^{2 \pi i}) \rangle \nn \\
\label{eid0}
&= &\langle \hg (\T^{(1)},\bz_1,z_1)  \cdots \hg (\T^{(N)},\bz_N,z_N)\rangle \\
&=  &E(\{T\})
\langle \hg (\T^{(1)},\bz_1,z_1)  \cdots \hg (\T^{(N)},\bz_N,z_N)
\rangle E(\{T\})^\ast \label{eid}
\end{eqnarray}
\begin{equation}
 E(\{T\}) = \otimes_{\r =1}^N E(T^{(\r)}) \sp
 E(\{T\})^\ast = \otimes_{\r =1}^N E(T^{(\r)})^\ast \ .
\end{equation}
\end{subequations}
Here the equality \eqref{eid0} follows from the correlators \eqref{npcornc},
while the equality \eqref{eid} follows from the monodromy \eqref{intfvop}
of the full twisted vertex operators. The agreement of the two forms follows
from the $\T$-selection rule \eqref{Tselg2} and the fact that
the $\T$'s appear quadratically in the correlators.

\section{The Charge Conjugation Orbifold on $\su(n)$ \label{sec3}}

\subsection{Charge conjugation}

As our second large example, we apply the general results of
Ref.~\cite{deBoer:2001nw} to study the (outer-automorphic%
\footnote{At the level of characters, useful references for outer-automorphically twisted affine Lie
algebras include [19--22].} 
) charge conjugation orbifold on $\su (n)$
\begin{equation}
\frac{A_{\su(n)}(\z_2)}{\z_2} \sp n \geq 3 \ .
\end{equation}
One can follow our development through for the charge
conjugation orbifold $A_{\su(2)}(\z_2)/\z_2$ as well, but in this case the automorphism
is an inner automorphism and all irreps are real representations.

For accessibility in physics we will work in the standard Cartesian basis of
$\su(n)$, with generalized $n\times n$ Gell-Mann matrices $\l_a\hc=\l_a$
 and real structure constants  $f_{ab}{}^c= f_{abc}$
\begin{subequations}
\begin{equation}
[\l_a,\l_b] = 2 i F_{abc} \l_c \sp f_{abc} = \sqrt{\psi^2} F_{abc} \sp
G_{ab} = k \de_{ab}
\end{equation}
\begin{equation}
{\rm Tr}(\l_a \l_b) = 2 \de_{ab} \sp {\rm Tr}\ \l_a = 0
\sp a,b =1 , \ldots , n^2-1
\end{equation}
\end{subequations}
where $\sqrt{\psi^2}$ is the root length of $\su(n)$ and $k$ is the level of
affine $\su(n)$. The standard iterative scheme is assumed for the generalized
Gell-Mann matrices of $\su(n \geq 4)$, so that
 e.g. $\l_9$ for $\su(4)$ has a 1 in the $(1,4)$ and $(4,1)$ entries.

In this basis, the OPEs of affine $\su(n)$ are
\begin{subequations}
\label{afope}
\begin{eqnarray}
J_A (z) J_B( w) & = & \frac{k \de_{AB}}{(z-w)^2} + \frac{if_{ABC} J_C(w)}{z-w}
+ \Ord (z-w)^ 0 \\
J_A (z) J_I( w) & = &  \frac{if_{AIJ} J_J(w)}{z-w}
+ \Ord (z-w)^ 0 \\
J_I (z) J_J( w) & = & \frac{k \de_{IJ}}{(z-w)^2} + \frac{if_{IJA} J_A(w)}{z-w}
+ \Ord (z-w)^ 0
\end{eqnarray}
\begin{equation}
A \in h  = \so(n) \sp I \in \frac{g}{h} = \frac{\su(n)}{\so(n)}
\end{equation}
\end{subequations}
where we have decomposed the affine algebra according to the symmetric space
\begin{equation}
\label{irem}
 \frac{\su(n)_x}{\so(n)_{2\tau x}} \sp
x= \frac{2k}{\psi^2} \sp \tau = \left\{
\begin{array}{ll}
2 & {\rm for}\,\; n =3 \\
1 & {\rm for}\,\; n \geq 4  \ .
\end{array} \right.
\end{equation}
We will choose in particular the $\so(n)$ subalgebra which corresponds to
the Cartan-Weyl generators $\{i(\tilde E_\a - \tilde E_{-\a}), \forall \; \a > 0  \}$,
so that for example:
\begin{subequations}
\begin{eqnarray}
\su(3) \quad & : & \quad A= 2,5,7 \sp I = 1,3,4,6,8 \\
\su(4) \quad & : & \quad A= 2,5,7,10,12,14 \sp I = 1,3,4,6,8,9,11,13,15 \ .
\end{eqnarray}
\end{subequations}
Then one finds for the fundamental representation of $\su(n)$
\begin{equation}
 T_a = \frac{\sqrt{\psi^2}}{2} \l_a \sp {\rm Tr} (T_a T_b) =
\frac{\psi^2}{2} \de_{ab} \sp
T_a\hc = T_a \sp T_A^\ast = - T_A
\sp T_I^\ast = T_I
\end{equation}
that the $\so(n)$ generators $\{T_A \}$
are proportional to the standard generators of Cartesian $\so(n)$ in the
vector representation
\begin{subequations}
\begin{equation}
(T_{ij})_{k}{}^{l} = i\frac{\sqrt{\psi^2}}{2}
(\de_{ik} \de_{j}^{l} - \de_{jk}\de_{i}^{l}) =
 i\sqrt{\frac{\tau \psi_{\so(n)}^2}{2}}
(\de_{ik} \de_{j}^{l} - \de_{jk}\de_{i}^{l})
\end{equation}
\begin{equation}
1 \leq i < j \leq n \sp k,l =1 \ldots n \sp
x_{\so(n)} = \frac{\psi^2}{\psi_{\so(n)}^2} x = 2 \tau x
\end{equation}
\end{subequations}
where $T_A \leftrightarrow -T_{i<j}$ is a relabelling of each $n\times n$
matrix $T_A$ by its non-zero entries. The quantity $\psi_{\so (n)}$ is
the highest root of $\so (n)$, and this choice of $\so(n)$ is
inner-automorphically
equivalent to any other $\so(n)$ subalgebra which
is irregularly embedded at the level shown in \eqref{irem}.

Note in particular that the commuting Cartan generators of $\su(n)$
\begin{subequations}
\begin{equation}
J_{\dot{a}} \sp \dot{a} \in {\rm Cartan}\, \su(n) \sp {\rm Cartan}\, \su(n)
\subset \frac{\su(n)}{\so(n)}
\end{equation}
\begin{equation}
\su(3) \;\;\;: \;\;\; \dot{a} = 3,8 \sp \su(4) \;\;\; : \;\;\; \dot{a} =3,8,15
\sp \su(5)  \;\;\; : \;\;\; \dot{a} =3,8,15,24
\end{equation}
\end{subequations}
are always in $g/h$. In this notation, the general matrix irrep $T$ of $\su(n)$
satisfies
\begin{subequations}
\begin{equation}
(T_a)_{\l^i}{}^{\l^j} = \langle \l^i(T) | J_a (0) | \l^j(T) \rangle \sp T_a\hc=T_a
\end{equation}
\begin{equation}
[ J_a (0), J_b (0) ] = if_{abc} J_c (0) \sp [T_a,T_b]= if_{abc} T_c
\end{equation}
\begin{equation}
(T_{\dot{a}})_{\l^i}{}^{\l^j} = \l^i_{\dot{a}} (T) \de_{\l^i}^{\l^j } \sp \dot{a} \in
{\rm Cartan}\,\su(n) \subset \frac{\su(n)}{\so(n)}
\end{equation}
\end{subequations}
where $J_a (0)$ are the generators of $\su(n)$ and
 $ \l^i (T)$, $i = 1 \ldots {\rm dim}\,T$ are the weights of irrep $T$.

We turn next to the non-trivial element ($\s=1$) of the charge conjugation
automorphism group, whose action on the currents is
\begin{subequations}
\label{data00}
\begin{equation}
J_a (z)' = \w_a{}^b J_b (z) \sp \w\hc = \w
\end{equation}
\begin{equation}
\label{oms}
\w_A{}^A = - \w_I{}^I = 1 \sp A \in \so(n)  \sp I \in \su(n)/\so(n)
\end{equation}
\end{subequations}
where $\w$ is a diagonal matrix whose diagonal elements are given in
\eqref{oms}. Because $\su(n)/\so(n)$ is a symmetric space it is trivial to
check that this is an automorphism of the affine OPEs \eqref{afope},
with invariant subalgebra $h=\so(n)$. To identify this automorphism as the Cartesian form of
charge conjugation we consider for each irrep $T$ the following two
closely-related representations of $\su(n)$
\begin{subequations}
\label{Tpr}
\begin{equation}
\bar{T}_a \equiv -T_a^{\rm T} \sp T_a{}' \equiv \w_a{}^b T_b
\end{equation}
\begin{equation}
[\bar T_a , \bar T_b] = if_{abc}\bar T_c  \sp [T_a{}',T_b{}'] = if_{abc} T_c{}'
\end{equation}
\end{subequations}
where superscript T is matrix transpose, $\bar{T}$ is the standard charge conjugate representation of $T$
and $T'$ is the automorphic transform of $T$. In fact, $T'$ is unitarily
equivalent to $\bar{T}$
\begin{equation}
\label{TpT}
T ' \simeq \bar{T}
\end{equation}
because both representations have the weights
\begin{equation}
(\bar{T}_{\dot{a}})_{\l^i,\l^j} = (T_{\dot{a}}{}')_{\l^i,\l^j}
= - \l_{\dot{a}}^i \de_{\l^i,\l^j} \sp \dot{a} \in {\rm Cartan}\,\su(n)
\end{equation}
of the charge conjugate representation. Our Cartesian automorphism $\w$ is
inner-automorphi- cally equivalent $\w \simeq \tau$ to a realization of the
action $\tau$ of the Dynkin automorphism. This is discussed explicitly for
$\su(3)$ in App.~\ref{Dyn}, which also remarks on the inner-automorphically
equivalent realization of $\tau$ with invariant subalgebra
$h = \mathfrak{c}_n$ for $\su(2n)$.

Following the observation \eqref{TpT}, we will say that a {\it real} representation of
$\su(n)$ satisfies
\begin{equation}
T \simeq \bar{T} \simeq T '
\end{equation}
while for a {\it complex} representation $\bar{T}$ or $T'$ is not equivalent to $T$.
As examples, the adjoint representation is real with
\begin{equation}
(T_a^{\rm adj})_b{}^c = -i f_{abc} \sp \bar{T}_a^{\rm adj} = T_a^{\rm adj}
\simeq T_a^{\rm adj}{}'
\end{equation}
while the fundamental representation $n$ is complex with
\begin{equation}
T_a =\frac{\sqrt{\psi^2}}{2} \l_a \sp
T_A^{\rm T} = -T_A \sp
T_I^{\rm T} = T_I
\sp  T_a{}' = \bar{T}_a \ .
\end{equation}
In this case the two equivalent matrix representations $T'$ and $\bar{T}$ of
$\bar{n}$ are exactly equal.

\subsection{The Cartesian form of $A_{n-1}^{(2)}$ }

Because the action $\w$ of charge conjugation is diagonal in the Cartesian
basis, the solution to the $H$-eigenvalue problem is trivial
\begin{subequations}
\label{dataH}
\begin{equation}
\label{dataH0}
\w U\hc = U\hc E \quad : \quad
\s =1 \sp \r = 2 \sp U\hc =1 \sp E =\w
\end{equation}
\begin{eqnarray}
\nb = 0 & &\leftrightarrow E_A = 1  \sp \forall \; A \in \so(n)  \\
\nb = 1 & & \leftrightarrow E_I = -1 \sp \forall \; I \in \su(n)/\so(n) \ .
\end{eqnarray}
\end{subequations}
Then the normalization $\chi =1$ gives the eigencurrents $\sj$ and the
non-zero entries of the basic duality transformations
\begin{subequations}
\label{bdrel}
\begin{equation}
\sj_{0A}(z) = J_A(z) \sp \sj_{1I}(z) = J_I (z) \qquad ; \qquad
\sj_{0A}(z)' =  \sj_{0A}(z) \sp \sj_{1I}(z) ' =-\sj_{1I}(z)
\end{equation}
\begin{equation}
\G_{0A;0B}  = k \de_{AB} \sp \G_{1I;-1,J} = k \de_{IJ}
\end{equation}
\begin{equation}
\F_{0A;0B}{}^{0C} = f_{ABC} \sp
\F_{0A;1I}{}^{1J} = f_{AIJ} \sp \F_{1I;-1,J}{}^{0A} = f_{IJA}
\end{equation}
\end{subequations}
and the principle of local isomorphisms \cite{deBoer:1999na} gives the
twisted current system:
\begin{subequations}
\label{monA0}
\begin{equation}
\hj_\nrm = \{ \hj_{0A} , \hj_{1I} \} \sp
\hj_{n(r) \pm 2,\m}(z) = \hj_{n(r)\m} (z) \sp
\hj_\nrm (z e^{2 \pi i}) = e^{- 2 \pi i \srac{n(r)}{\r(\s)}} \hj_\nrm (z)
\end{equation}
\begin{equation}
\hj_{0A} (z)\hj_{0B}(w) = \frac{ k \d_{AB}}{(z-w)^2}
+ \frac{ i f_{ABC} \hj_{0C} (w)}{z-w} + \Ord (z-w)^0
\end{equation}
\begin{equation}
\hj_{0A} (z)\hj_{1I}(w) =
\frac{ i f_{AIJ} \hj_{1J} (w)}{z-w} + \Ord (z-w)^0
\end{equation}
\begin{equation}
\hj_{1I} (z)\hj_{1J}(w) =\frac{ k \d_{IJ}}{(z-w)^2}
+ \frac{ i f_{IJA} \hj_{2A} (w)}{z-w} + \Ord (z-w)^0
\end{equation}
\begin{equation}
\label{monA} \hj_{0A} (z e^{2 \pi i}) =   \hj_{0A} (z) \sp
 \hj_{1I} (z e^{2 \pi i}) =  -\hj_{1I} (z) \sp A \in \so(n) \sp I \in \su(n)/\so(n) \ .
\end{equation}
\end{subequations}
Finally the monodromies \eqref{monA} give the twisted mode expansions
\begin{equation}
\label{monA1}
\hj_{0A} (z) = \sum_{m \in \sz} \hj_{0A} (m) z^{-m-1}
\sp
\hj_{1I} (z) = \sum_{m \in \sz} \hj_{1I} (m + \srac{1}{2}) z^{-(m+\srac{1}{2})-1}
\end{equation}
and one finds from \eqref{monA0}, \eqref{monA1}
that the twisted modes satisfy the Cartesian form of the
outer-automorphically twisted affine Lie algebra $A_{n-1}^{(2)}$:
\begin{subequations}
\label{A22}
\begin{equation}
[\hj_{0A} (m),\hj_{0B} (n)] = if_{ABC} \hj_{0C} (m+n) + k m \d_{AB} \d_{m+n,0}
\end{equation}
\begin{equation}
[ \hj_{0A} (m), \hj_{1I} (n + \srac{1}{2})] = i f_{AIJ} \hj_{1J} (m +n+ \srac{1}{2})
\end{equation}
\begin{equation}
[\hj_{1I} (m + \srac{1}{2}),\hj_{1J} (n+\srac{1}{2})] = if_{IJA} \hj_{2A} (m+n+1) +
k ( m + \srac{1}{2}) \d_{IJ} \d_{m+n+1,0}
\end{equation}
\begin{equation}
\hj_{ \pm 2,A}(m) = \hj_{0A}(m) \sp
\hj_{-1,I}( m -\srac{1}{2}) = \hj_{1I} (m - 1 + \srac{1}{2})
\end{equation}
\begin{equation}
\label{dagrel}
\hj_{0A} (m)\hc = \hj_{0A} (-m) \sp
\hj_{1I}(m+\srac{1}{2})\hc = \hj_{-1,I}(-m-\srac{1}{2}) = \hj_{1I}(-m-1+\srac{1}{2})
\end{equation}
\begin{equation}
\nn  A \in \so(n)  \sp I \in \su(n)/\so(n) \ .
\end{equation}
\end{subequations}
This twisted current algebra (and \eqref{monA0}) can alternately be obtained
by substitution of the data \eqref{dataH}, \eqref{bdrel} into the general
twisted current algebra of Ref.~\cite{Halpern:2000vj}, and the
 dagger relations in \eqref{dagrel} follow from the general orbifold
adjoint operation \cite{Halpern:2000vj} in this case.
App.~\ref{Dyn} gives a translation
dictionary between our Cartesian form of $A_2^{(2)}$ and its more conventional
form in mathematics.

We also note that for {\it even levels} the outer-automorphically twisted
affine Lie algebra \eqref{A22} is a subalgebra of the order-two orbifold affine
algebra \cite{Borisov:1997nc} on $\su(n)$
\begin{subequations}
\begin{equation}
[\hj_a^{(r)} ( m + \srac{r}{2}), \hj_b^{(s)} ( n + \srac{s}{2}) ]
= i f_{abc} \hj_c^{(r+s)} (m+ n + \srac{r+s}{2}) + \hat k \de_{ab}
(m + \srac{r}{2})  \de_{m + n + \srac{r+s}{2},0}
\end{equation}
\begin{equation}
\hat k = 2 k \sp \bar r, \bar s = 0,1 \sp a,b,c = 1 , \ldots , n^2-1
\end{equation}
\end{subequations}
which lives in the twisted sector of $\z_2$ cyclic permutation orbifolds
on $\su(n)_k \oplus \su (n)_k$.

Similarly, one reads from Ref.~\cite{Halpern:2000vj} the twisted left-mover
affine-Sugawara construction of sector $\s=1$
\begin{subequations}
\label{aslm0}
\begin{equation}
\label{aslm}
{\cL}_{\gfrakh} ^{0A;0B} = \frac{1}{2k +Q_g} \d^{AB} \sp
{\cL}_{\gfrakh} ^{1I;-1,J} = \frac{1}{2k +Q_g} \d^{IJ} \sp
\hat c = c = \frac{x (n^2-1)}{x + n}
\end{equation}
\begin{eqnarray}
\hat T(z)  &=&  {\cL}_{\gfrakh}^{\nrm ;\mnrn}  : \hj_\nrm (z)\hj_\mnrn (z) : \\
      &=& \frac{1}{2k +Q_g} \left(: \sum_A \hj_{0A} (z) \hj_{0A} (z)
+ \sum_I \hj_{1I} (z) \hj_{-1,I}(z) : \right)
\end{eqnarray}
\end{subequations}
where $\gfrakh  \equiv \gfrakh (\s=1)$, $Q_g$ is the quadratic Casimir of
$\su(n)$ and $:\cdot :$ is operator product normal ordering.

\subsection{The twisted representation matrices \label{repsec}}

To study the representation theory of the charge conjugation orbifold in the
twisted sector, we need to determine the {\it twisted representation matrices}
 $\T(T)$ corresponding to each untwisted representation $T$.

In order to do this, we must first find the action $W(T)$ of the automorphism
in representation $T$, which satisfies the {\it linkage relation}
 \cite{deBoer:2001nw}
\begin{equation}
\label{linkreal}
W\hc (T) T_a W(T) = T_a{}' = \w_a{}^b T_b
\end{equation}
and then we must solve the {\it extended $H$-eigenvalue problem}
\cite{deBoer:2001nw} for $W(T)$
\begin{equation}
\label{eHpr}
W(T) U\hc(T) = U\hc(T) E(T) \ .
\end{equation}
Given this solution (see below) and the eigendata \eqref{dataH} of the
$H$-eigenvalue problem, we read from
Ref.~\cite{deBoer:2001nw} the form and properties of the twisted
representation matrices  $\T$:
\begin{subequations}
\begin{equation}
\label{TAIrel}
\T_{0A} (T)  =   \T_{\pm 2, A} (T) =  U(T) T_A U\hc(T)
 \sp \T_{1I} (T)  =   \T_{-1, I} (T)  =  U(T) T_I U\hc(T)
\end{equation}
\begin{eqnarray}
 {[} {\cal{T}}_{0A} (T)  , \T_{0B} (T)] & = &  i f_{ABC} \T_{0C}  (T)
 \label{TAB} \\
 {[} {\cal{T}}_{0A} (T)  , \T_{1I} (T)] & = &  i f_{AIJ} \T_{1J} (T)
 \label{TAI} \\
  {[} {\cal{T}}_{1I} (T)  , \T_{1J} (T)] & = &  i f_{IJA} \T_{0A} (T)
  \label{TIJ}
\end{eqnarray}
\begin{equation}
\label{TAIsel} \T_{0A} (T) = E(T) \T_{0A} (T) E(T)^\ast \sp
\T_{1I} (T) = -E(T) \T_{1I} (T) E(T)^\ast \ .
\end{equation}
\end{subequations}
Here the relations \eqref{TAIsel} are the $\T$-selection rules, which are the
orbifold dual of the linkage relation.

To go further we must distinguish between real and complex representations
of $\su (n)$. For any real representation, the solution $W(T)$ of the linkage
relation \eqref{linkreal} is guaranteed because $T' \simeq T$ for real representations.
As an example, we know \cite{deBoer:2001nw} for the adjoint representation
$ (T_a^{\rm adj})_{b}{}^{c} = - if_{ab}{}^{c}$:
\begin{subequations}
\label{data0}
\begin{equation}
\label{winv0}
W (T^{\rm adj}) = \w\sp U (T^{\rm adj})= U =1 \sp
E(T^{\rm adj}) = E = \w \sp  R(T^{\rm adj}) = \r = 2
\end{equation}
\begin{equation}
\label{winv}
W\hc(T^{\rm adj}) T_a^{\rm adj} W(T^{\rm adj}) = \w_a{}^b T_b^{\rm adj}
\quad \leftrightarrow \quad \w_a{}^d \w_b{}^e f_{de}{}^f (\w\hc)_f{}^c =
f_{ab}{}^c \ .
\end{equation}
\end{subequations}
Indeed, the relation in \eqref{winv} reminds us \cite{deBoer:2001nw} that
$W(T^{\rm adj}) = \w$ for the adjoint in all current-algebraic orbifolds.
Using formulae \eqref{TAIrel} above, we find the twisted
representation matrices for the adjoint
\begin{equation}
\label{twradj}
\T_{0A}(T^{\rm adj})=T_A^{\rm adj} \sp
\T_{1I}(T^{\rm adj})=T_I^{\rm adj}  \ .
\end{equation}
The selection rules \eqref{TAIsel} are easily checked explicitly in this case,
for example
\begin{equation}
\T_{1I}(T^{\rm adj})_J{}^K = -\T_{1I}(T^{\rm adj})_J{}^K
\end{equation}
is satisfied as $0=0$ because $f_{IJK} =0$ for a symmetric space.

For complex representations  $T^{\rm (c)}$ of $\su(n)$, we choose the
{\it reducible} representation \cite{deBoer:2001nw}
\begin{subequations}
\begin{equation}
\label{comrep}
T_a =\left( \begin{array}{cc}
T_a^{\rm (c)}& 0 \\ 0 & T_a^{\rm (c)}{}' \end{array} \right)
\end{equation}
\begin{equation}
T_a{}' =\w_a{}^b T_b =\left( \begin{array}{cc}
T_a^{\rm (c)}{}' & 0 \\ 0 & T_a^{\rm (c)}{} \end{array} \right)
\end{equation}
\end{subequations}
where $T^{\rm (c)}{}'$ in Eqs.~\eqref{data00}, \eqref{Tpr}
is the automorphic transform of $T^{\rm (c)}$.
The choice in \eqref{comrep} results in considerable simplification over the
unitarily-equivalent choice with $T^{\rm (c)}{}' \rightarrow \bar{T}^{\rm (c)}$.
In particular we then find for all complex representations that the solution
 of the linkage relation \eqref{linkreal} is very simple
\begin{equation}
W(T) = \exp \left( i \frac{\pi}{2} \left(
\begin{array}{cc}
0 & \one  \\ \one & 0 \end{array} \right) \right) =
i \left( \begin{array}{cc}
0 & \one \\ \one & 0
\end{array} \right) \sp W\hc(T) W(T) =\left( \begin{array}{cc}
\one & 0 \\ 0 & \one
\end{array} \right) \ .
\end{equation}
Moreover, the extended $H$-eigenvalue problem \eqref{eHpr} is also easy to solve
\begin{equation}
\label{exHval}
U(T) = U\hc(T) = \frac{1}{\sqrt{2}}
\left( \begin{array}{cc}
\one & \one \\ \one & -\one
\end{array} \right)
\sp U\hc(T) U(T) =\left( \begin{array}{cc}
\one & 0 \\ 0 & \one
\end{array} \right) \sp E(T)= i \left( \begin{array}{cc}
\one & 0 \\ 0 & - \one
\end{array} \right) \ .
\end{equation}
Then we find for the twisted representation matrices
\begin{subequations}
\label{twrfun}
\begin{equation}
\T_{0A} (T)= U(T)\left( \begin{array}{cc}
T_A^{\rm (c)} & 0 \\ 0 & T_A^{\rm  (c)}{}' \end{array} \right) U\hc(T) =
\left( \begin{array}{cc} T_A^{\rm  (c)} & 0 \\ 0 & T_A^{\rm (c)} \end{array} \right)
\end{equation}
\begin{equation}
\T_{1I} (T) = U(T)\left( \begin{array}{cc}
T_I^{\rm (c)} & 0 \\ 0 & T_I^{\rm (c)}{}' \end{array} \right) U\hc(T) =
\left( \begin{array}{cc} 0 & T_I^{\rm (c)} \\ T_I^{\rm (c)} & 0 \end{array} \right)
\end{equation}
\end{subequations}
and these explicit forms allow us to verify that the
$\T$-selection rules \eqref{TAIsel}
are identities.

\subsection{Twisted left-mover KZ system  for charge conjugation orbifolds
\label{kzsec}}

We consider here only the $\so(n)$-invariant or scalar twist-field state,
which satisfies
\begin{subequations}
\label{stwf}
\begin{equation}
\hj_{0A} ( m \geq 0) | 0 \rangle = \langle 0 | \hj_{0A} (m\leq 0) =0 \sp
A \in \so(n)
\end{equation}
\begin{equation}
\hj_{1I} ( m+\srac{1}{2} \geq 0) | 0 \rangle = \langle 0 | \hj_{1I}
(m + \srac{1}{2}\leq 0) =0 \sp I \in \su(n)/\so(n)
\end{equation}
\end{subequations}
in the twisted sector of the charge conjugation orbifold. Then the
$M$-ordered (mode-ordered) expressions of
Refs.~\cite{Halpern:2000vj,deBoer:2001nw} are immediately useful.
For example, the left-mover Virasoro generators and the conformal weight
of the scalar twist-field state
\begin{subequations}
\begin{eqnarray}
L_\s(m) & = &  \frac{1}{2k+ Q_g} \sum_{p \in \sz} : \sum_A \hj_{0A}(p) \hj_{0A}(m-p)
+\sum_I \hj_{1I}(p+\srac{1}{2} ) \hj_{-1,I}(m-p-\srac{1}{2}):_M \nn \\
 &   &  \hskip 1cm + \de_{m,0} \hat \Delta_0
 \sp [ L_\s (m \geq 0) - \de_{m,0} \hat \Delta_0 ] |0\rangle = 0
\end{eqnarray}
\begin{eqnarray}
\hat \Delta_0 & = &  \lr^{\nrm ; -\nrn} (\s)
\frac{\nb}{2 \r(\s)} \left(1 - \frac{\nb}{\r (\s)} \right)
\G_{\nrm ; - \nrn} (\s ) \label{gendel} \\
& = &  \frac{{\rm dim}(g/h)}{16}\frac{x}{x+n} =
\frac{(n-1)(n+2)}{32} \frac{x}{x+n}
\end{eqnarray}
\end{subequations}
are easily read from \eqref{dataH}, \eqref{bdrel},
\eqref{aslm0} and the general form given in Eq.~(6.8b) of
Ref.~\cite{deBoer:2001nw}.

We turn next to the left-mover {\it twisted affine primary fields}
$\hgp (\T,z)$, $\T = \T (T)$ of the charge conjugation orbifold. From
Ref.~\cite{deBoer:2001nw}, we read the OPEs
\begin{subequations}
\begin{equation}
\hj_{0A} (z) \hgp (\T,w) = \frac{ \hgp(\T,w)}{z-w} \T_{0A} (T)+ \Ord (z-w)^0
\end{equation}
\begin{equation}
\hj_{1I} (z) \hgp (\T,w) = \frac{ \hgp(\T,w)}{z-w} \T_{1I} (T)+ \Ord (z-w)^0
\end{equation}
\begin{equation}
\hat T(z) \hgp (\T,w ) = \left( \frac{\Delta}{(z-w)^2} +
\frac{1}{z-w} \partial_w \right) \hgp  (\T,w) +\Ord (z-w)^0
\end{equation}
\begin{equation}
\Delta \equiv \left\{
\begin{array}{cl}
\Delta (T) & \mbox{for real reps}\; T \\
\Delta( T^{\rm (c)})
\left( \begin{array}{cc}
1 & 0 \\ 0 & 1 \end{array} \right)
& \mbox{for complex reps}\; T^{\rm (c)}
\end{array} \right.
\end{equation}
\end{subequations}
where $\Delta (T)$ and $\Delta( T^{\rm (c)})$ are the conformal weight
of the representation under the (untwisted) affine-Sugawara construction
\cite{Bardakci:1971nb,Halpern:1971ay,Dashen:1975hp,Knizhnik:1984nr} on
$\su (n)$.

Similarly, using \eqref{dataH0}, \eqref{aslm} and
Eq.~(6.13a) of Ref.~\cite{deBoer:2001nw}, we find the
twisted left-mover vertex operator equation
\begin{eqnarray}
\label{twlsu}
\partial \hgp(\T,z) &= &  \frac{2}{2k+Q_g} \left[
\sum_A : \hj_{0A}(z) \hgp (\T,z) :_M \T_{0A} \right. \nn \\
& & \hskip 1.5cm \left. + \sum_I \left( :\hj_{1I} (z) \hgp (\T,z) :_M - \frac{1}{2z}
\hgp (\T,z) \T_{1I} \right) \T_{1I} \right]
\end{eqnarray}
for the twisted affine primary fields.
$M$-ordering for the twisted vertex operator equation is defined in
Ref.~\cite{deBoer:2001nw} (see also Eq.~\eqref{vopt}).
In combination with the twisted current algebra \eqref{A22} and the ground
state conditions \eqref{stwf}, the twisted vertex operator equation
\eqref{twlsu} suffices to derive a twisted KZ system for the twisted sector of
 the charge conjugation orbifolds.

We prefer however to include this computation as an example of a more general
result. For any twisted current algebra with an action on a
{\it general scalar twist-field  state}
$| 0 \rangle_\s$
\begin{equation}
\hj_\nrm (m + \srac{n(r)}{\r(\s)} \geq 0) | 0\rangle_\s =
{}_\s \langle 0| \hj_\nrm (m + \srac{n(r)}{\r(\s)} \leq 0)   = 0
\end{equation}
one finds that the general expression for the conformal weight
$\hat \Delta_0 (\s) $ of the scalar twist-field state is that
given in Eq.~\eqref{gendel}.
Moreover, we may use the general twisted current algebra and the general twisted
vertex operator equation of Ref.~\cite{deBoer:2001nw} to find  the
{\it general twisted left-mover  KZ system}
\begin{subequations}
\label{tlmkzeq}
\begin{equation}
\hat A_+ (\T,z,\s) \equiv {}_\s\langle 0|  \hgp(\T^{(1)},z_1,\s) \hgp(\T^{(2)},z_2,\s)
\cdots \hgp(\T^{(N)},z_N,\s) |0 \rangle_\s
\end{equation}
\begin{equation}
\part_\kappa \hat A_+(\T,z,\s) = \hat A_+ (\T,z,\s) \hat W_\kappa (\T,z,\s) \sp
\sp \kappa = 1 \ldots N \sp \s = 0, \ldots ,N_c -1
\end{equation}
\begin{equation}
\label{kzct2}
\hat W_{\kappa}(\T,z,\s) = 2 \lr^{n(r)\m;-n(r),\n} (\s)
 \left[ \sum_{\r \neq \k}\left( \frac{z_\r}{z_\k}
\right)^{ \srac{\bar n(r)}{\r(\s)}} \frac{1}{z_{\k \r }}
\T_{n(r)\m}^{(\r)} \T_{-n(r),\n}^{(\k)}
- \srac{\bar n(r)}{\r(\s)} \frac{1}{z_{\k}}
\T_{n(r)\m}^{(\k)} \T_{-n(r),\n}^{(\k)} \right]
\end{equation}
\begin{equation}
\label{gwig}
\hat A_+ (\T,z,\s) \left(\sum_{\r =1}^N \T_{0 \m}^{(\r)}\right) = 0 \sp \forall \;\m
\end{equation}
\end{subequations}
for the correlators in the scalar twist-field states of any
WZW orbifold.
The global Ward identity \eqref{gwig} follows from the residual symmetry
\begin{equation}
{}_\s \langle 0| [ \hj_{0\m} (0), \hgp(\T^{(1)},z_1,\s) \cdots
\hgp(\T^{(N)},z_N,\s)] | 0 \rangle_\s = 0 \sp \forall \; \m
\end{equation}
generated by the zero modes of the integral affine subalgebra of the
general twisted current algebra. The so-called ground state
\cite{Halpern:2000vj,deBoer:2001nw} in sector $\s$ of any WZW permutation
orbifold is in fact a scalar twist-field state, so
the general twisted KZ system \eqref{tlmkzeq}
includes as a special case the twisted KZ system obtained for the WZW
permutation orbifolds in Ref.~\cite{deBoer:2001nw}.

Moreover, a scalar twist-field state exists for each twisted sector of
every outer-automor- \linebreak phically
twisted affine Lie algebra, so the general system \eqref{tlmkzeq} includes
a twisted KZ system for every outer-automorphic WZW orbifold. In these
cases, the residual symmetry expressed by the global Ward identity
\eqref{gwig} is that of the invariant subalgebra $h \subset g$ of each
automorphism group.

In particular, we obtain the twisted KZ system for the left-mover sector
of the charge conjugation orbifold on $\su (n)$
\begin{subequations}
\label{outkz}
\begin{equation}
\hat A_+ (\T,z) \equiv \langle 0 | \hgp (\T^{(1)},z_1) \cdots \hgp(\T^{(N)},z_N)
| 0 \rangle
\end{equation}
\begin{equation}
\label{twkzeq}
\part_\mu \hat A_+(\T,z) = \hat A_+ (\T,z) \hat W_\mu (\T,z) \sp
\m = 1 \ldots N
\end{equation}
\begin{eqnarray}
\hat W_{\mu}(\T,z)  &= &  \frac{2}{2k+Q_g}
 \left[ \sum_{\nu \neq \mu} \frac{1}{z_{\mu \nu}}\left( \sum_A \T_{0A}^{(\n)} \T_{0A}^{(\m)}
+ \left( \frac{z_{\nu}}{z_{\mu}} \right)^{\srac{1}{2}}
\sum_I \T_{1I}^{(\n)} \T_{1I}^{(\m)} \right)
\right. \nn \\  & & \hskip 2cm \left.
- \frac{1}{2 z_{\mu}} \sum_I \T_{1I}^{(\mu)} \T_{1I}^{(\mu)} \right]
\label{kzcc}
\end{eqnarray}
\begin{equation}
\label{glwieq}
\hat A_+ (\T,z) \left( \sum_{\m=1}^N \T_{0A}^{(\m)} \right) = 0 \sp
\forall \; A \in \so(n)
\end{equation}
\end{subequations}
where $|0 \rangle$ is the $\so (n)$-invariant twist-field state in
\eqref{stwf}. For this case, the explicit form of the twisted
representation matrices $\T = \T (T)$ is given in Subsec.~\ref{repsec}.
Although it is guaranteed by the construction, we check explicitly in
App.~\ref{flch} that the twisted connection \eqref{kzcc} is flat.

We emphasize here that the flatness check of App.~\ref{flch} uses
only the symmetric-space form ($f_{IJK}=0$) of the algebra
\eqref{TAB}, \eqref{TAI}, \eqref{TIJ} of the twisted
representation matrices. All $\z_2$-type outer automorphism groups are
similarly associated to symmetric spaces, so we expect that
the twisted KZ systems of all
$\z_2$-type outer-automorphic orbifolds will have the same form
(with $\so(n)$ replaced by the relevant invariant subalgebra) as
that given in \eqref{outkz}.

\subsection{Rectification and the twisted right-mover KZ system \label{recsec}}

The form of the general twisted right-mover current algebra was given in
Ref.~\cite{deBoer:2001nw}: As an example of the general phenomenon discussed
there, we know that the twisted right-mover modes
$\hjb_{0A}(m)$ and $\hjb_{1I}(m +\srac{1}{2})$ satisfy the same algebra
as $A_{n-1}^{(2)}$ in \eqref{A22}, but now with $k\rightarrow -k$.
This situation is rectifiable%
\footnote{The rectification problem is the question whether the twisted
right- and left-mover current algebras are isomorphic. The question has been
answered in the affirmative for inner-automorphic orbifolds on simple $g$ and
for permutation orbifolds in Ref.~\cite{deBoer:2001nw}. Except for the
triality orbifold on $\so(8)$, all outer automorphism groups of simple $g$ are
$\z_2$'s, and we know that the non-trivial element of $\z_2$ satisfies
$h_1^{-1}=h_1$. According to the discussion of Ref.~\cite{deBoer:2001nw}
this means that the twisted right-mover current algebra of all these
orbifolds is similarly rectifiable, as seen here for the charge conjugation
orbifold on $\su(n)$. Thus the rectification problem is solved except for the
triality orbifold on $\so(8)$, to which we expect to return elsewhere.}
 into a copy of $A_{n-1}^{(2)}$ by the mode-reversed definition
\begin{equation}
\hjbb_{0A} (m) \equiv \hat{\bar{J}}_{0A}(-m)
\sp \hjbb_{1I} (m + \srac{1}{2}) \equiv \hat{\bar{J}}_{-1,I}(-m-\srac{1}{2})
= \hjb_{1I}(-m-1 + \srac{1}{2})
\ .
\end{equation}
The rectified right-mover modes  $\hjbb$ also satisfy a copy of the left-mover
conditions
\begin{equation}
\hjbb_{0A} ( m \geq 0) | 0 \rangle = \langle 0 | \hjbb_{0A} (m\leq 0) =0 \sp
\hjbb_{1I} ( m+\srac{1}{2} \geq 0) | 0 \rangle = \langle 0 | \hjbb_{1I}
(m + \srac{1}{2}\leq 0) =0
\end{equation}
on  the scalar twist-field state, and we find that the
right-mover Virasoro generators are also a copy of the left movers
\begin{subequations}
\begin{eqnarray}
\bar L_\s(m) & = &  \frac{1}{2k+ Q_g} \sum_{p \in \sz} : \sum_A \hjbb_{0A}(p)
\hjbb_{0A}(m-p) +\sum_I \hjbb_{1I}(p+\srac{1}{2} ) \hjbb_{-1,I}(m-p-\srac{1}{2}):_M \nn \\
 & & \hskip 2cm + \de_{m,0} \hat{\bar{\Delta}}_0  \sp
 \hat{\bar{c}} = \hat c = \frac{x (n^2-1)}{x + n}
\end{eqnarray}
\begin{equation}
[ \bar L_\s (m \geq 0) - \de_{m,0} \hat{\bar{\Delta}}_0 ] |0\rangle = 0 \sp
\hat{\bar{\Delta}}_0 = \hat \Delta_0 =
\frac{(n-1)(n+2)}{32} \frac{x}{x+n}
\end{equation}
\end{subequations}
with the same action on the scalar twist-field state.

Following Ref.~\cite{deBoer:2001nw}, we also find the twisted KZ system
for the right movers
\begin{subequations}
\label{trmkzeq}
\begin{equation}
\hat A_- (\T,\bz) \equiv \langle 0 |  \hgm(\T^{(1)},\bz_1)
\cdots \hgm(\T^{(N)},\bz_N) |0 \rangle
\end{equation}
\begin{equation}
\bar \part_\m \hat A_-(\T,\bz) =  \hat{\bar{W}}_\m (\T,\bz) \hat A_- (\T,\bz)\sp
\m = 1 \ldots N
\end{equation}
\begin{eqnarray}
\hat{\bar{W}}_{\mu}(\T,\bz)  &= &  \frac{2}{2k+Q_g}
 \left[ \sum_{\nu \neq \mu} \frac{1}{\bz_{\mu \nu}}\left( \sum_A \T_{0A}^{(\m)}
 \T_{0A}^{(\n)} + \left( \frac{\bz_{\nu}}{\bz_{\mu}} \right)^{\srac{1}{2}}
\sum_I \T_{1I}^{(\m)} \T_{1I}^{(\n)} \right)
\right. \nn \\  & & \hskip 2cm \left.
- \frac{1}{2 \bz_{\mu}} \sum_I \T_{1I}^{(\mu)} \T_{1I}^{(\mu)} \right]
\label{kzccr}
\end{eqnarray}
\begin{equation}
\left( \sum_{\m=1}^N \T_{0A}^{(\m)} \right) \hat A_-(\T,\bz)  = 0 \sp
\forall \; A \in \so(n) \ .
\end{equation}
\end{subequations}
In this case $\hat{\bar{W}}_\m$ is just $\hat W_\m$ with
$\{ z_\m \} \rightarrow \{ \bz_\m \} $ because the order of the matrices
is reversible.

\subsection{The classical theory of charge conjugation orbifolds}

The classical action formulation of each sector of all WZW orbifolds, in
terms of appropriate {\it group orbifold elements} with definite
monodromy, was given in
Ref.~\cite{deBoer:2001nw}. The group orbifold elements are the classical
limit of the twisted affine primary fields of the WZW orbifolds. Before
discussing solutions of our twisted KZ systems above,
it will be helpful to work out the details of this classical
formulation in the case of the charge conjugation orbifolds.

Reading from Subsec.~5.7 of Ref.~\cite{deBoer:2001nw}, we find for all
representations that the group elements $g$, the eigengroup elements $\sg$
and the group orbifold elements $\hg$ satisfy
\begin{subequations}
\label{rel1}
\begin{equation}
\label{grel}
g(T,\xi) = e^{i \be^a (\xi) T_a} \sp
\be^a (\xi)' = \be^b(\xi) (\w\hc)_b{}^a =  \be^b(\xi) \w_b{}^a
\end{equation}
\begin{equation}
g(T,\xi)' = W(T) g(T,\xi) W\hc(T) =e^{i \be^a (\xi)' T_a} = e^{i \be^a(\xi)T_a{}'}
\end{equation}
\end{subequations}
\begin{subequations}
\label{rel2}
\begin{equation}
\sg (\T,\xi) = U(T) g(T,\xi) U\hc (T) =
e^{i (\be^{0A} (\xi)\T_{0A}(T) +\be^{1I} (\xi)\T_{1I}(T))}
\end{equation}
\begin{equation}
\be^{0A} (\xi) \equiv \be^A (\xi) ,\;\; \be^{0A} (\xi) ' =\be^{0A} (\xi)
\qquad ; \qquad
\be^{1I} (\xi) \equiv \be^I (\xi) \sp
 ,\;\; \be^{1I} (\xi) ' =-\be^{1I} (\xi)
\end{equation}
\begin{equation}
\sg (\T,\xi)' = E(T) \sg (\T,\xi) E(T)^\ast
\end{equation}
\end{subequations}
\begin{subequations}
\label{rel3}
\begin{equation}
\label{hgrel}
\hg (\T,\xi)  =
e^{i (\hb^{0A} (\xi)\T_{0A}(T) +\hb^{1I} (\xi)\T_{1I}(T))} \sp
\hb^{0A} (\xi + 2\pi) = \hb^{0A} (\xi ) \sp
\hb^{1I} (\xi + 2\pi) = -\hb^{1I} (\xi )
\end{equation}
\begin{equation}
\label{hgmon}
\hg (\T,\xi + 2\pi) = E(T) \hg (\T,\xi) E(T)^\ast \ .
\end{equation}
\end{subequations}
Here $\xi$ is the spatial coordinate on the cylinder, $W(T)$ is the
action of charge conjugation in rep $T$,
$\{ U(T),E(T)\}$ is the eigendata of the extended $H$-eigenvalue
problem \eqref{eHpr}
and $\be '$, $g '$, $\sg '$ are the responses of the untwisted objects to
charge conjugation. The $\T$-selection rules \eqref{TAIsel} guarantee the
consistency of these responses, as well as the consistency of the
monodromies of $\hat \be$ and $\hat g$.

As an example, Eqs.~\eqref{oms}, \eqref{winv0}, \eqref{twradj} and
\eqref{hgmon} give the monodromy of the group orbifold elements
\begin{subequations}
\begin{equation}
\hg (\T(T^{\rm adj}),\xi + 2\pi)_A{}^B =\hg (\T(T^{\rm adj}),\xi)_A{}^B
\sp
\hg (\T(T^{\rm adj}),\xi + 2\pi)_I{}^J =\hg (\T(T^{\rm adj}),\xi)_I{}^J
\end{equation}
\begin{equation}
\hg (\T(T^{\rm adj}),\xi + 2\pi)_A{}^I =-\hg (\T(T^{\rm adj}),\xi)_A{}^I
\sp
\hg (\T(T^{\rm adj}),\xi + 2\pi)_I{}^A =-\hg (\T(T^{\rm adj}),\xi)_I{}^A
\end{equation}
\begin{equation}
A \in \so(n) \sp I \in \su(n)/\so(n)
\end{equation}
\end{subequations}
in the twisted adjoint representation $\T (T^{\rm adj})$.

For general twisted complex representations, we may use
Eqs.~\eqref{exHval} and \eqref{twrfun} to write out the results
\eqref{rel1}, \eqref{rel2} and \eqref{rel3}
in terms of the complex untwisted representation matrices
$T^{\rm (c)}$:
\begin{subequations}
\begin{equation}
\label{grelc}
g (T,\xi)
=\exp \left( i \be^a (\xi) \left( \begin{array}{cc}
T_a^{\rm (c)} & 0 \\ 0 & T_a^{\rm (c)}{}' \end{array} \right) \right) \sp
g(T,\xi) ' =
\exp \left( i \be^a (\xi) \left( \begin{array}{cc}
T_a^{\rm (c)}{}' & 0 \\ 0 & T_a^{\rm (c)} \end{array} \right) \right)
\end{equation}
\begin{equation}
\sg (\T,\xi)
= \exp \left( i \left( \begin{array}{cc}
\be^{0A} (\xi) T_A^{\rm (c)} & \be^{1I} (\xi) T_I^{\rm (c)} \\
 \be^{1I} (\xi) T_I^{\rm (c)}  & \be^{0A} (\xi) T_A^{\rm (c)} \end{array} \right) \right)
\end{equation}
\begin{equation}
\label{hgrelc}
\hg (\T,\xi) = \exp \left( i \left( \begin{array}{cc}
\hb^{0A} (\xi) T_A^{\rm (c)} & \hb^{1I} (\xi) T_I^{\rm (c)} \\
 \hb^{1I} (\xi) T_I^{\rm (c)}  & \hb^{0A} (\xi) T_A^{\rm (c)}
  \end{array} \right) \right)
\end{equation}
\begin{eqnarray}
\label{for1}
\hg (\T,\xi + 2\pi)
& = & \exp \left( i \left( \begin{array}{cc}
\hb^{0A} (\xi) T_A^{\rm (c)} & -\hb^{1I} (\xi) T_I^{\rm (c)} \\
 -\hb^{1I} (\xi) T_I^{\rm (c)}  & \hb^{0A} (\xi) T_A^{\rm (c)}
\end{array} \right) \right) \\
&= & E(T) \hg(\T,\xi) E(T)^\ast \sp E(T) = i
\left( \begin{array}{cc} 1 & 0 \\ 0 & -1 \end{array} \right) \ .
\label{for2}
\end{eqnarray}
\end{subequations}
The consistency of the forms in \eqref{for1} and \eqref{for2} is easily checked.

We turn next to the action formulation of the charge conjugation orbifolds.
The untwisted charge-conjugation-invariant
WZW action for any representation $T$ of $\su(n)$ can be taken as
\begin{subequations}
\begin{eqnarray}
 S_{\rm WZW}[g(T)] & = & -\frac{k}{\epsilon y(T)} \left( \frac{1}{8 \pi}
 \int d^2\xi\0b {\rm Tr}\Big{(} \ginv (T)\pl_+
 g(T)\0b\ginv(T)\pl_-g(T)\Big{)} \right. \nn \\
   & &  \hskip 1.5cm \left. +\frac{1}{12\pi}\int_{\Gamma} {\rm Tr}
\Big{(}  (\ginv(T)dg(T))^3\0b\Big{)}  \right)
\label{action-a}
\end{eqnarray}
\begin{equation}
{\rm Tr}(T_a T_b) = y(T) \de_{ab}
\sp \epsilon = \left\{ \begin{array}{ll}
1 & \mbox{for real reps}\,\; T \\
2 & \mbox{for complex reps}\;\, T^{\rm (c)} \end{array} \right.
\end{equation}
\begin{equation}
S_{\rm WZW} [ g(T)'] = S_{\rm WZW}[g(T)] \sp
g(T,\xi)' = W(T) g(T,\xi) W\hc (T) \ .
\end{equation}
\end{subequations}
For complex
representations $T^{\rm (c)}$, the block-diagonal group element \eqref{grelc} and
the fact that $y(T^{\rm (c)}{}')=y(\bar T^{\rm (c)}) = y(T^{\rm (c)})$ tell us that
the action \eqref{action-a} is in fact one half
the sum of the WZW actions for $T^{\rm (c)}$ and for
$\bar T^{\rm (c)} \simeq T^{\rm (c)}{}'$.
Moreover, since the WZW action is numerically independent of $T$, the choice
\eqref{action-a} is in fact equal to the WZW action for either $T^{\rm (c)}$ or $\bar T^{\rm (c)}$.

Then following Ref.~\cite{deBoer:2001nw}, one finds
the action for any twisted representation $\T$ in twisted sector $\s =1$ of the charge conjugation orbifold
on $\su(n)$
\begin{subequations}
\begin{eqnarray}
\hat S [\hg (\T)]  & = & -\frac{k}{\epsilon y(T)} \left(
\frac{1}{8\pi}\int d^2\xi
 \0b {\rm Tr}\big{(}\;\hat{g}^{-1} (\st,\s)
\pl_+\hat{g}(\st,\s)\0b\hat{g}^{-1}
(\st,\s)\pl_-\hat{g}(\st,\s)\;\big{)} \right. \nn \\
\label{orbact}
 & & \hskip 1.5cm \left.
  +\frac{1}{12\pi}\int_{\Gamma} {\rm Tr}\big{(}\;(\;\hat{g}^{-1}
  (\st,\s) d\hat{g}(\st,\s)\;)^3\,\big{)} \right)
\end{eqnarray}
\begin{equation}
\hat S [ \hg (\T,\xi + 2 \pi) ] = \hat S [ \hg (\T,\xi) ] \sp
\hg (\T,\xi + 2\pi) = E(T) \hg (\T,\xi) E(T)^\ast \ .
\end{equation}
\end{subequations}
The equations of motion of this action give twisted
left- and right-mover classical matrix currents proportional to  $\hg^{-1} \partial_+ \hg$ and
$\hg \partial_- \hg^{-1}$ whose monodromies agree with the monodromies of
the quantum currents above.

Two further remarks about the case of complex representations are relevant
here. We note first that the block-diagonal group elements $g$ in
\eqref{grelc} and the matrix
\begin{equation}
\Big( \begin{array}{cc} 0 & 1 \\ 1 & 0 \end{array}
\Big) =-i W(T)
\end{equation}
 generate a {\it non-connected Lie group} \cite{Wendt:2001}.
Second, we emphasize that the corresponding group orbifold elements $\hg$ are also
reducible. To see this we consider the so-called group orbifold elements
with twisted boundary conditions $\hat{\sg} (T,\xi)$
\begin{subequations}
\begin{equation}
\label{mondec}
\hg (\T,\xi) = U(T) \hat{\sg} (T,\xi) U\hc (T)
\end{equation}
\begin{equation}
\hat{\sg} (T,\xi) =\left( \begin{array}{cc}
 \hat{\sg}_+ (T,\xi) & 0  \\ 0  & \hat{\sg}_- (T,\xi)
  \end{array} \right)
  \sp \hat{\sg}_{\pm} (T,\xi) =
\exp \left( i \Big( \hb^{0A} (\xi) T_A^{\rm (c)} \pm
\hb^{1I}(\xi) T_I^{\rm (c)} \Big) \right)
\end{equation}
\begin{equation}
\hat{\sg} (T,\xi + 2 \pi) = W(T) \hat{\sg}(T,\xi) W\hc(T) \qquad
\leftrightarrow \qquad
\hat{\sg}_{\pm} (T,\xi + 2 \pi) = \hat{\sg}_{\mp} (T,\xi)
\end{equation}
\end{subequations}
which are block-diagonal with mixed monodromy: As seen in \eqref{mondec},
the group orbifold element $\hg$ is the monodromy decomposition of
$\hat{\sg}$. Then the orbifold action \eqref{orbact} decomposes into the sum
of two terms $\hat{s} (\hat{\sg}_+) + \hat{s} (\hat{\sg}_-)$ which mix
under monodromy transformations.

\subsection{Correlator examples and undetermined parameters}

In this subsection we consider some correlator examples, pointing out and
interpreting the existence of certain undetermined parameters in the solutions
of the twisted KZ systems \eqref{outkz}, \eqref{trmkzeq} for complex representations.

We begin with  the solution of \eqref{outkz} for the
twisted left-mover one-point correlators
\begin{equation}
\label{gwie}
\langle 0 |\hgp (\T,z) |0\rangle = C_+ (\T)z^{-\frac{1}{2k + Q_g} \sum_I \T_{1I} (T)
\T_{1I} (T) } \sp
 C_+  (\T)\T_{0A} (T) = 0 \sp \forall \; A \in \so (n) \ .
\end{equation}
We have checked for real and complex representations that the only solution
for non-trivial $\T$ is
\begin{equation}
\langle 0|\hgp (\T,z)  |0\rangle = 0
\end{equation}
because there is no non-trivial solution to the global Ward identity
in this case. This makes sense because our scalar twist-field state
$| 0 \rangle$ is an $\so (n)$-singlet, and the same result
$ \langle \hgm (\T,\bz)\rangle=0$ is found for the
twisted right movers when $\T$ is non-trivial.

We have also solved the twisted left-mover KZ equations for the two-point
correlator in the case when both representations
$T^{\rm (c)}{}^{(1)} \equiv T^{(1)}$, $T^{\rm (c)}{}^{(2)} \equiv T^{(2)}$
are taken as the fundamental representation of $\su (n)$. The result is
\begin{subequations}
\label{2pcor}
\begin{equation}
  \langle 0|\hgp (\T^{(1)},z_1)\hgp (\T^{(2)},z_2)|0 \rangle
= C_+(\T) (z_1 z_2)^{-\Delta_n}
\left( \frac{z_1 z_2}{z_{12}^2} \right)^{\alpha_n}
 \left( \frac{ \sqrt{z_1} - \sqrt{z_2} }{\sqrt{z_1} +\sqrt{z_2}}
\right)^{M(1,2)}
\end{equation}
\begin{equation}
C_+ (\T)\left( \begin{array}{cc}
 T_A^{(1)} + T_A^{(2)} & 0   \\  0  & T_A^{(1)} + T_A^{(2)}\end{array} \right)
= 0 \sp \forall \; A \in \so (n)
\end{equation}
\begin{equation}
\Delta_n \equiv \frac{n^2-1}{2n (x+n)} \sp \alpha_n \equiv
\frac{n-1}{4(x+n)}
\end{equation}
\begin{equation}
M(1,2) \equiv \frac{2}{2k +Q_g}\sum_I
\left( \begin{array}{cc}
0 & T_I^{(2)}  \\ T_I^{(2)} & 0 \end{array} \right) \otimes
\left( \begin{array}{cc}
0 & T_I^{(1)}  \\ T_I^{(1)} & 0 \end{array} \right)
\end{equation}
\end{subequations}
where the quantity $\Delta_n$ is the conformal weight of the fundamental
representation in the symmetric theory.
To obtain this result, we used the explicit form \eqref{twrfun} of the
twisted representation matrices and the identities
\begin{subequations}
\begin{equation}
\frac{1}{2k + Q_g} \sum_A \T_{0A} (T) \T_{0A} (T) =
\frac{1}{2k + Q_g}\one \sum_A T_A T_A = \alpha_n \one \otimes \one_n
\sp \one =\left( \begin{array}{cc}
1& 0  \\ 0 & 1 \end{array} \right)
\end{equation}
\begin{equation}
\frac{1}{2k + Q_g} \sum_I \T_{1I}(T) \T_{1I}(T) =
\frac{1}{2k + Q_g} \one \sum_I T_I T_I =(\Delta_n-\alpha_n) \one \otimes \one_n
\end{equation}
\begin{equation}
[ M(1,2), \T_{0A}^{(1)} + \T_{0A}^{(2)} ] = 0 \sp \forall \; A \in \so (n)
\end{equation}
\end{subequations}
where $1$ is the $2\times 2$ unit matrix in the doubled space and $1_n$
is the $n\times n$ unit matrix in the space of the generalized Gell-Mann matrices.

We have further solved \eqref{trmkzeq} for the twisted right-mover two-point
correlator in this case, solved the full set of global Ward identities
\begin{subequations}
\begin{equation}
C(\T) = C_- (\T) C_+ (\T)
\end{equation}
\begin{equation}
\label{ctgl}
C (\T)\left( \begin{array}{cc}
 T_A^{(1)} + T_A^{(2)} & 0   \\  0  & T_A^{(1)} + T_A^{(2)}\end{array} \right)
=\left( \begin{array}{cc}
 T_A^{(1)} + T_A^{(2)} & 0   \\  0  & T_A^{(1)} + T_A^{(2)}\end{array} \right)
C (\T) =0
, \; \forall \; A \in \so (n)
\end{equation}
\end{subequations}
and combined these results with \eqref{2pcor} to
obtain the non-chiral two-point correlator for the full twisted vertex
operators $\hg = \hgm \hgp$. The final result is
\begin{subequations}
\begin{equation}
  \langle 0|\hg (\T^{(1)},\bz_1,z_1)\hg (\T^{(2)},\bz_2,z_2)|0 \rangle
= C(\T) |z_1 z_2|^{-2\Delta_n}
\left| \frac{z_1 z_2}{z_{12}^2} \right|^{2\alpha_n}
 \left| \frac{ \sqrt{z_1} - \sqrt{z_2} }{\sqrt{z_1} +\sqrt{z_2}}
\right|^{2M(1,2)}
\end{equation}
\begin{equation}
C(\T) =
\left( \begin{array}{cc}
 \gamma_1 D & \gamma_2 D   \\  \gamma_3 D  & \gamma_4 D \end{array} \right)
\sp [ C (\T), M(1,2) ] = 0
\end{equation}
\begin{equation}
(T_a^{(1)})_{\a_1}{}^{\be_1} \; ,\; (T_a^{(2)})_{\a_2}{}^{\be_2} \;\;\;\; :
\;\;\;\; D_{\be_1 \be_2}{}^{\a_1 \a_2} =\de_{\be_1 \be_2} \de^{\a_1 \a_2 }
\end{equation}
\end{subequations}
where $\gamma_1 \ldots \gamma_4$ are arbitrary constants found in  the
general  solution of the global Ward identities \eqref{ctgl}.

Further determination of these constants is beyond the scope of this paper.
However, we note here that the constants can be interpreted as undetermined
couplings among the blocks of the quantum analogue $\hat{\sg} (T,\bz,z)$
of the group orbifold elements with twisted boundary  conditions:
\begin{subequations}
\begin{equation}
\hat{\sg} (T,\bz,z) \equiv U(T) \hg (\T,\bz,z) U\hc (T) \sp
[ U (T^{(1)}) \otimes U (T^{(2)}), C(\T) ] =0
\end{equation}
\begin{equation}
  \langle 0|\hat{\sg} (T^{(1)},\bz_1,z_1)\hat{\sg} (T^{(2)},\bz_2,z_2)|0 \rangle
=\left( \begin{array}{cc}
 \gamma_1 D & \gamma_2 D   \\  \gamma_3 D  & \gamma_4 D \end{array} \right)
|z_1 z_2|^{-2\Delta_n}
\left| \frac{z_1 z_2}{z_{12}^2} \right|^{2\alpha_n}
 \left| \frac{ \sqrt{z_1} - \sqrt{z_2} }{\sqrt{z_1} +\sqrt{z_2}}
\right|^{2 N (1,2)}
\end{equation}
\begin{equation}
N(1,2) \equiv \frac{2}{2k +Q_g}\sum_I
\left( \begin{array}{cc}
T_I^{(2)} & 0  \\ 0 & - T_I^{(2)}  \end{array} \right) \otimes
\left( \begin{array}{cc}
T_I^{(1)} & 0   \\ 0 & -T_I^{(1)}  \end{array} \right) \ .
\end{equation}
\end{subequations}
This suggests that a natural boundary condition is $\gamma_2 = \gamma_3=0$,
i.e. a block diagonal solution for the $\hat{\sg}$ correlator. Such a
boundary condition could in principle be implemented in general because,
for any number of $\hat{\sg}$'s in any set of complex representations,
we find the block-diagonal twisted KZ system:
\begin{subequations}
\begin{equation}
\hat{\cal{A}} (T,\bz,z) \equiv \langle 0| \hat{\sg} (T^{(1)},\bz_1,z_1)
\cdots \hat{\sg} (T^{(N)},\bz_N,z_N) | 0 \rangle
\end{equation}
\begin{equation}
\partial_\m \hat{\cal{A}} (T,\bz,z) = \hat{\cal{A}} (T,\bz,z)
\hat{\cal{W}}_\m (T,z) \sp
\bar{\partial}_\m \hat{\cal{A}} (T,\bz,z) = {\hat{\bar{\cal{W}}}}_\m (T,\bz)
\hat{\cal{A}} (T,\bz,z)
\end{equation}
\begin{equation}
\hat{\cal{W}}_\m (T,z) = \hat{W}_\m (\T,z) \vert_{ \T \rightarrow T}
\sp
{\hat{\bar{\cal{W}}}}_\m (T,\bz) = {\hat{\bar{W}}}_\m (\T,\bz) \vert_{ \T \rightarrow T}
\end{equation}
\begin{equation}
\hat{\cal{A}} (T,\bz,z) \left( \sum_{\m=1}^N T_A^{(\m)} \right) =
\left(\sum_{\m=1}^N T_A^{(\m)} \right) \hat{\cal{A}} (T,\bz,z) = 0
\sp \forall \; A \in \so (n) \ .
\end{equation}
\end{subequations}
Here the twisted connections $\hat W$ and $\hat{\bar{W}}$ are given in
\eqref{tlmkzeq}, \eqref{trmkzeq} and the
untwisted representation matrices $T$ have the block-diagonal
form  in \eqref{comrep}.

\vskip 1cm
\noindent {\bf Acknowledgements}

For helpful discussions we thank J. de Boer, C. Helfgott, J. Lepowsky,
C. Schweigert, F. Wagner and N. Warner.
MBH and NO thank the Niels Bohr Institute for hospitality and support.
The work of MBH was supported in part by the Director, Office of Energy Research,
Office of High Energy and Nuclear Physics, Division of High Energy Physics of
the U.S. Department of Energy under Contract DE-AC03-76SF00098 and in part by
the National Science Foundation under grant PHY95-14797.
The work of NO is supported in part by the stichting FOM and
the European Commission RTN programme HPRN-CT-2000-00131.

\appendix

\section{The $H$-eigenvalue problem for permutation groups \label{Heigp} }

In this appendix we use the modified notation of Subsec.~\ref{permsec} to solve
the $H$-eigenvalue problem \cite{deBoer:1999na,Halpern:2000vj,deBoer:2001nw}
for all permutation groups $H \subset S_N$.
The solution of the
$H$-eigenvalue problem also gives an explicit derivation of various results
which were argued more abstractly for the WZW permutation orbifolds
in Ref.~\cite{deBoer:2001nw}.

We begin with the permutation-invariant system
\begin{subequations}
\label{pisys}
\begin{equation}
\oplus_{I=0}^{K-1} \gfrak^I \sp \gfrak^I \simeq \gfrak \sp K \leq N
\end{equation}
\begin{equation}
J_{aI}(z) J_{bJ} (w) = \de_{IJ} \left\{ \frac{k\eta_{ab}}{(z-w)^2}
+ \frac{if_{ab}{}^cJ_{cI}(w)}{z-w} \right\} +\Ord (z-w)^0
\end{equation}
\begin{equation}
T (z) =\frac{1}{2k + Q_\sgbn} \sum_{I=0}^{K-1} \eta^{ab}: J_{aI}(z) J_{bI}(z):
\sp a ,b= 1 \ldots {\rm dim}\,\gfrak
\end{equation}
\end{subequations}
where the automorphism group acts on the currents and stress tensor as
$J_{aI}{}' = \w(h_\s)_I{}^J J_{aJ}$, $h_\s \in H ({\rm permutation})
 \subset S_N$ and $T(z)' = T(z)$. The eigencurrents
 \cite{deBoer:1999na,Halpern:2000vj,deBoer:2001nw}
\begin{subequations}
\label{copeig}
\begin{equation}
\sj_{n(r)aj}(z) = \chi_{n(r)aj} (\s) U (\s)_{n(r)j}{}^I J_{aI} (z)
\end{equation}
\begin{equation}
\sj_{n(r) aj} (z)' = E_{n(r)} (\s) \sj_{n(r)aj} (z) \sp
E_{n(r)} (\s) = e^{- 2\pi i \srac{n(r)}{\rho(\s)}}
\end{equation}
\end{subequations}
have diagonal response under $H$, where $\w (h_\s) U\hc (\s)=U\hc (\s) E(\s)$,
$\srange$ is the $H$-eigenvalue problem
 \cite{deBoer:1999na,Halpern:2000vj,deBoer:2001nw} and the $\chi$'s are
 normalization constants.

To study the $H$-eigenvalue problem most efficiently, it is convenient to
introduce a new basis. For each $h_\s \in H({\rm permutation})$, we may
relabel the copies $I$ to obtain a $\s$-dependent {\it cycle basis} of the type constructed for
$H=S_N$ in Ref.~\cite{Halpern:2000vj}
\begin{subequations}
\begin{equation}
 I \quad \rightarrow \quad  \hjs j
\sp J_{aI} \quad \rightarrow \quad \tilde J_{a\hjs j}
\end{equation}
\begin{equation}
\tilde J_{a\hjs j}(z) \tilde J_{b\hls l}(w) = \de_{jl} \de_{\hjs-\hls,\,0\rmod
f_j(\s)} \left\{ \frac{k\eta_{ab}}{(z-w)^2}
+ \frac{if_{ab}{}^c\tilde J_{c\hjs j }(w)}{z-w} \right\}+\Ord (z-w)^0
\end{equation}
\begin{equation}
T (z) = \frac{1}{2k + Q_\sgbn}\sum_j \sum_{\hjs=0}^{f_j(\s)-1}
 \eta^{ab}: \tilde J_{a\hjs j}(z) \tilde J_{b \hjs j}(z):
\end{equation}
\begin{equation}
\label{sjjtr}
\sj_{n(r) aj} (z) =\chi_{n(r)aj} (\s) U (\s)_{n(r)aj}{}^{\hls l} \tilde J_{a\hls l} (z)
\end{equation}
\begin{equation}
\label{jlpb}
\bar{\hjs} = 0, \ldots , f_j(\s)-1 \sp \bar{\hls} =0, \ldots , f_l(\s)-1 \sp
\sum_j f_j (\s) = K \ .
\end{equation}
\end{subequations}
Here the unhatted indices $j$ or $l$ label the disjoint cycles (of size and order
$f_j(\s)$ or $f_l(\s)$) of each
$h_\s \in H$ and the hatted indices $\hjs$ or $\hls$ run inside the
disjoint cycle $j$ or $l$. The labelling inside each disjoint cycle is
periodic $\hjs \rightarrow \hjs \pm f_j(\s)$ and the barred quantities
$\bar{\hjs}$, $\bar{\hls}$ in \eqref{jlpb} are the pullbacks of $\hjs$ and
$\hls$ into the fundamental range. The cycle basis
can be chosen so that each $h_\s$ has matrix representation
\begin{equation}
\w (h_\s)_{a\hjs j}{}^{b \hls l} = \de_a^b \w(h_\s)_{\hjs j}{}^{\hls l}
\sp \w(h_\s)_{\hjs j}{}^{\hls l} = \de_j^l \de_{\hjs +1, \, \hls\,\rmod f_j(\s)}
\sp \srange \end{equation}
that is so that the automorphism is a cyclic permutation $\hjs \rightarrow
\hjs +1$ in each disjoint cycle.

\un{Examples}. \nl
a) The cycle basis is discussed explicitly for the permutation groups
$S_N$ in Ref.~\cite{Halpern:2000vj}, with
\begin{equation}
K= N \sp f_j(\s) = \s_j \sp \s_{j+1} \leq \s_j \sp j = 0 ,\ldots  ,
n(\vec{\s}) -1 \sp \sum_{j=0}^{n(\vec\s)-1} \s_j = N \ .
\end{equation}
This choice gives one element $h_\s \in S_N$ in each conjugacy class of $S_N$. \newline
b) For the cyclic permutation groups $\z_\l$ one has
\begin{equation}
K = \l \sp f_j(\s) = \r (\s) \sp \bar{\hjs} = 0 , \ldots , \r (\s) -1 \sp
 j = 0 , \ldots , \srac{\l}{\r (\s)} -1
\end{equation}
where $\rho (\s)$ is the order of $h_\s \in \z_\l$ and $\s = 0, \ldots ,
\l -1$. In this case we give some simple examples of the relabelling $ I \rightarrow
\hjs j$: For $\l$=prime, one has  $\r(\s) = \l$, $\s = 1,\ldots , \l-1$ and
a single disjoint cycle $j=0$. Choosing also  $\s=2$ we maintain a shift
$\hjs \rightarrow  \hjs +1$ in the single disjoint cycle by relabelling
\begin{subequations}
\begin{equation}
\hjs = (0,1,\ldots,\l-1) \quad \leftrightarrow
I =( 0 , 2, \ldots , \l-1, 1, 3, \ldots, \l -2 )
\end{equation}
\begin{equation}
\label{exsn}
\tilde J_{a\hjs} = \tilde J_{a \hjs 0} = ( J_{a0}, J_{a2}, \ldots ,J_{a,\l-1},
J_{a1},J_{a3}, \ldots ,J_{a,\l-2} ) \ .
\end{equation}
\end{subequations}
For $\l=4$, $\s=2$ we have $\r(2)=2$ and two disjoint cycles $j=0$ and 1,
and we may relabel as follows:
\begin{equation}
\tilde J_{a \hjs 0} = ( J_{a0}, J_{a2} ) \sp
\tilde J_{a \hjs 1} =(J_{a1},J_{a3} )
\end{equation}
where $\hjs =0$ and 1 in each disjoint cycle.

We consider now the $H$-eigenvalue problem and its solution in the cycle
basis:
\begin{subequations}
\label{Usolperm}
\begin{equation}
\sum_l \sum_{\hls =0}^{f_l(\s) -1} \w (h_\s)_{\hjs j}{}^{\hls l}
U\hc (\s)_{\hls l}{}^{\hms m} = U\hc(\s)_{\hjs j}{}^{\hms m} E_{\hms}^m(\s)
\end{equation}
\begin{equation}
U\hc(\s)_{\hjs j}{}^{\hls l} = \frac{\d_j^l}{\sqrt{f_j(\s)}} e^{- 2\pi i
\srac{\hjs \hls}{f_j(\s)}} \sp
U(\s)_{\hjs j}{}^{\hls l} = \frac{\d_j^l}{\sqrt{f_j(\s)}} e^{ 2\pi i
\srac{\hjs \hls}{f_j(\s)}} \sp E_\hjs^j = e^{-2\pi i \srac{\hjs}{f_j(\s)}}
\end{equation}
\begin{equation}
\sum_l \sum_{\hls =0}^{f_j(\s)-1} U\hc(\s)_{\hjs j}{}^{\hls l} U(\s)_{\hls l}
{}^{\hms m} = \d_j^m \d_{\hjs -\hms,\,0 \rmod f_j(\s)}
\end{equation}
\begin{equation}
\label{sprange}
\bar{\hjs} = 0, \ldots , f_j(\s)-1 \sp
\bar{\hls} = 0, \ldots , f_l(\s)-1 \sp \bar{\hms} = 0, \ldots , f_m(\s)-1 \ .
\end{equation}
\end{subequations}
Here the relabelling of the spectral integers
\begin{equation}
n(r) j \rightarrow \hjs j \sp U\hc (\s)_{n(r)j}{}^{\hls l} \rightarrow
U\hc (\s)_{\hjs j}{}^{\hls l}
\end{equation}
is dictated by the form of the eigenvalue problem, and the correspondence
\begin{equation}
\label{coreig}
\frac{n(r)}{\r(\s)} = \frac{\hjs }{f_j(\s)}
\sp \bar{\hjs} = 0 , \ldots , f_j(\s)-1
\end{equation}
is obtained by comparing the eigenvalues $E_{\hjs}^j$ to the standard form
of the eigenvalues in \eqref{copeig}. In what follows we use the
correspondence \eqref{coreig} to relabel all the spectral integers
$n(r)  \rightarrow \hjs $.

Returning to the eigencurrents $\sj_{n(r)aj} \rightarrow \sj_{\hjs aj}$
we choose the normalization
\begin{equation}
\label{chinor}
\chi_{n(r) aj}(\s) \rightarrow \chi_{\hjs aj} (\s)= \sqrt{f_j(\s)} \ .
\end{equation}
This choice of normalization is in agreement with the prior conventions
\cite{deBoer:1999na,Evslin:1999ve,Halpern:2000vj,deBoer:2001nw}
adopted for $\z_\l$ and $S_N$.
Then, combining \eqref{sjjtr}, \eqref{Usolperm} and \eqref{chinor}
we find an explicit form for the eigencurrents
\begin{subequations}
\begin{equation}
\label{sjperm}
 \sj_{\hjs a j}(z) = \sqrt{f_j(\s)} \ U(\s)_{\hjs j}{}^{\hls l}
\tilde J_{a \hls l} (z) = \sum_{\hjs'=0}^{f_j(\s)-1}
e^{2 \pi i \srac{\hjs \hjs'}{f_j(\s)}} \tilde J_{a \hjs' j} (z) \sp
\sj_{\hjs a j}(z)' = e^{- 2 \pi i \srac{\hjs}{f_j(\s)}} \sj_{\hjs a j}(z)
\end{equation}
\begin{equation}
\tilde  J_{a \hjs j}(z) = \frac{1}{f_j(\s)}\sum_{\hjs'=0}^{f_j(\s)-1}
e^{-2 \pi i \srac{\hjs \hjs'}{f_j(\s)}} \sj_{a \hjs' j} (z) \sp
\sum_j \sj_{0 a j} (z) = \sum_j \sum_{\hjs =0}^{f_j(\s)-1} \tilde J_{a \hjs j}(z)
= \sum_{I=0}^{K-1} J_{aI}(z)
\end{equation}
\end{subequations}
and the OPEs of the eigencurrents are
\begin{subequations}
\begin{equation}
\sj_{\hjs a j}(z) \sj_{\hls bl}(w)
=\frac{ \G_{\hjs a j; \hls b l} (\s) }{(z-w)^2} + i \F_{\hjs a j; \hls b l}
{}^{\hjs + \hls,cm} (\s) \frac{\sj_{\hjs + \hls,cm}(w) }{z-w} + \Ord(z-w)^0
\end{equation}
\begin{equation}
\G_{\hjs a j ;\hls b l}(\s) = \eta_{ab} \hat k_j (\s) \de_{jl}
\de_{\hjs + \hls,\,0 \rmod f_j(\s)} \sp \hat k_j(\s) = k f_j (\s)
\end{equation}
\begin{equation}
\F_{\hjs a j; \hls b l} {}^{\hms cm} (\s) = f_{ab}{}^c \de_{jl} \de_l^m
\de_{\hjs + \hls-\hms,\,0 \rmod f_j(\s)} \ .
\end{equation}
\end{subequations}
We may also rewrite the stress tensor in terms of the eigencurrents
\begin{subequations}
\begin{equation}
\label{Lrew}
\lr^{\hjs aj; \hls b l} (\s) =\frac{k}{2k +Q_\sgbn} \G^{\hjs a j ;\hls b l}(\s)
\sp \G^{\hjs a j ;\hls b l}(\s) = \eta^{ab} \hat k_j^{-1}(\s) \de^{jl}
\de_{\hjs + \hls,\,0 \rmod f_j(\s)}
\end{equation}
\begin{equation}
T (z)  = \sum_j \lr^{\hjs aj; \hls b l} (\s) : \sj_{\hjs aj} (z) \sj_{\hls bl} (z) :
 =   \frac{1}{2k +Q_\sgbn} \sum_j \frac{1}{f_j(\s)}
\sum_{\hjs =0}^{f_j(\s)-1}\eta^{ab} : \sj_{\hjs aj} (z) \sj_{-\hjs,bj} (z) :
\end{equation}
\end{subequations}
where $:\cdot :$ is operator product normal ordering
\cite{Evslin:1999qb,deBoer:1999na,Evslin:1999ve,Halpern:2000vj,deBoer:2001nw}.
Under the relabelling $n(r) \rightarrow \hjs $, these results for
$\G$, $\F$ and ${\cL}$ are equivalent  to those given in
Ref.~\cite{deBoer:2001nw}.

Then using local isomorphisms $\sj \dual \hj$ \cite{deBoer:1999na} we obtain
for sector $\s$ of the general WZW permutation orbifold
\begin{subequations}
\begin{equation}
\hj_{\hjs a j} (z) \hj_{\hls b l}(w) = \frac{\G_{\hjs a j; \hls b l}(\s)}{(z-w)^2}
+ i \F_{\hjs a j; \hls b l}
{}^{\hjs + \hls,cm} (\s) \frac{\hj_{\hjs + \hls,cm}(w) }{z-w} + \Ord(z-w)^0
\end{equation}
\begin{equation}
\hj_{\hjs a j}(ze^{2 \pi i}) = e^{-2 \pi i \srac{\hjs}{f_j(\s)}} \hj_{\hjs aj}(z)
\sp
\hj_{\hjs a j}(z) = \sum_{m \in \sz} \hj_{\hjs a j}( m + \srac{\hjs}{f_j(\s)})
z^{-(m +\srac{\hjs}{f_j(\s)})-1}
\end{equation}
\begin{eqnarray}
[ \hj_{\hjs a j}(m + \srac{\hjs}{f_j(\s)}),
\hj_{\hls b l}(n + \srac{\hls}{f_j(\s)})] & = &
\d_{jl} \BIG( i f_{ab}{}^c \hj_{\hjs + \hls} (m + n + \srac{\hjs+\hls}{f_j(\s)})
 \nn \\
& & + \hat{k}_j (\s) \eta_{ab} (m + \srac{\hjs}{f_j(\s)})
\d_{m+n+\srac{\hjs+\hls}{f_j(\s)},0} \BIG)
\end{eqnarray}
\begin{equation}
\hat T_\s (z)  =   \lr^{\hjs aj;\hls bl} (\s) : \hj_{\hjs aj}(z) \hj_{\hls b l} (z):
 =  \frac{1}{2k+Q_\sgbn} \sum_j \frac{1}{f_j(\s)} \sum_{\hjs =0}^{f_j(\s)-1}
\eta^{ab} : \hj_{\hjs aj} (z) \hj_{-\hjs b j} (z) :  \;\; .
\end{equation}
\end{subequations}
Under the relabelling $n(r) a j \rightarrow \hjs a j$, these results are
also equivalent to those given in Ref.~\cite{deBoer:2001nw}.

For the special case $H = \z_\l$ another form of the eigencurrents
\begin{equation}
\label{eigcurz}
\sj_{r aj } (z) = \sum_{s=0}^{\r(\s) -1}
e^{2 \pi i \srac{N(\s) rs}{\r(\s)}} J_{a, \srac{\l}{\r (\s)}s + j} (z)
\sp \bar{\hjs} = \bar{r} = 0 ,\ldots , \r(\s)-1 \sp
j = 0 ,\ldots, \srac{\l}{\r(\s)} -1
\end{equation}
was given in Eq.~(3.28) of Ref.~\cite{deBoer:1999na}, where the integers
$N(\s)$ are also defined. This is the form taken by the $\z_\l$
eigencurrents when expressed in terms of the untwisted currents $J_{aI}$
of the original $I$ basis. The two expressions \eqref{eigcurz} and
\eqref{sjperm} for $H=\z_\l$ are  equal because the eigencurrents are
independent of the basis choice $I$ or $\hjs j$. To see how this works
in a simple example, consider the case $\l=3$, $\s=2$ with
$\r(2)=3$, $N(2)=2$ and $j=0$:
\begin{equation}
\label{eigcs}
\sj_{ra0}(z) = \sum_{s=0}^ 2 e^{4 \pi i \srac{rs}{3}} J_{as}
=\sum_{s=0}^2 e^{2\pi i \srac{rs}{3}} \tilde J_{as}\sp\bar{r} = 0,1,2
\sp \tilde J_{as} = (J_{a0},J_{a2},J_{a1} ) \ .
\end{equation}
The form of $\tilde J_{as}$ in \eqref{eigcs} is a special case of the
relabelling \eqref{exsn}.

For the extended $H$-eigenvalue problem, we have the solution
\cite{deBoer:2001nw}
\begin{equation}
U\hc (T,\s)_{\a I}{}^{n(r) \be j} \quad \rightarrow U\hc(T,\s)_{\a \hjs j}
{}^{\hls \be l} = \d_{\a}^\be U\hc (\s)_{\hjs j}{}^{\hls l} \sp
\a, \be = 1 \ldots {\rm dim}\,T
\end{equation}
where $T$ is any irrep of $\gfrak$ and
$U\hc(\s)$ is the solution \eqref{Usolperm} to the $H$-eigenvalue problem.
Then we may evaluate Eq.~(7.7e) of Ref.~\cite{deBoer:2001nw} to find the
factorized form of the twisted representation matrices $\T$
\begin{subequations}
\label{facform}
\begin{equation}
\T_{n(r) j}(T,\s) \quad \rightarrow \quad \T_{\hjs a j}(T,\s) \sp
[\T_{\hjs a j}(T,\s) ,\T_{\hls b l}(T,\s) ] = i \de_{jl} f_{ab}{}^c
\T_{\hjs +\hls, cj}(T,\s)
\end{equation}
\begin{equation}
\label{trepa}
\T_{\hjs a j} (T,\s)= T_a t_{\hjs j} (\s) \sp [T_a,T_b] = i f_{ab}{}^c T_c
\sp t_{\hjs j }(\s)_{\hls l}{}^{\hms m} =\d_{jl} \d_l^m
\d_{\hjs + \hls - \hms,0\,\rmod f_j (\s)}
\end{equation}
\end{subequations}
in agreement with the result given in Ref.~\cite{deBoer:2001nw}.
The formulae \eqref{trepp}, \eqref{ttrep} and \eqref{ttex} are the abelian
limit of \eqref{facform}, as expected.

\section{More about the WZW permutation orbifolds \label{nakz}}

We continue with the WZW permutation orbifolds, using the spectral index
relabelling
\begin{equation}
n(r) aj \quad \rightarrow \hjs a j \sp N(r) \a j \quad \rightarrow
\quad \hjs \a j \sp a = 1 \ldots {\rm dim}\, \gfrak \sp
\a = 1 \ldots {\rm dim}\, T
\end{equation}
to rewrite more of the results of Ref.~\cite{deBoer:2001nw}. First we
rewrite the results of Ref.~\cite{deBoer:2001nw} for the left- and
right-mover ground state conformal weights in each sector $\s$ of all
WZW permutation orbifolds \cite{deBoer:2001nw}
\begin{subequations}
\label{nacw}
\begin{eqnarray}
\hat{\Delta}_0 (\s) =\hat{\bar{\Delta}}_0 (\s) & = &
\frac{ c_\sgbn}{2}
\sum_{r j } \frac{\nb}{2 \rho (\s)} \left( 1 - \frac{\nb}{\rho (\s)} \right)
=\frac{ c_\sgbn}{2} \sum_j \sum_{\hjs = 1}^{f_j(\s)-1}
 \frac{\hjs}{2 f_j(\s)} \left( 1 - \frac{\hjs}{f_j(\s)} \right)
 \nn \\
&=&\frac{c_\sgbn}{24} \sum_j \left(f_j (\s) - \frac{1}{f_j(\s)} \right)
=\frac{c_\sgbn}{24} \left( K - \sum_j  \frac{1}{f_j(\s)} \right)
\label{cwn}
\end{eqnarray}
\begin{equation}
c_\sgbn = \frac{x_\sgbn {\rm dim}\,\gfrak}{x_\sgbn + \tilde{h}_\sgbn}
\sp x_\sgbn \ \equiv \frac{2k}{\psi_\sgbn^2} \sp \tilde{h}_\sgbn \equiv
\frac{Q_\sgbn}{\psi_\sgbn^2}
\end{equation}
\end{subequations}
where $K$ is defined in \eqref{pisys} and
$c_\sgbn$ is the central charge of the affine-Sugawara construction
\cite{Bardakci:1971nb,Halpern:1971ay,Dashen:1975hp,Knizhnik:1984nr}
on affine $\gfrak$. Eq.~\eqref{cwper} is
the abelian limit of \eqref{nacw}, as expected.

To go further it will be useful to have the identities
\begin{subequations}
\begin{equation}
\lr^{\hjs aj; - \hjs, bl}(\s) \T_{\hjs aj} \T_{-\hjs,bl} =
 \Delta_{\sgbn} (T) \one \sp \frac{(\eta^{ab} T_a T_b)_\a{}^{\be}}{2k + Q_{\sgbn}} =
\Delta_{\sgbn} (T) \de_\a^\be
\end{equation}
\begin{equation}
\lr^{0 aj; 0 bl}(\s) \T_{0aj} \T_{0bl} =
\Delta_{\sgbn} (T)  \sum_j \frac{1}{f_j (\s)} t_{0j}(\s)
\end{equation}
\begin{eqnarray}
2\lr^{n(r) aj; - n(r) bl}(\s) \srac{\nb}{\r(\s)}\T_{n(r)aj} \T_{-n(r),bl} & = &
 \lr^{n(r) aj; - n(r) bl}(\s) \T_{n(r)aj} \T_{-n(r),bl} (1- \d_{\nb,0}) \;\;\;\;\;
\\ && =\Delta_{\sgbn} (T) \left( \one- \sum_j \frac{1}{f_j (\s)} t_{0j}(\s) \right)
\end{eqnarray}
\end{subequations}
where we have  used \eqref{Lrew} and
the factorized form $\T = T t$ of the twisted representation
matrices. These results are analogous to those given for the abelian case in
Subsec.~\ref{permsec}.

Then the twisted left-mover KZ systems of the WZW permutation orbifolds
take the form \cite{deBoer:2001nw}
\begin{subequations}
\begin{equation}
\partial_\m \hat A_+ (\s)= \hat A_+  (\s) \hat W_\m (\s) \sp
\hat A_+ (\s) \equiv \hat A_+(\T,z,\s) \sp \s = 0, \ldots, \r (\s) -1
\end{equation}
\begin{eqnarray}
\hat W_{\mu} (\s)  & = &  \frac{2}{2k + Q_{\sgbn}}
\sum_{\nu \neq \mu} \eta^{ab} \frac{T_a^{(\nu)} T_b^{(\mu)}}{z_{\mu
\nu}} \sum_j \sum_{\hjs=0}^{f_j (\s) -1}
\left( \frac{z_\nu}{z_\mu} \right)^{\srac{\hjs}{f_j (\s )}}
\frac{t_{\hjs j }^{(\nu)} (\s) t_{-\hjs , j}^{(\mu)} (\s) }{f_j (\s)} \nn \\
 && -
\frac{\Delta_{\sgbn}
(T^{(\mu)})}{z_{\mu}} \left(\one - \sum_j \frac{1}{f_j(\s)} t_{0j}^{(\m)}(\s) \right)
\label{azlprop} \end{eqnarray}
\begin{equation}
\hat A_+ (\s) \sum_{\m =1}^N T_a^{(\m)} t_{0j}^{(\m)} (\s) = 0 \ .
\end{equation}
\end{subequations}
Finally we note the form of the
general left-mover two-point correlators of Ref.~\cite{deBoer:2001nw} in this notation
\begin{subequations}
\label{2ptper}
\begin{eqnarray}
& & \!\!\!\!\! \hat A_+(1,2) = \langle \hgp (\T^1,z_1,\s) \hgp (\T^2,z_2,\s) \rangle_\s \nn
\\
 & &  =  C_+(\T,\s)
z_1^{- \Delta_{\sgbn}(T^{(1)})
( \one- \sum_j \srac{1}{f_j (\s)} t_{0j}^{(1)}(\s) )}
z_2^{- \Delta_{\sgbn}(T^{(2)})
( \one- \sum_j \srac{1}{f_j (\s)} t_{0j}^{(2)}(\s) )}
 z_{12}^{ - 2\Delta_{\sgbn} (T^{(1)})\sum_j \srac{1}{f_j(\s)} t_{0j}^{(1)}(\s)}
 \nn \\
 & & \hskip 1cm \times
\exp \left( \frac{2}{2k+Q_{\sgbn}} T_a^{(2)} \eta^{ab}
 T_b^{(1)}\sum_j \frac{1}{f_j(\s)}\sum_{\hjs=1}^{f_j(\s)-1}
  t_{\hjs j}^{(2)} (\s) t_{-\hjs, j}^{(1)} (\s)
 I_{\srac{\hjs}{f_j (\s)}} ( \srac{z_1}{z_2})  \right) \;\;\;
\end{eqnarray}
\begin{equation}
C_+ (\T,\s)( T_a^{(1)} t_{0j}^{(1)}(\s) + T_a^{(2)} t_{0j}^{(2)} (\s) ) =0
\end{equation}
\end{subequations}
where the indefinite integrals $I$ were evaluated in Ref.~\cite{deBoer:2001nw}.
The abelian limit of \eqref{2ptper} is in agreement with the result
\eqref{Nptaper} of the text.

\section{An identity for the full correlators of the abelian
orbifolds \label{idnc} }

In this appendix, we verify the symmetry $F(\k,\r) = F(\r,\k)$ stated in
Eq.~\eqref{symid0} of the text. The steps followed here  parallel those used in
Ref.~\cite{deBoer:2001nw} to prove $1 \leftrightarrow 2$ symmetry for the
non-chiral two-point correlators of the WZW cyclic permutation orbifolds.

We start with the definition ($\r \neq k$)
\begin{eqnarray}
F(\r,\k) & \equiv & \G^{\nrm; \mnrn} (\s)( 1 - \d_{\nb ,0} ) \nn \\
& & \times \left( \T_\nrm^{(\r)} \T_\mnrn^{(\k)}
I_{\srac{\nb}{\r (\s)}} \left( \srac{z_\r}{z_\k},\infty \right)
+\T_\nrm^{(\k)} \T_\mnrn^{(\r)}
I_{\srac{\nb}{\r (\s)}} \left( \srac{\bz_\r}{\bz_\k},\infty \right) \right) \ .
\end{eqnarray}
Then we compare this to $F(\k,\r)$, following the steps
\begin{subequations}
\begin{eqnarray}
F(\k,\r) & = & \G^{n(r)-\r(\s),\m;\r(\s)-n(r),\n} (\s)( 1 - \d_{\nb ,0} )
 \nn \\
& & \hskip -1cm \times \left( \T_{\r(\s)-n(r),\n}^{(\r)} \T_{n(r)-\r(\s),\m}^{(\k)}
I_{\srac{\r(\s)-\nb}{\r (\s)}} \left( \srac{z_\r}{z_\k},0 \right)
+\T_{\r(\s)-n(r),\n}^{(\k)} \T_{n(r)-\r(\s),\m}^{(\r)}
I_{\srac{\r(\s)-\nb}{\r (\s)}} \left( \srac{\bz_\r}{\bz_\k},0 \right) \right)
\nn
 \\ \label{st1}\\
 & = & \G^{-n(r),\m;n(r)\n} (\s)( 1 - \d_{\nb ,0} )
 \nn \\
& & \times \left( \T_{n(r)\n}^{(\r)} \T_{-n(r),\m}^{(\k)}
I_{\srac{\nb}{\r (\s)}} \left( \srac{z_\r}{z_\k},0 \right)
+\T_{n(r)\n}^{(\k)} \T_{-n(r),\m}^{(\r)}
I_{\srac{\nb}{\r (\s)}} \left( \srac{\bz_\r}{\bz_\k},0 \right) \right)
\label{st2} \\
 & = & \G^{n(r),\m;-n(r),\n} (\s)( 1 - \d_{\nb ,0} )
 \nn \\
& & \times \left( \T_{n(r)\m}^{(\r)} \T_{-n(r),\n}^{(\k)}
I_{\srac{\nb}{\r (\s)}} \left( \srac{z_\r}{z_\k},0 \right)
+\T_{n(r)\m}^{(\k)} \T_{-n(r),\n}^{(\r)}
I_{\srac{\nb}{\r (\s)}} \left( \srac{\bz_\r}{\bz_\k},0 \right) \right)
\label{st3} \\
& = & F(\r,\k) + \Delta \label{st4}
\end{eqnarray}
\begin{equation}
\label{st5}
\Delta \equiv \G^{\nrm; \mnrn} (\s)( 1 - \d_{\nb ,0} )
\left( \T_\nrm^{(\r)} \T_\mnrn^{(\k)} +\T_\nrm^{(\k)} \T_\mnrn^{(\r)} \right)
I_{\srac{\nb}{\r (\s)}} (\infty,0) \ .
\end{equation}
\end{subequations}
Here we have used the identity \eqref{Iid} and the periodicity
$n(r) \rightarrow n(r) \pm \r(\s)$ of $\G$ and $\T$ to obtain \eqref{st1}
and the variable change $ n' \equiv \r - n  $ to obtain \eqref{st2}.
Finally \eqref{st3}, \eqref{st4} and \eqref{st5} are obtained with the
symmetry of $\G$ and a $\m \leftrightarrow \n$ relabelling.

The same steps on the first term of $\Delta$ are used to show that $\Delta=0$
\begin{subequations}
\begin{eqnarray}
& & \G^{\nrm; \mnrn} (\s)( 1 - \d_{\nb ,0} )
  \T_\mnrn^{(\k)} \T_\nrm^{(\r)} I_{\srac{\nb}{\r (\s)}} (\infty,0)\nn \\
= & &  \G^{n(r)-\r(\s),\m;\r(\s)-n(r),\n} (\s)( 1 - \d_{\nb ,0} )
  \T_{\r(\s)-n(r),\n}^{(\k)} \T_{n(r)-\r(\s),\m}^{(\r)}
I_{\srac{\r(\s)-\nb}{\r (\s)}} ( 0,\infty ) \\
= & &\G^{-n(r),\m; \nrn} (\s)( 1 - \d_{\nb ,0} )
  \T_\nrn^{(\k)} \T_{-n(r) ,\m}^{(\r)} I_{\srac{\nb}{\r (\s)}} (0,\infty) \\
= & & - \G^{\nrm ; \mnrn} (\s)( 1 - \d_{\nb ,0} )\T_\nrm^{(\k)}
 \T_{\mnrn}^{(\r)} I_{\srac{\nb}{\r (\s)}} (\infty,0)
\end{eqnarray}
\end{subequations}
which establishes the symmetry of $F$.

\section{Relation to the Dynkin automorphisms \label{Dyn} }

Our first exercise in this appendix is to show explicitly that the
standard action $\tau$ of the Dynkin diagram automorphism of $\su(3)$
\begin{subequations}
\label{su3cw}
\begin{equation}
\tau (\a_1) =\a_2 \sp \tau (\a_2) = \a_1 \sp
\tau (\a \cdot H) = \a \cdot \tau (H) = \tau (\a)\cdot H
\end{equation}
\begin{equation}
\tau (E_{\pm \a_1}) = E_{\pm \a_2} \sp \tau (E_{\pm \a_2}) = E_{\pm \a_1} \sp
\tau (E_{\pm (\a_1 +\a_2) }) = -E_{\pm (\a_1+\a_2)}
\end{equation}
\begin{equation}
\label{su3cwd}
\a_1 = ( 1,0) \sp \a_2 = \srac{1}{2} ( -1, \sqrt{3})
\end{equation}
\end{subequations}
is inner-automorphically equivalent $\tau \simeq \w$ to the action of the outer automorphism
$\w$ in the Cartesian basis of the text. As seen in \eqref{su3cwd}, we take
$\a^2=1$ for our discussion of $\su(3)$.

We present this equivalence in a set of relations
\begin{subequations}
\label{su3cur}
\begin{eqnarray}
(-) \;\;\;\; J_1 &=& \tilde J_5 =\srac{1}{i\sqrt{2}}(E_{\a_1+\a_2}-
E_{-(\a_1+\a_2)}) \\
(+) \;\;\;\; J_2  &=& \srac{1}{2}(\tilde J_3 + \sqrt{3} \tilde J_8)
=(\a_1+\a_2) \cdot H \\
(-) \;\;\;\; J_3 & =&\tilde J_4 =\srac{1}{\sqrt{2}}(E_{\a_1+\a_2}+
E_{-(\a_1+\a_2)}) \\
(-) \;\;\;\; J_4 &=& \srac{1}{\sqrt{2}} (\tilde J_7 -\tilde J_2) =\srac{i}{2}
(E_{\a_1}- E_{\a_2}-
(E_{-\a_1}- E_{-\a_2} )) \\
(+) \;\;\;\; J_5 &=& \srac{1}{\sqrt{2}} (\tilde J_1 +\tilde J_6) =\srac{1}{2}
(E_{\a_1}+ E_{\a_2}+
E_{-\a_1}+ E_{-\a_2} ) \\
(-) \;\;\;\; J_6 &=& \srac{1}{\sqrt{2}} (\tilde J_1 -\tilde J_6) =\srac{1}{2}
(E_{\a_1}- E_{\a_2}+
E_{-\a_1}- E_{-\a_2} ) \\
(+) \;\;\;\; J_7 &=& \srac{1}{\sqrt{2}} (\tilde J_2 +\tilde J_7) =\srac{1}{2i}
(E_{\a_1}+ E_{\a_2}-(
E_{-\a_1}+ E_{-\a_2}) ) \\
(-) \;\;\;\; J_8 & = & \srac{1}{2}(\sqrt{3} \tilde J_3 - \tilde J_8)
=\frac{1}{\sqrt{3}} (\a_1-\a_2) \cdot H
\end{eqnarray}
\end{subequations}
whose derivation and meaning we discuss below.

In \eqref{su3cur} the generators $J\hc = J$ are the Cartesian generators
of the text, while the generators $\tilde J\hc
= \tilde J$ are another set of  Cartesian generators related to the
Cartan-Weyl generators as follows:
\begin{subequations}
\begin{equation}
E_{\pm \a_1} = \srac{1}{\sqrt{2}}(\tilde J_1 \pm i \tilde J_2) \sp
E_{\pm \a_2} = \srac{1}{\sqrt{2}}(\tilde J_6 \pm i \tilde J_7) \sp
E_{\pm (\a_1 + \a_2)} =\srac{1}{\sqrt{2}}(\tilde J_4 \pm i \tilde J_5)
\end{equation}
\begin{equation}
\sp H = (\tilde J_3,\tilde J_8)
\sp E_\a\hc= E_{-\a} \sp H\hc = H \ .
\end{equation}
\end{subequations}
The $\tilde J \leftrightarrow (E,H)$ relations  in \eqref{su3cur} give
the $\tilde J$ form of the hermitian eigenstates of the Cartan-Weyl generators
under $\tau$. The signs $(\pm)$ shown on the far left of \eqref{su3cur}
are the eigenvalues of all the combinations under $\tau \simeq \w$.

The relation between the $J$'s and the $\tilde J$'s is in fact an
inner automorphism of $\su(3)$. To see this explicitly, consider
the unitary matrix
\begin{equation}
\label{unitm}
U = \frac{e^{-i \pi/6}}{\sqrt{2}}
\left( \begin{array}{ccc}
i & 1 & 0 \\ 0 & 0 & \sqrt{2} \\ i & -1 & 0 \end{array} \right)
\sp U\hc U =\one \sp {\rm det}\,U = 1
\end{equation}
Then we find by explicit computation that
\begin{eqnarray}\label{jjtrel}
 & & U\hc \left\{ \frac{\l_1 + \l_6}{\sqrt{2}} ,
\frac{\l_2 + \l_7}{\sqrt{2}} ,
 \frac{\l_3 + \sqrt{3} \l_8}{2}, \l_4 ,
\frac{\sqrt{3} \l_3 - \l_8}{2}, \frac{\l_1 - \l_6}{\sqrt{2}},
\frac{\l_7 - \l_2}{\sqrt{2}} , \l_5 \right\} U \nn \\
  & & \hskip 2cm = \{ \l_5,\l_7,\l_2,
 \l_3,\l_8,\l_6,\l_4, \l_1 \}
 \end{eqnarray}
where $\l_a$, $a=1\ldots 8$ are the $3\times 3$ Gell-Mann matrices. This result
shows that the  relation in \eqref{su3cur} between the $J$'s and $\tilde J$'s
is an automorphism, and moreover  this automorphism is an inner automorphism
because $U$ in \eqref{unitm} is an element of the Lie group $SU(3)$.

To see this for all representations of $\su(3)$, note that the matrix $U$ can be
written in the form $U = \exp (i \be \cdot T)$ for $T$ the fundamental of
$\su(3)$. Then the relations \eqref{jjtrel} imply the same forms as
operator relations with $U \rightarrow \tilde U = \exp (i \beta \cdot J)$
and $\l_a \rightarrow J_a$ for any set of Cartesian generators,
including $J$ or $\tilde J$.

Two further comments are in order. We mention first that the
matrix $U$ in \eqref{unitm} is nothing but a particular matrix of
eigenvectors of $\l_4$. Repeating the computation with the most
general matrix of eigenvectors of $\l_4$ gives a more general
form of \eqref{jjtrel} which shows the inner-automorphic
equivalence  of all the irregularly embedded $\so(3)$'s in
$\su(3)$. Second, the $J \leftrightarrow (E,H)$ relations in \eqref{su3cur}
can also be used to check the
equivalence of our Cartesian form of $A_2^{(2)}$ in \eqref{A22}
with the conventional form of $A_2^{(2)}$ obtained directly from
the action of the Dynkin automorphism. One need only hat  the
operators in the $J \leftrightarrow (E,H)$ relations
\begin{subequations}
\begin{equation}
J \rightarrow \hj \sp E_{\pm(\a_1 + \a_2)} \rightarrow \hat E_{\pm(\a_1 + \a_2)}
\sp H \rightarrow \hat H
\end{equation}
\begin{equation}
E_{\pm \a_1} + E_{\pm \a_2} \rightarrow \widehat{E_{\pm \a_1} + E_{\pm \a_2}}
\sp
E_{\pm \a_1} -E_{\pm \a_2} \rightarrow \widehat{E_{\pm \a_1} - E_{\pm \a_2}}
\end{equation}
\end{subequations}
being careful as shown to maintain the eigenstates of $\tau$.

For $\su(2n+1)$ it is well known \cite{KacB} that there is a
realization of the Dynkin automorphism with invariant subalgebra $h=\so(2n+1)$.
This is easily checked starting from the trivial phase
$\tau (E_{\a_i})= E_{\tau (\a_i)}$ for all simple roots.

For $\su(2n)$ however it is known \cite{KacB} that
there are two inner-automorphically
equivalent realizations of the Dynkin automorphism, resulting in
a choice of invariant subalgebra $h=\so(2n)$ or $h=\mathfrak{c}_n$.
This is not difficult to check explicitly, starting from the positive roots
\begin{equation}
\a_{ij} = e_i - e_j \sp e_i \cdot e_j = \de_{ij} \sp \a_{ij}^2 =2 \sp 1 \leq i < j \leq 2n
\end{equation}
(with simple roots $\a_{i,i+1}$) and their corresponding generators
$E_{ij} \equiv E_{\a_{ij}}$:
\begin{equation}
[E_{ik}, E_{kj} ] = E_{ij} \sp
\tau (\a_{i,i+1}) = \a_{2n-i,2n-i+1} \sp \tau(\a_{ij}) = \a_{2n-j+1,2n-i+1} \ .
\end{equation}
The $n$ invariant positive roots are $\a_{i,2n-i+1}$, $i= 1 \ldots n$,
including the invariant simple root $\a_{n,n+1}$.

By comparison with $[\tau(E_{ik}),\tau(E_{kj})]= \tau(E_{ij})$ we find that
the general automorphism has the form
\begin{equation}
\label{gauto}
\tau(E_{ij})= \xi_{ij} E_{2n-j+1,2n-i+1} \sp \xi_{ij} = (-)^{j-i-1}
\left( \prod_{l=1}^{j-1} \xi_{l,l+1} \right)
\end{equation}
where $\xi_{i,i+1}$ is the phase of the simple root generator $E_{i,i+1}$.
Moreover, the form \eqref{gauto} is unique given the ``initial condition''
$\{ \xi_{i,i+1} \}$. The two realizations of the Dynkin automorphism
are then found by distinguishing the sign of the invariant simple root
\begin{equation}
\label{xich}
\begin{array}{rclcl}
{\rm I.} & \forall & \xi_{i,i+1} = 1 & \rightarrow & h = \mathfrak{c}_n \\
{\rm II.} & \forall & \xi_{i,i+1} = 1 \;\, \mbox{except}\,\; \xi_{n,n+1}=-1
& \rightarrow & h = \so(2n) \ .
\end{array}
\end{equation}
As far as counting is concerned, the difference between the two cases is the
sign of the $n$ invariant positive roots
\begin{equation}
\tau(E_{i,2n-i+1}) = \xi_{i,2n-i+1} E_{i,2n-i+1} \sp
\xi_{i,2n-i+1} = \left\{ \begin{array}{ll}
+1 & \mbox{for}\;\, {\rm I} \\
-1 & \mbox{for}\;\, {\rm II}
\end{array} \right.
\sp i = 1 \ldots n \ .
\end{equation}
The same is found for the $n$ invariant negative roots, and the relation
\begin{equation}
{\rm dim}(\mathfrak{c}_n) = {\rm dim}(\so(2n))+ 2n = n (2n+1)
\end{equation}
reflects this difference between cases I and II.
Finally, it is easily checked with \eqref{xich} and the fundamental
representation of $\su(2n)$ that the composition of automorphisms I and II is an inner
automorphism of $\su(2n)$.

\section{Flat connections in the charge conjugation orbifolds \label{flch}}

To check explicitly that the twisted KZ connection \eqref{kzcc} is flat, we will
need only the symmetric-space form ($f_{IJK} =0$) of
the algebra \eqref{TAB}, \eqref{TAI}, \eqref{TIJ} of the twisted representation matrices.
In further detail, we will use only the following list of simple identities
\begin{subequations}
\begin{equation}
\label{ida}
\frac{1}{2 z_{\m\n} \sqrt{z_\m z_\n}} \pm \frac{1}{z_{\m\n}^2}
\left( \frac{z_\n}{z_\m} \right)^{\pm \srac{1}{2}} =
\pm \frac{z_\m + z_\n}{2 z_{\m\n}^2 \sqrt{z_\m z_\n}}
\end{equation}
\begin{equation}
\label{idb}
[ \T_{0A}, \sum_I \T_{1I} \T_{1I} ] =0
\end{equation}
\begin{equation}
\label{idc}
\frac{1}{z_{\m \r} z_{\n \r}} =\frac{1}{z_{\m \r} z_{\n \m}} +
\frac{1}{z_{\m \n} z_{\n \r}}
\end{equation}
\begin{equation}
\label{idd}
\frac{1}{z_{\m \r} z_{\n \r}} \frac{z_\r}{\sqrt{z_\m z_\n}}
-\frac{1}{z_{\m \r} z_{\n \m}} \left( \frac{z_\m}{z_\n} \right)^{\srac{1}{2}}
- \frac{1}{z_{\m \n} z_{\n \r}}\left( \frac{z_\n}{z_\m} \right)^{\srac{1}{2}}=0
\end{equation}
$$
\sum_{IJA} i f_{IJA} \Big(\T_{1I}^{(\n)} \T_{1J}^{(\n)} \T_{0A}^{(\m)}
- (\m \leftrightarrow \n) \Big)
= \frac{1}{2} \sum_{IJA} i f_{IJA} \Big( [ \T_{1I}^{(\n)} ,\T_{1J}^{(\n)}]
\T_{0A}^{(\m)} -(\m \leftrightarrow \n) \Big)
$$
\begin{equation}
= \frac{1}{2} \sum_{IJAB} i f_{IJA}  i f_{IJB} \Big( \T_{0B}^{(\n)}
\T_{0A}^{(\m)} -(\m \leftrightarrow \n) \Big) =0 \label{ide}
\end{equation}
$$
 \sum_{IJA} i f_{AIJ} \Big(\T_{0A}^{(\m)} \T_{1J}^{(\n)} \T_{1I}^{(\m)}
+\T_{1J}^{(\m)} \T_{1I}^{(\n)} \T_{0A}^{(\m)} - (\m \leftrightarrow \n) \Big)
 =   \sum_{IJA} i f_{AIJ} \Big( [ \T_{0A}^{(\m)} , \T_{1J}^{(\n)} \T_{1I}^{(\m)}]
-(\m \leftrightarrow \n) \Big)
$$
\begin{equation}
    =  \sum_{IJAK} i f_{AIJ}  i f_{AIK} \Big(  \T_{1J}^{(\n)} \T_{1K}^{(\m)}
-(\m \leftrightarrow \n) \Big) =0  \label{idf}
\end{equation}
\end{subequations}
where $\m,\n$ and $\r$ are distinct. The identity \eqref{idd} follows from
\eqref{idc} and many identities of this form were needed
\cite{deBoer:2001nw} in the explicit check of flatness of the twisted KZ
connections of the WZW cyclic permutation orbifolds. The identities
\eqref{idb}, \eqref{ide} and \eqref{idf} are simple consequences of the
symmetric-space form of the algebra of twisted representation matrices.

Then by differentiation and \eqref{ida} we find that
\begin{equation}
\partial_\m \hat W_\n - \partial_\n \hat W_\m = \frac{2}{2k+Q_g}
\left[ \sum_A \T_{0A}^{(\m)} \T_{0A}^{(\n)}
+ \frac{z_\m + z_\n}{2 \sqrt{z_\m z_\n}} \sum_I \T_{1I}^{(\m)} \T_{1I}^{(\n)}
\right]- (\m \leftrightarrow \n)  =0
\end{equation}
In what follows, we sketch the more difficult check that
\begin{equation}
[ \hat W_\m , \hat W_\n] =0 \ .
\end{equation}
For this computation, we introduce the symbolic notation
\begin{equation}
\hat W_\m = h_\m + (g/h)_\m + e_\m
\end{equation}
where $h$ denotes the untwisted subalgebra terms and so on. Then we find using
\eqref{ida}, \eqref{idb} and \eqref{idc} that
\begin{equation}
[e_\m , e_\n] = [ e_\m,h_\n] = [h_\m, e_\n] = [h_\m,h_\n] = 0 \ .
\end{equation}
The last relation in this list is not surprising because $h$ is an untwisted subalgebra.

After some algebra, the rest of the terms can be divided into summed terms
$\sum_{\r \neq \m,\n}$ and unsummed terms which involve only $z_\m$ and
$z_\n$. For example one finds that the commutator
\begin{eqnarray}
[ (g/h)_\m,(g/h)_\n] & =  &\sum_{IJA} i f_{IJA} \sum_{\r \neq \m,\n}
\left\{
\frac{1}{z_{\m \r} z_{\n \m}} \left( \frac{z_\r}{z_\n} \right)^{\srac{1}{2}}
\T_{1I}^{(\r)} \T_{0A}^{(\m)} \T_{1J}^{(\n)} \right. \nn \\
 & & \left.+ \frac{1}{z_{\m \r} z_{\n \r}} \frac{z_\r}{\sqrt{z_\m z_\n}}
\T_{0A}^{(\r)} \T_{1J}^{(\n)} \T_{1I}^{(\m)}
+ \frac{1}{z_{\m \n} z_{\n \r}}\left( \frac{z_\r}{z_\m} \right)^{\srac{1}{2}}
\T_{1J}^{(\r)} \T_{0A}^{(\n)} \T_{1I}^{(\m)} \right\} \hskip 1cm \label{comex}
\end{eqnarray}
has no unsummed terms, which have cancelled according to \eqref{ide}. Using \eqref{idc}
and \eqref{idd} the summed terms \eqref{comex} are found to cancel against the
summed terms from $[h_\m,(g/h)_\n] + [(g/h)_\m,h_\n]$.
This leaves the unsummed terms from this sum of commutators plus those
from $[e_\m,(g/h)_\n] + [(g/h)_\m,e_\n]$. Using \eqref{idf}, we find
that all these terms sum to zero
\begin{equation}
-\frac{1}{z_{\m\n}} \frac{z_\m +z_\n}{2 \sqrt{z_\m z_\n}}
\sum_{IJA} i f_{IJA}( \T_{0A}^{(\m)} \T_{1J}^{(\n)} \T_{1I}^{(\m)}
+\T_{1J}^{(\m)} \T_{1I}^{(\n)} \T_{0A}^{(\m)} ) =0
\end{equation}
which completes the check of flatness.

One may also check  the consistency relation between the twisted KZ equation
\eqref{twkzeq} and the global Ward identity \eqref{glwieq}
\begin{equation}
[ \hat W_\m (\T,z), \sum_{\m=1}^N \T_{0A}^{(\m)}] =0 \sp
\forall \; A \in \so(n)
\end{equation}
but this is left as an exercise for the reader.

\vskip .5cm
\addcontentsline{toc}{section}{References}

\renewcommand{\baselinestretch}{.4}\rm
{\footnotesize

\providecommand{\href}[2]{#2}\begingroup\raggedright\endgroup

}
\end{document}